\documentclass[12pt]{article}

\usepackage{natbib}
\usepackage{graphicx}
\usepackage{authblk}
\usepackage[margin=0.5in]{geometry}

\begin{document}


\title{Surveying the reach and maturity of machine learning and artificial intelligence in astronomy}



\author[{1,2}]{Christopher J.~Fluke}
\author[2]{Colin Jacobs}


\affil[1]{Advanced Visualisation Laboratory, DRICP, Swinburne University of Technology, Hawthorn, Victoria, 3122, Australia,
cfluke@swin.edu.au}
\affil[2]{Centre for Astrophysics \& Supercomputing, Swinburne University of Technology, Hawthorn, Victoria, 3122, Australia}




\maketitle

\begin{abstract}
Machine learning (automated processes that learn by example in order to classify, predict, discover or generate new data) and artificial intelligence (methods by which a computer makes decisions or discoveries that would usually require human intelligence) are now firmly established in astronomy.  Every week, new applications of machine learning and artificial intelligence are added to a growing corpus of work.  Random forests, support vector machines, and neural networks (artificial, deep, and convolutional) are now having a genuine impact for applications as diverse as discovering extrasolar planets, transient objects, quasars, and gravitationally-lensed systems, forecasting solar activity, and distinguishing between signals and instrumental effects in gravitational wave astronomy.  This review surveys contemporary, published literature on machine learning and artificial intelligence in astronomy and astrophysics.  Applications span seven main categories of activity: classification, regression, clustering, forecasting, generation, discovery,  and the development of new scientific insight. These categories form the basis of a hierarchy of maturity, as the use of machine learning and artificial intelligence emerges, progresses or becomes established. 
\end{abstract}

\section{Introduction}
\label{sct:introduction}

Astronomy has a rich history of data gathering and record keeping \citep{Brunner2002,Feigelson2012,Jaschek68,Jaschek78,Zhang15}.  Data about and from celestial objects is collected using an assortment of telescopes, photon detectors and particle detectors.  While all of the electromagnetic spectrum is of interest, the bulk of observational data comes from the visible/infrared (wavelengths from 400 nm to 1 mm) and radio (wavelengths from 1 cm to 1 km) portions of the spectrum.  Much of this data is recorded in the form of two-dimensional pixel-based images and one-dimensional spectra.    Secondary data products are derived from observational data, often as catalogues of individual source properties: position, size, mass, chemical composition, and so forth.   Observational data can be recorded and analyzed at a single epoch, or the properties of astronomical sources -- especially brightness and position -- can be monitored over time.

Complementing observational data gathering are the dual fields of numerical simulation and astrophysical theory, although there is a great deal of overlap between the two.  While some branches of theory do not in themselves produce large quantities of data, focusing instead on mathematical descriptions of cosmic phenomena, computer simulations generate data that can be used to model, predict, and support analysis of the observational data.   

Modern astronomical data is measured in Terabytes, Petabytes and, soon, Exabytes.  When astronomy crossed the 100 Terabyte scale near the end of the 20th century \citep[e.g.,][]{Brunner2002,Szalay2001} a new data-driven astronomy emerged: where data mining was mooted as the likely future \citep{Bell2009,Ivezic2014,Szalay-Gray-2006}. The expectation was an increased reliance on automated systems to locate, classify, and characterise objects.  At the same time, new fundamental relationships between derived properties might be found, by allowing clever algorithms to search through complex, multi-dimensional data catalogues \citep[e.g.][]{Graham13}.

\subsection{Data mining}
The background and early history of data mining in astronomy is covered in some detail by \citet{borne_scientific_2009}, \cite{Ball2010}, and the collection of articles in \citet{Way2012}.   An early emphasis of data mining was to find new samples of rare sources, by applying workflows that gathered data from large, often online, repositories. \citet{Lepine05} applied a software blink comparator to 615,800 sub-fields downloaded from the Digitized Sky Survey, identifying and cataloging 61,977 stars with high proper motions (Dec > 0$^\circ$). Targeting the planned SkyMapper \citep{Keller07} Southern Sky Survey, \citet{Walsh07} mined the Sloan Digitzed Sky Survey (SDSS) Data Release 5  \citep[DR5;][]{Adelman06} for stellar over-densities, uncovering a new Milky Way dwarf galaxy satellite (Bo{\"o}tes II). \citet{Gonzalez08} opened up their INT/WFC Photometric H$\alpha$ Survey of the Northern Galactic Plane (IPHAS) dataset through the AstroGrid Virtual Observatory Desktop,\footnote{{\tt http://www.astrogrid.org}}  with the goal of making  their 200 million-object photometric catalog available for data mining. Discoveries resulting from exploration of the IPHAS initiative included new samples of young stars \citep{Vink08}, planetary nebulae \citep{Viironen09}, and galactic supernova remnants \citep{Sabin13}. Virtual Observatory infrastructure was also utilized by \citet{Chilingarian09} in a workflow to identify a sample of compact elliptical galaxies.  Candidates were selected by leveraging a combination of resources including imaging data from the VizieR Catalogue Service\footnote{{\tt http://vizier.u-strasbg.fr}} at the Centre de Données Astronomiques de Strasbourg, the NASA/IPAC (National Aeronautics and Space Administration/Infrared Processing and Analysis Center) Extragalactic Database (NED)\footnote{{\tt http://nedwww.ipac.caltech.edu}}, the Hubble Legacy Archive\footnote{{\tt http://hla.stsci.edu}}, and photometric and spectroscopic results from SDSS Data Release 7 \citep[DR7;][]{Abazajian09}. SDSS catalogues also played a role in projects such as the identification of dwarf novae candidates \citep{Wils10}, found by cross-matching DR7 with an astrometric catalogue from the Galaxy Evolution Explorer (GALEX) space mission \citep{Martin05}.  The need for practical data mining infrastructure led to the development of tools such as DAMEWARE \citep[DAta Mining \& Exploration Web Application and REsource;][]{Brescia14,Brescia16}.
See also \citet{Ivezic2014} and the AstroML Python module\footnote{{\tt https://www.astroml.org}} for a selection of data mining implementations, with an emphasise on large-scale observational surveys.

\subsection{The emergence of machine learning and artificial intelligence in astronomy}
The value of automated data mining as an approach to knowledge discovery in astronomy has been firmly  established across a broad range of sub-disciplines of astronomical interest.  Within many fields of astronomy, though, the discussion of data mining is evolving rapidly to focus almost exclusively on {\em machine learning} (ML; automated processes that learn by example in order to classify, predict, discover or generate new data) and, to a lesser extent,  {\em artificial intelligence} (AI; methods by which a computer makes decisions or discoveries that would usually require human intelligence).

The developing use of ML and AI in astronomy has mirrored the broader use in computer science and the scientific community. Traditional statistical techniques found application first. Principal component analysis (PCA) was used, for instance: in the 1980s for morphological classification of spiral galaxies \citep{whitmoreObjectiveClassificationSystem1984}; in the 1990s for quasar detection \citep{francisObjectiveClassificationScheme1992} and stellar spectral classification \citep{singhStellarSpectralClassification1998}; and in the 2000s for galaxy classification \citep{conseliceFundamentalPropertiesGalaxies2006} and quasar detection in the Sloan Digital Sky Survey \citep{yipSpectralClassificationQuasars2004}. PCA is now a standard technique, which continues to be used in hundreds of astronomy projects and papers per year. 

By the early 1990s, astronomers begin to take advantage of more complex methods requiring labelled training sets. In the 1990s, decision Trees (DTs) began to be employed for tasks such as star-galaxy separation \citep{weirSKICATSystemProcessing1995} and galaxy morphology classification \citep{ AutomatedMorphologicalClassification,owensUsingObliqueDecision1996}. By the 2000s, use of the technique proliferated and random forests (RFs) begin to dominate, with a key application being photometric redshift estimation  
 \citep{carrascokindTPZPhotometricRedshift2013}.
Boosted decision tree techniques, such as AdaBoost, appeared in more recent years and continue to be used, including for assignment of
photometric redshifts \citep{hoyleFeatureImportanceMachine2015} and for
star-galaxy separation \citep{sevilla-noarbeEffectTrainingCharacteristics2015}.

Support Vector Machines (SVMs) also found application in the 2000s and beyond, for instance in the detection of red variable stars \citep{p.r.wozniakIdentifyingRedVariables2004}, determination of photometric redshifts \citep{wadadekarEstimatingPhotometricRedshifts2004}, prediction of solar flares \citep{qahwajiAutomaticShortTermSolar2007}, star-galaxy separation \citep{fadelySTARGALAXYCLASSIFICATIONMULTIBAND2012}, and noise analysis in gravitational wave detection \citep{biswasApplicationMachineLearning2013}.

One of the dominant machine learning techniques, the artificial neural networks (ANN), appeared in the field at the end of the 1980s \citep{angelAdaptiveOpticsArray1990, rosenthalApplyingArtificialIntelligence1988} and by the 1990s was applied across a broad range of problems in astronomy. Early applications included star-galaxy separation \citep{odewahnAutomatedStarGalaxy1992}, galaxy morphology classification \citep{lahavNeuralComputationTool1996,storrie-lombardiMorphologicalClassificationGalaxies1992}, and object detection in the staple astronomical software Source Extractor \citep[SExtractor;][]{bertinSExtractorSoftwareSource1996a}. By the 2000s, ANNs were playing a key role in photometric redshift estimation \citep{collister04,firthEstimatingPhotometricRedshifts2003,  vanzellaPhotometricRedshiftsMultilayer2004}, galaxy classification \citep{ballGalaxyTypesSloan2004} and the detection of gamma ray bursts  \citep[GRBs;][]{ballGalaxyTypesSloan2004}. Paving the way for the ``Deep Learning'' era, the use of ANNs in astronomy has accelerated over the last decade, for instance in the analysis of asteroid composition \citep{leonObservationsCompositionalPhysical2010}, pulsar detection \citep{eatoughSelectionRadioPulsar2010}, and finding gravitationally lensed quasars \citep{agnelloDataMiningGravitationally2015a}.

Two strongly linked occurrences have had a significant impact on the growth of adoption of ML and AI in astronomy.   First was the appearance of graphics processing units (GPUs) as affordable, massively parallel computational accelerators, with applicability to a wide range of computationally-demanding problems [see, for example, \citet{Barsdell2010,Fluke2011} for adoption strategies in astronomy].  Secondly was the emergence of deep neural networks and convolutional neural networks.  These approaches -- extensions to the ``vanilla'' ANN -- benefit from GPU acceleration to perform computationally arduous calculations  in parallel in a reasonable time and at relatively low cost.  For data-rich fields, such as astronomy, the predictive performance of these deep learning networks improves as more data is provided for training and tuning. 

In the field of computer vision -- the computational analysis of image data -- the use of deep neural networks also accelerated after the spectacular demonstration by \citet{krizhevskyImageNetClassificationDeep2012} of the power of convolution neural networks applied to classifying images of millions of everyday objects. Astronomers were quick to take advantage of this revolution, with \citet{dieleman_rotation-invariant_2015} and \citet{huertas-companyCATALOGVISUALLIKEMORPHOLOGIES2015} achieving human-level performance on galaxy morphology classification and \citet{hoyleMeasuringPhotometricRedshifts2016a} demonstrating the possibility of estimating photometric redshifts directly from images using convolutional neural networks.

Recent applications in astronomy utilizing ML and AI include: the discovery of extrasolar planets \citep{Pearson18,Shallue18} and gravitationally-lensed systems \citep{Jacobs17,Lanusse18,Pourrahmani18}; discovery and classification of transient objects \citep{Connor18,Farah18,Mahabal19}; forecasting solar activity \citep{Florios18,Inceoglu18,Nishizuka17}; assignment of photometric redshifts within large-scale galaxy surveys \citep{Bilicki2018,Ruiz18,Speagle17}; and the classification of gravitational wave signals and instrumental noise \citep{George18b,George18a,Powell17}.

\subsection{Scope and structure}
\label{sct:scope}

This advanced review surveys progress in the wide-scale adoption of machine learning and artificial intelligence within astronomy, as evidenced by a collection of recently published works.  
Techniques are not explained and referenced in detail, expect with respect to their particular adoption in astronomy.    Most advances in astronomical and astrophysical knowledge have relied on a relatively small number of general methods (see Section \ref{sct:methods}).

The primary avenue for identifying relevant literature was NASA's Astrophysics Data System\footnote{http://ui.adsabs.harvard.edu} \citep{Chyla15,Kurtz00} -- a knowledge discovery tool without equal. 
To gather an extensive and representative  current sample of publications, abstracts of published, peer reviewed journal articles were probed for key words such as ``machine learning'', ``artificial intelligence'', ''neural networks'', and ``data mining''.  Bayesian methods are intentionally omitted, as these align more naturally with traditional statistical methods \citep{Heck1985}.  Such an approach does miss important research results, so no claim is made as to the exhaustiveness or completeness of the review.  Progress in the adoption of ML and AI in astronomy is occurring rapidly.  However, through the qualitative examination of $\sim 200$ refereed publications from 2017 to February 2019, using an approach sharing elements with Grounded Theory,\footnote{Grounded Theory is a qualitative analysis strategy, used, for example, in education research and social sciences, where a sequence of reader-assigned codes is applied to allow identification and tracking of themes within a sample of relevant literature. Codes do not have to be selected in advance, but are defined dynamically in multiple iterations through the literature.} a broad collection of astronomy applications has been assessed such that common themes have emerged regarding the reach and maturity of ML and AI in astronomy.  

Indicative examples are drawn from the recent published literature to highlight how ML and AI techniques are used across seven categories of activity (Section \ref{sct:methods}) and three phases of maturity (Section \ref{sct:applications}).   It is important to remember that a successful use of a machine learning algorithm is more likely to be reported than one where a method failed to work.  Counter examples or cautionary tales \citep[e.g.][]{Connor18}, where an algorithm may not have performed as well as hoped, are rare.

\section{Machine learning and artificial intelligence in astronomy}
\label{sct:methods}
Data-driven scientific discovery occurs through a combination of statistical methods, machine learning and artificial intelligence techniques, and the use of database systems.\footnote{As proposed by \citet{Ball2010}, a sufficiently flexible definition for a database in astronomy is ``{\em any machine-readable astronomical data}''.}  Scientific discovery requires techniques for identifying patterns within datasets (the original scope of {\em data mining}) as part of a multi-stage process for selecting, cleaning, processing, and transforming raw data into useful knowledge (i.e., {\em knowledge discovery in databases}, as described in \citet{Fayyad1996}).  

As highlighted in Section \ref{sct:introduction}, global astronomy data collections are approaching the Exabyte scale.  A key motivation for many applications of ML and AI to astronomical data is the need to prepare for the data streams expected from near-term observatories and space missions.  The Large Synoptic Survey Telescope \citep{Ivezic19,LSST09}; the Euclid satellite \citep{Laureijs11,Amendola18}; MeerKAT \citep{Booth09}; the Australian Square Kilometre Array Pathfinder \citep{Johnston07,Johnston08}; and the Square Kilometre Array \citep{Dewdney09}, among others, will all generate datasets on scales (volumes and velocities) that vastly exceed the discovery capabilities of humans.  In the interim, the Sloan Digital Sky Survey \citep[SDDS;][]{York00,Stoughton02,Abazajian09}, 

the Panoramaic Survey Telescope and Rapid Response System \citep[Pan-STARRS;][]{Kaiser04}, the Catalina Real-Time Transient Survey  \citep[CRTS;][]{Drake09,Mahabal11} and the Zwicky Transient Facility \citep[ZTF;][]{Bellm19},
the Kilo Degree Survey \citep[KiDS;][]{dejong13} and the Fornax Deep Survey \citep{Iodice16}, both using the the VLT Survey Telescope\footnote{These last two survey projects provide data products for the SUrvey Network for Deep Imaging Analysis \& Learning (SUNDIAL) which is building  inter-disciplinary teams of astronomers, computer scientists and industry partners.  See {\tt https://www.astro.rug.nl/~sundial/}} ,
LOFAR \citep{VanHaarlem13}, the Solar Dynamic Observatory \citep[SDO;][]{Lemen12,Pesnell12}, the Kepler Planet-Detection Mission \citep{Borucki10}, and the GAIA space mission \citep{GAIA16b,GAIA16a,GAIA18}, are generating data with which ML and AI has enabled classification, regression, forecasting, and discovery, leading to new knowledge and new insights.    

\subsection{The nature of the data}
\label{sct:nature}
The applicability and efficacy of any ML or AI technique depends on the nature of the data.  \citet{Brunner2002} classified astronomical data into five domains, extended slightly here to allow a clearer connection to the specific applications of ML and AI within astronomy.   In this section, references are given to recent works that apply ML and AI to each of the data types, rather than to the originator(s) of the data type.

{\em Images} are pictures of astronomical objects (usually as a pixel grid of numerical intensity values), such that the appearance informs a classification \citep[e.g.,][]{Aniyan17,Xin17,Dominguez18, Kuminski18,Ma2019} or provides insight about physical processes that are occurring \citep{Muller18}.  For visible/infrared (IR) observations (i.e. optical astronomy), light passes through, and is focused by, a telescope's optical system to be captured on a charge-coupled device (CCD).  Various filters are used to select only specific regions of the visible/IR spectrum.  For radio observations, it is common to refer to the frequency bandwidth over which flux is recorded from a particular location in the sky, with most radio images created from interferometers using the technique of aperture synthesis.
    
{\em Spectroscopy} refers to wavelength-dependent numerical intensity measurements over a finite range of wavelengths (or frequencies), but often with very high resolution.  Spectroscopy provides information on the atomic and molecular composition, from which other physical properties (temperature, density, metallicity, etc.) can be inferred \citep[e.g.,][]{Li18,Marquez18,Miettinen18,Ucci18}.

A special case at the intersection of imaging and spectroscopy is the spectral cube \citep[e.g.,][]{Araya18,Bron18}.  This is a volumetric dataset comprising a sequence of images, each captured over a very narrow wavelength or frequency range.  When looking for structures within a spectral cube, it is treated as an image.  When extracting a spectrum at a fixed spatial location, it is treated as a spectroscopic data product.  Spectroscopic data cubes, with their high dimensionality, may prove a challenge for established machine learning methods to handle. Convolutional neural networks have proved to work robustly on astronomical image data in several photometric bands, so there is no theoretical obstacle to extending this to thousands of spectroscopic frequencies. However, the practical challenges are yet to be fully explored.

{\em Photometry} is concerned with accurate measurements of the brightness (i.e. intensity, luminosity, flux) of an object recorded through a filter.  It is a secondary, numerical data product derived from a calibrated image.  Comparisons between photometric measurements through different filters are often used as an alternative to detailed spectroscopic observations, with specific application to determining the distance to a celestial source 
[e.g. \citet{Cavuoti17b,Morrison17,Beck18,Bilicki2018}].

Images, spectroscopic, and photometric measurements can be made as a function of time.    For optical astronomers, a {\em light curve} is time-based photometry, where the variation in intensity of a source over time helps with the identification and classification of a variety of variable star types \citep[e.g.][]{Cohen17,Naul18,Papageorgiou18} or indicates the presence of otherwise unseen objects, such as an extrasolar planet 
\citep[e.g.][]{Mislis18,Pearson18,Shallue18}.   In this review, other time-based measurements in radio astronomy [e.g. pulsar and transient object searches - \citet{Connor18,Michilli18,Pang18,Tan18,Farah18}], and the emerging field of observational gravitational wave astronomy \citep{Powell17,Zevin17,George18b,George18a}, are categorized as {\em time series}.

The end product of many data gathering programs is a {\em catalogue}, which comprises one or more numerical or categorical data types. Some may be derived from the standard data gathering approaches introduced above, while others are calculated or otherwise derived -- including through ML \citep[e.g.][]{Marchetti17,Tachibana18}.  Within a catalogue, {\em astrometry} refers to the accurate measurement of the spatial locations of objects,on the celestial sphere or with respect to an alternative coordinate system \citep{Castro-Ginard18,Gao18a,Gao18b}. Most objects are reported with at least one measurement of position. Some local objects, such as the Solar System's planets and minor planets, and the growing number of stars within the reach of the GAIA space mission, move with respect to the coordinate system over time [\cite{Chen18,Lin18}]. A {\em morphological classification} places a particular type of object into an object-based category where a common physical process (or set of processes) is thought to drive the appearance of an object \citep[e.g.][]{Aniyan17,Dominguez18,Kuminski18,Ma2019}.

Finally, although a single catch-all name does not do such a diverse field justice, {\em simulation} will be used to describe the data products from any numerical or computational method.  For example, cosmological simulations \citep[e.g.][]{Agarwal18,Hui18,Lucie18,Nadler18,Rodriguez18} follow the gravity-induced formation and growth of structures, requiring approximations to various physical mechanisms, a suitable choice of initial conditions, and a strategy for time-based evolution (down to some minimum level of accuracy).

\subsection{From classification to insight}
\label{sct:7methods}
For the science of astronomy to progress through the use of ML and AI, it must be possible to demonstrate that the outcome is not merely an automated classification or a numerical prediction, but that astronomers are using the data mining phase to discover new objects or generate new insights into the underlying physical processes and relationships.   In assessing the recently published literature, seven categories emerge pertaining to how ML and AI are used in astronomy (cf. \citet{Fayyad1996} and \citet{Zhang15}, who identify similar categories).    The last two categories (discovery and insight) are those where a higher order scientific outcome arises.

\begin{enumerate}
    \item {\bf Classification}: Categories or labels are applied to objects or features.   Based on a training set (labelled or unlabelled), the machine learning algorithm learns the characteristics that relate an instance to a category.  When applied to a new instance, the algorithm assigns the most likely category label.

    \item {\bf Regression}: Assignment of a numerical value (or values) based on the characteristics that are learnt or otherwise predicted by the machine learning algorithm.  As with classification, a training set may be used or the characteristics may be inferred from the dataset.

    \item {\bf Clustering}: These algorithms determine whether an object or a feature is part of (i.e. a member of) something.  This might be a physical structure or association -- as in the more familiar usage of the term in astronomy as applied to open, globular, or galactic clusters -- or a region within an $N$-dimensional parameter space.   

    \item {\bf Forecasting}:  The purpose of the machine learning algorithm is to learn from previous events, and predict or forecast that a similar event is going to occur.  There is an implicit time-dependence to the prediction.
    
    \item {\bf Generation and Reconstruction}: Missing information is created, expected to be consistent with the underlying truth. 
    The cause of the missing information might be due to the presence of noise, processing artefacts, or additional astronomical phenomena, all of which conspire to obscure the required signal.
    
    \item {\bf Discovery}: New celestial objects, features or relationships are identified as a consequence of the application of a ML or AI method.  
    
    \item {\bf Insight}: Moving beyond the discovery of celestial objects, new scientific knowledge is demonstrated as a consequence of applying machine learning or AI.  This includes cases where insight is gained into the suitability of applying machine learning, choice of data set, hyperparameters, and comparisons with human-based classification.
\end{enumerate}

Classification, regression and clustering processes are often presented as a comparison with a similar human-centred approach -- but with a need to ``scale-up'' in terms of either the size of the dataset to be explored or ``speed-up'' the time taken to achieve the task.   Classification and regression outcomes can either be the end-point of an investigation, or the input to a forecasting, generation, discovery or insight process.  In Section \ref{sct:applications}, these categories will be used to make an assessment of the maturity of adoption of ML and AI in various sub-fields of astronomy.

A subset of discovery is the field of anomaly or outlier detection. Many of the most exciting discoveries to come are likely to lie among the ``unknown unknowns'' in new areas of parameter space captured within the Petascale and Exascale datasets of the future. New methods are being developed to find anomalous objects in astronomical datasets, such as the work by \citet{baronWeirdestSDSSGalaxies2016} using an unsupervised Random Forest to find outliers amongst SDSS galaxies.
Promising avenues involve a combination of unsupervised learning methods, such as isolation forests \citep{liuIsolationForest2008}; dimensionality reduction, such as PCA, t-Distributed Stochastic Neighbor Embedding \citep[t-SNE;][]{maatenVisualizingDataUsing2008}, e.g. \citet{reisDetectingOutliersLearning2018a} and \citet{nakonecznyCatalogQuasarsKiloDegree2019}, self-organizing maps \citep[SOMs;][]{kohonenSelforganizingMap1990}, e.g. \citet{carrascokindSOMzPhotometricRedshift2014} and \citet{armstrongK2VariableCatalogue2016}, or the latent space of a variational encoder \citep[e.g.][]{yangAutoencoderStellarSpectra2015, maMachineLearningBased2019}. Novel visualization techniques are also contributing.
For instance, \citet{mastersMAPPINGGALAXYCOLOR2015}, using self-organizing maps to vizualize the distribution of galaxies in photometric color space, were able to identify regions that were under-sampled spectroscopically and develop an optimal strategy for the Euclid mission's photometric redshift calibration efforts.

\begin{table}
\caption{ From a qualitative examination of a sample of $\sim$ 200 refereed publications from 2017 to February 2019, a mapping emerges between the nature of astronomical data and the way that machine learning and artificial intelligence is actively been pursued.  The table presents a qualitative summary of the categories of ML/AI algorithms and the most common types of astronomical data in the sample of publications.} 

  \label{tab:natureandtype}
    \centering
    \begin{tabular}{c|c|c|c|c|c|c|c}
{\bf Nature}/{\bf Type}	&	\rotatebox{90}{{\bf Classification}} &
\rotatebox{90}{{\bf Regression}} &
\rotatebox{90}{{\bf Clustering}} &
\rotatebox{90}{{\bf Forecasting}} &
\rotatebox{90}{{\bf Generation}} &
\rotatebox{90}{{\bf Discovery}} &\rotatebox{90}{{\bf Insight}} \\ \hline
Image	&	\textbullet	&	\textbullet	&	\textbullet	&	\textbullet	&	\textbullet	&	\textbullet	&	\textbullet	\\
Spectroscopy	&	\textbullet	&	\textbullet	&	\textbullet	&		&		&	\textbullet	&	\textbullet	\\
Photometry	&	\textbullet	&	\textbullet	&	\textbullet	&	\textbullet	&		&	\textbullet	&	\textbullet	\\
Light curve	&	\textbullet	&	\textbullet	&		&		&		&	\textbullet	&	\textbullet	\\
Time Series	&	\textbullet	&	\textbullet	&	\textbullet	&		&		&	\textbullet	&	\textbullet	\\
Catalogue	&	\textbullet	&	\textbullet	&	\textbullet	&	\textbullet	&		&	\textbullet	& \textbullet		\\
Simulation	&	\textbullet	&	\textbullet	&		&		&	\textbullet	&	\textbullet	&		\\ \hline
\end{tabular}
\end{table}

Based on a qualitative examination of a sample of $\sim$ 200 refereed publications from 2017 to February 2019, Table \ref{tab:natureandtype} summarises the mapping between the most common astronomical data types with the seven categories of ML/AI methods.   Classification and regression algorithms are being applied to all of the data types.  Although clustering activities span a number of data types, overall they were not common outside of studies of stellar clusters \citep{Castro-Ginard18,Gao18a,Gao18b} or segmentation processes \citep{Bron18,Yang18}.

Forecasting outcomes were mostly confined to images [e.g. \citet{Mukkavilli18} with Mars images; \citet{Liu17b}, \cite{Nishizuka17}, \citet{Inceoglu18}, and \citet{Florios18} using Solar magnetograms], photometric measurements [\citet{French18} predicted likely tidal disruption events in post-starburst galaxies using a random forest algorithm] and catalogue data [forecasts of coronal mass ejections from the Sun based on $\sim 180$ similar events using a support vector machine (SVM) \citet{Liu18}].   Generation methods, in particular generative adversarial networks, have been used with images from observations \citep{Vavilova18}, and to simplify or remove the need for expensive numerical simulation  \citep[e.g.][]{Diakogiannis19, Rodriguez18,Fussell19}.   

With regards to the role of ML and AI in advancing knowledge in astronomy, there was clear evidence  from the sample of recent publications that discovery tasks are being performed with all of the data types: images \citep{Hartley17,Ciuca17,Gomez18,Pourrahmani18,Lanusse18,Wan18,Morello18,Jacobs17}; spectroscopy \citep{Bu17,Li18}; photometry \citep{Timlin18,Vida18,Ostrovski17};  light-curves \citep{Armstrong18,Pena18,Cohen17,Hedges18,Heinze18,vanRoestel18,Giles19}; time-series \citep{Michilli18,Tan18,Pang18,Connor18,Farah18,Morello19}; catalogues \citep{Nguyen18,Yan18,Lin18,Marchetti17}; and simulation: \citep{Xu17a,Nadler18,Kuntzer17}. 

For most data types,  outcomes have progressed to the insight phase, for example: enhanced understanding of human biases in classification of Galaxy Zoo project images \citep{Peng18,Cabrera18}; determination of the evolution of the effective radius and stellar mass of Kilo Degree Survey \citep[KiDS;][]{dejong13} galaxies based on photometric redshifts derived from ML \citep{Roy18}; and 
new relationships between physical and envrionmental properties of galaxies by applying an SVM to the results of a cosmological simulation \citep{Hui18}.

\subsection{Techniques}  
Machine learning algorithms are usually classified as being either supervised or unsupervised.  Supervised methods rely on a pre-labeled dataset, which is used to help train and tune the algorithm.  This learning allows for new instances to be assigned a label (classification) or numerical value (regression).  Unsupervised methods allow the data to speak for itself, but do not necessarily make use of any existing knowledge.   Although there is no shortage of data in astronomy, there is often a paucity of relevant pre-labeled data.  For example, when the discovery of rare events is the target of an observational program, it is very difficult to train a network on sufficient examples and counter-examples.  Moreover, astronomical discovery does rely on serendipity -- anomalous cases that are potentially unlike anything that has previously been examined, and hence no exemplars exist \citep{Norris17}.  

The lion's share of machine learning in astronomy is performed with five classes of algorithms: artificial neural networks; convolutional neural networks; decision trees; random forests; and support vector machines.  These are primarily used as supervised learning algorithms.    Since the \citet{Ball2010} review, convolutional neural networks are the only new method  amongst these five to emerge and reach wide-spread usage in astronomy.

\textbf{Artificial neural networks} \citep[ANNs;][]{Rosenblatt1957, Fukushima1980} are the key technique behind the recent AI boom, but date back to the 1950s. They are designed by analogy to a biological neuron, with signals from multiple inputs weighted and added together; a biological neuron sends an electrical signal if these weighted input signals cross a certain activation threshold. In the case of an artificial neuron (``perceptron''), the inputs and trainable weights are vectors of real valued numbers, and the output is a scalar value. A single artificial neuron can be employed as a classifier, however they are typically combined in ANNs where the outputs of an array (layer) of neurons form the inputs to a subsequent layer. At the output layer, the values of one or more neurons are interpreted according to the problem domain. Optimization of the weights using a labelled training set follows the gradient descent paradigm with the backpropagation algorithm \citep{LeCun1989}.

\textbf{Support vector machines} \citep[SVMs;][]{Cortes1995}, similarly to ANNs, learn non-linear decision boundaries in spaces of arbitrary dimension, finding a hyperplane in a space of arbitrary dimensionality that distinctly separates the supplied data. The algorithm works by finding the hyperplane with the maximum separation between extreme examples (support vectors). A linear SVM (LSVM) finds the optimal hyperplane for the supplied input features, but using the so-called `kernel trick' a SVM projects the data into higher dimensions where the data is linearly separable.

\textbf{Decision trees} \citep[DTs;][]{Quinlan1986} perform classification by recursive binary splitting of the data, learning through the training process a series of decisions based on features of the input data. The root node (the entire data space) is repeatedly split into two child nodes based on the most discriminative feature of the data, until at the leaf node a category is determined.
\textbf{Random forests} \citep[RFs;][]{Ho1995, Breiman2001} are an extension of decision trees, improving accuracy by constructing an ensemble of decision trees, trained on subsets of the training data \citep[bagging;][] {Breiman1996} and/or feature set (feature randomness), and using the median or mode of the ensemble as the final output value. \textbf{AdaBoost} \citep{Freund1995} is another ensemble method that weights the contribution of each decision tree based on misclassifications; similarly,  \textbf{gradient boosting} \citep{Friedman2001} uses decision trees trained in sequence on the residual errors of other DTs.

A convolutional neural network (CNN) is an extension of the simple ANN but with many hidden layers (i.e. a deep neural network).  CNNs are characterized by their use of convolutional layers, which are sensitive to specific features -- usually within images -- that may have undergone transformations through translation, rotation, or scaling.   Working in conjunction with pooling layers, which reduce the spatial size of image features within the network, the final stage of a CNN is often a fully-connected ANN to generate a classification or numerical prediction.

CNNs have now been used in astronomy for a variety of image-based classification, regression and discovery activities. They appear in the literature as: {\em binary classifiers} [\citet{Gieseke17, Jacobs17} and \citet{Shallue18}], where the training sets comprise two distinct categories representing ``present'' and ``not present'' examples; {\em morphological classifiers} [\citet{Aniyan17,Dominguez18,Gonzalez18,Huertas18} and \citet{Ma2019}], where there are multiple categories that have been determined previously, usually by human inspection, but also using other machine learning approaches \citep{Kim17}; and for {\em detection}, with applications to discovery of exoplanet candidates in light-curves \citep{Pearson18}, or real-time discovery of transient objects \citep{Connor18} and gravitational wave events \citep{George18b,George18a}.

Outside of this core group, are several unsupervised methods: $k$-nearest neighbours ($k$-NN), $k$-means clustering, and the DBSCAN method.  The latter has been used as a discovery tool to improve efficiency at detecting exoplanet transits from light curves, based on the recovery of artificial transit signatures \citep{Mislis18}, and in partnership with an ANN to identify open clusters \citep{Castro-Ginard18} in the GAIA DR2 \citep{GAIA18} -- an example of a clustering process.  

Other techniques that have been investigated, often in conjunction with one or more the above methods, include:
AdaBoost \citep{Xin17,Bethapudi18}; genetic algorithms \citep{Sarro18}; self-organizing maps \citep{Armstrong17,Suveges17,Armstrong18};  recurrent neural networks \citep{Naul18}; auto-encoders \citep{Vincent08,Sedaghat18};
and transfer learning \citep{Benavente17}.
 Falling within the generation and reconstruction category (Section \ref{sct:7methods}), generative adversarial networks (GANs) are likely to be the next most significant machine learning approach for astronomy.  Early applications of GANs include generating dark matter structures in cosmological simulations \citep{Rodriguez18,Diakogiannis19}, the creation of realistic images of galaxies as an input to weak gravitational lensing analysis \citep{Fussell19}, and deblending overlaps between foreground and background galaxies in highly-crowded images \citep{Reiman19}.

Machine learning has also been used to identify the astrophysical features most significant for classification. For example, in the area of photometric redshift estimation, \citet{polstererImprovingPerformancePhotometric2014} used GPUs to conduct an exhaustive feature search of over 341,000 feature combinations to identify the four most significant ones; and  \citet{hoyleFeatureImportanceMachine2015a}, who used Random Forests with AdaBoost to select the top photometric features to increase the performance of ANN-based redshift estimators. \citet{Frontera17} used denoising autoencoders for unsupervised feature learning from galaxy spectral energy distributions (SEDs).

\begin{table}
    \caption{
   From a qualitative examination of a sample of $\sim$ 200 refereed publications from 2017 to February 2019, a mapping emerges between the nature of astronomical data and the types of machine learning and artificial intelligence algorithms that are being applied.  The table presents a summary of the types of astronomical data and the algorithms that appeared most regularly.  The purpose of the table is to provide a convenient starting point for selecting an algorithm that has been used successfully for each data type. }
    \label{tab:dataandmethod}
    \centering
    \begin{tabular}{c|c|c|c|c|c|c|c|c|c}
{\bf Data/Method}	&	{\bf ANN}	&	{\bf CNN}	&	{\bf GAN}	&	{\bf SVM}	&	{\bf DT}	&	{\bf RF}	&	{\bf DBSCAN}	&	{\bf k-NN} & {\bf k-M}	\\ \hline 
Image	&	\textbullet	&	\textbullet	&	\textbullet	&	\textbullet	&	\textbullet	&	\textbullet	&		&	\textbullet	 & \\  
Spectroscopy	&	\textbullet	&	\textbullet	&			&	\textbullet	&		&	\textbullet	&		&	& \textbullet	\\  
Photometry	&	\textbullet	&		&		&		&	\textbullet	&	\textbullet	&	\textbullet	&		&	 \textbullet	\\  

Light curve	&		&		\textbullet	&		&		&		&	\textbullet	&		&	&	\\  
Time Series	&	\textbullet	&	\textbullet	&		&		&	\textbullet	&	\textbullet	&	\textbullet	&	&	\\  
  Catalogue&	\textbullet			&		&		&	\textbullet	&	\textbullet	&	\textbullet	&	\textbullet	&	\textbullet	 & \\  
Simulation	&	\textbullet		&	\textbullet	&	\textbullet	&	\textbullet	&		&	\textbullet	&		&	 &	\\  
\hline
    \end{tabular}

ANN = Artificial Neural Network; CNN = Convolutional Neural Network; 
GAN = Generative\\
Adversarial Network; SVM = Support Vector Machine; DT = Decision Tree; RF = Random Forest;\\
DBSCAN = Density-based spatial clustering of applications with noise; k-NN = k-Nearest Neighbours; \\
k-M = k-means clustering
\end{table}

Table \ref{tab:dataandmethod} summarises the relationships found between the main types of astronomical data and specific techniques in the sample of  recently-published papers. It is expected that specific techniques have been applied to other data types,  and it is important to remember that some fields will have trialled particular methods and moved on as new alternatives appear.  The Table's purpose is to emphasise areas of current activity and interest only, and thus provides a starting point for astronomers wishing to adopt a ML/AI approach by matching the data types to the methods.   

ANNs \citep{Ciuca17,Marchetti17,Bethapudi18,Bilicki2018,Fujimoto18,Ho19}, random forest methods  \citep{Schindler17,Goulding18,Hedges18,Reis18,Pang18,Tachibana18,Nadler18}, and SVM algorithms  \citep{Hartley17,Hui18,Kong18,Yan18,Zhang18} have been used extensively across most data types. CNNs are more suitable for image-style data (see above), although they have been used successfully with one-dimensional light curves \citep[][identification and ranking of transiting exoplanet candidates in Kepler light curves, including the discovery of two new exoplanets]{Shallue18}  and  time series  \citep[][identification of gravitational wave signatures within noisy time series data -- a solution that scales better than template matching as the number of templates grows ]{George18b,George18a}.   DBSCAN \citep{Castro-Ginard18} and k-NN \citep{Smirnov17} have been used to find structures in multi-dimensional catalogues.  Given the prevalence of imaging data in astronomy, it is not surprising that images are being analyzed with the largest range of ML/AI methods. 

There is still plenty of scope for studies  that perform structured comparisons between multiple methods.  This can occur more easily when reference datasets are made accessible to the community.  For example, the availablility of the PHoto-z Accuracy Testing datasets \citep{Hildebrandt10} allowed \citet{Cavuoti12} and \citet{Brescia13} to establish the efficacy of a multi-layer perceptron (i.e. neural network) method coupled with the Quasi Newton Algorithm (MLPQNA) at assigning photometric redshifts for galaxies and quasars respecitvely.  Training on the PHAT-1 spectroscopic sample, MLPQNA out-performed alternative statistical and neural network-based methods
[e.g. ANNz; \citet{collister04}] with regards to bias\footnote{the mean of $\Delta z \equiv \left(z_{\rm spec} - z_{\rm phot}\right)/\left(1 + z_{\rm spec}\right)$, where $z_{\rm spec}$ and $z_{\rm phot}$ are the known spectroscopic and predicted photometric redshifts respectively.} for all objects, bright objects, and distant vs near objects in the PHAT-1 sample \citep{Cavuoti12}.  Studies into the accuracy and validity of ML-based photometric redshifts continues today -- see, for example, \citet{Almosallam16}, \citet{Cavuoti17b} and \citet{Amaro19}.

Looking more broadly, these systematic comparisons tend to be occurring more often in solar astronomy than in other disciplines [e.g. \citet{Nishizuka17,Florios18} and \citet{Inceoglu18}], although see \citet{Ksoll18,Paschenko18}, and \citet{Zhang18} for examples pertaining to stellar and variable star classifications.   While certain disciplines have adopted specific methods, experimentation with emerging techniques is on-going [e.g. probabilistic random forests and transfer learning \citet{Reis19}].

\section{Assessing the maturity of adoption}
\label{sct:applications}
The seven categories introduced in Section \ref{sct:7methods} allow an assessment of the maturity of the use of machine learning and artificial intelligence within a sub-field of astronomy, as they represent a loose hierarchy of sophistication. 
The common starting point is to apply a machine learning technique to perform a classification, regression or  clustering task.   Once established as being comparable to, or exceeding, a more traditional approach, machine learning can be used to forecast likely future outcomes [e.g. solar flares \citep{Nishizuka17,Florios18} or coronal mass ejections from the Sun \citep{Inceoglu18}] or make new discoveries [e.g. classification schemes for stellar types permitting the identification of new candidates of rare objects as in \citet{Bu17}, \citet{vanRoestel18}, and \citet{Zhang18}].  The most mature disciplines move beyond classification and discovery as ends in their own to that of gaining insight -- where new physical knowledge is identified, often for the first time, because a machine learning approach was used.   

The hierarchy of categories is used to assess the maturity of ML and AI within a set of sub-fields of astronomy as one of {\em  emerging}, {\em progressing}, or {\em established}.  In all cases, the reader should refer to the highlighted works in order to understand the scientific background, historical context for the establishment of a particular method, and the technical details of the data mining, machine learning or artificial intelligence approach that was applied.  

\subsection{Emerging}
The emerging stage is applied to sub-fields of astronomy and astrophysics that are starting to investigate the use of ML and AI, often by tackling the ``low-hanging fruit''.   This includes a problem that requires a classification or regression approach, or through a comparison between machine learning and an alternative, established method.  While some of the emerging disciplines show evidence of reaching the discovery and insight phases, the approaches are not as firmly established, or the size of the community is small.  Emerging fields include: 
\begin{itemize}
    \item {\em Planetary studies}. ML-based identification and classification of clouds, dust storms and surface features on Mars \citep{Gichu19}, with potential to forecast future dust storms \citep{Mukkavilli18}, and the discovery of previously unknown impact craters \citep{Xin17} using the AdaBoost algorithm.
    \item {\em Non-stellar components of the Milky Way}.   The primary components of the Milky Way are stars (see below), dust and gas, which can be concentrated in atomic and molecular clouds or more diffusely in the interstellar medium.   \citet{Ucci18} developed the {\tt GAME} (GAlaxy Machine learning for Emission lines) code to study physical properties of the interstellar medium.   Segmentation and clustering algorithms are used to identify individual components of clouds of atomic and molecular gas within the Milky Way.   \citet{Bron18} used the Meanshift clustering algorithm to identify regions within molecular clouds that based on physical/chemical properties, instead of seeking purely spatial connections; \citet{Denes18} used ML to determine individual Guassian components of the Riegel-Crutcher cloud, based on 21-cm neutral hydrogen (H{\sc i}) observations of extra-galactic continuum sources behind the cloud complex; and SVM was used by \citet{Yan18} to classify, and hence select, H{\sc ii} regions (gas clouds comprised mostly of singly-ionized hydrogen) from the Infrared Astronomical Satellite (IRAS) Point Source Catalogue.  Time-consuming human classifications of Milky Way ``bubbles', caused by stellar feedback within molecular clouds, was enhanced through ML \citep{Xu17a}. A random forest method was used by \citet{Chen19} to determine the amount of dust reddening of more than 56 million stars, with application to future GAIA datasets.  Using a GAN, \citet{Vavilova18} demonstrated how to generate missing parts of the cosmological large-scale structure that are obscured by the Milky Way's zone of avoidance.  
    \item {\em Stellar clusters}.  Application of the density-based DBSCAN algorithm to the Gaia Data Release 2 (DR2) has lead to the discovery of new open clusters \citep{Castro-Ginard18}, with an ANN used to separate real clusters from spatial over-densities of stars.  Multi-dimensional clustering processes have been used to determine the components of several open clusters, including M67 \citet{Gao18a} and NGC188 \citep{Gao18b}.
    \item {\em Instrumentation}. \citet{Li18b} presented an AI solution for identifying faults for a telescope drive system.  Through an automated expert system, a series of self-healing decisions are made until an appropriate solution is found.   The knowledge base is updated in real-time with human input for faults that have not previously been diagnosed or corrected.   
\end{itemize}

Other emerging disciplines include: information retrieval  systems, matching queries regarding specific instruments \citep{Mukund18}; identification of cosmic strings in all-sky maps [e.g. \citet{Ciuca17} using a neural network; \citet{Vafaei18}]; and the detection  and classification of cosmic ray events \citep{Krause17,Zhao18}.

\subsection{Progressing}
Characteristics of disciplines identified as progressing in their use of ML and AI include a broader variety of techniques being applied, or a particular technique is used multiple times, or there is an immediate move to the forecasting, discovery or insight phases.  Sub-fields at this stage of maturity include:
\begin{itemize}
    \item {\em Solar System objects}.  Due to their relative proximity to the Earth, the motion of Solar System objects is key to their discovery.  ML and AI have enabled removal, or reduction, of false detections from the moving object detection pipeline in the Subaru/Hyper-Suprime-Cam Strategic Survey Program \citep{Lin18}, with applications to Trans-Neptunian Objects \citep{Chen18};  and detection and classification of asteroids \citep{Erasmus17,Erasmus18,Smirnov17}.  \citet{Duev19} trained a CNN to discover fast-moving candidates from ZTF observations in order to more reliably identify potentially hazardous near-Earth objects.

      \item {\em Active galactic nuclei and quasars}.  A common theme in this field is the need for classification and detection methods, including assigning morphological types to radio-detected active galactic nuclei with a CNN \citep{Ma2019}, identifying blazar candidates in the Fermi-LAT (3LAC) Clean Sample \citep{Kang19}, detecting rare high-redshift, extremly luminous quasars \citep{Schindler17}, and discriminating populations of broad absorption line quasars (BALQs) from non-BALQs in  SDSS data releases \citep{Yong18}.
      \item {\em Cosmological simulations}.  ML is providing new methods for examining the outputs of cosmological simulations, leading to new insights about the connections between physical properties of galxies, dark matter halos and the cosmic environment. Examples include the use of an ANN to aid in determining the total mass of the Milky Way and the Andromeda Galaxy from the Small MultiDark simulation \citep{McLeod17}, and both classification of sub-halos \citep{Nadler18} and assignment of galaxies to halos \citep{Agarwal18} in dark matter-only simulations.
\end{itemize}

\subsection{Established}
In the established sub-fields, the use of ML and AI have become essential, a substantial body of literature exists, and the focus is mostly on forecasting, discoveries, or insight.   Here, there is no longer a need to evaluate a suitability of machine learning -- its usage has become ingrained.  Established sub-fields include:
\begin{itemize}

\item {\em Solar astronomy}.  Machine learning has been used for
classification of solar flares [e.g., \citet{Liu17}, \citet{Liu17b}, and \citet{Benvenuto18} via a hybrid method using both supervised and unsupervised methods]; clustering [e.g. \citet{Yang18} presented the simulated annealing genetic (SAG) AI method to distinguish between the umbra, penumbra and solar photosphere through a segmentation approach]; and forecasting of coronal mass ejections with a SVM \citep{Liu18}, and SVM and multilayer perceptrons \citep{Inceoglu18}.  A number of  systematic comparison studies have been conducted in order to assess which ML methods perform best at forecasting solar events.  \citet{Nishizuka17} compared three ML methods to forecast solar activity from time-based features using data from the SDO and the Geostationary Operational Environmental Satellite (GEOS) system. They determined that k-NN performed more effectively than SVM or extremely randomized trees.  For a different data set from the SDO, \citet{Florios18} found that random forests provided greater prediction accuracy.

\item {\em Extra-solar planets}. The Kepler Mission, amongst other search programs, has been a rich source of light curve data. ML and AI have enhanced and improved candidate selection \citep{Armstrong18}, and classification of light-curves, by removing false positives \citep{Armstrong17}, raising detection efficiency \citep{Mislis18} and accuracy \citep{Pearson18}, and identifying anomalies \citep{Giles19}.  ML techniques have allowed discovery of fainter candidates than were accessible with existing methods using neural network and random forest algorithms \citep{Gomez18} and a CNN \citep{Shallue18}.   Insight into candidate and confirmed extrasolar planets is also being achieved with ML and AI, such as through the determination of a ``habitability score'' for extra-solar planets \citep{Saha18} and improved model-fitting of atmospheric composition \citep{Marquez18}.

\item {\em Stars and stellar products}.  Two key activities in stellar astronomy are spectral classification [e.g. \citet{Wang17}; \citet{Garcia18} with k-means clustering;  \citet{Kong18}; classification of young stellar objects with eight different methods by \citet{Miettinen18}] and photometric classification [e.g., \citet{Ksoll18}; \citet{Zhang18} with SVM, RF and Fast Boxes].  Many new examples of specific stellar classes have been discovered, such as Wolf-Rayet stars \citep{Morello18}, blue horizontal branch stars \citep{Wan18}, hot sub dwarf stars \citep{Bu17}, and rare hypervelocity stars \citep{Marchetti17}.  ML/AI have also led to the discovery of unresolved binary stars in simulated catalogues using RF and ANN algorithms \citep{Kuntzer17}, and new pulsars, and fewer false postives, from the LOFAR Tied-Array All-Sky Survey \citep{Michilli18,Tan18}.  \citet{Fujimoto18} used an ANN to gain new insights into neutron star equation of state from numerical simulations.

\item {\em Variable stars}. Time-based photometric observations of variable stars provide extensive data sets of many millions of individual objects for mining.  There has been highly productive use of ML and AI for classification [e.g. \citet{Benavente17}, \citet{Naul18}, and \citet{Papageorgiou18}] and discovery activities [e.g. comparative study with a variety of techniques, including SVM, k-NN, neural networks and random forests by \citet{Paschenko18}; discovery of new EL CVn-type binaries from the Palomar Transient Factory \citep{vanRoestel18} supported with data from other large-scale surveys]. \citet{Valenzuela18} used an unsupervised method to overcome the limitations of available training samples, particularly when approaching new survey data, by identifying similariaties between light curves rather then relying on pre-classified examples.

\item {\em Transient object detection.} While not yet the preferred method for all observational programs, the applicability of ML and AI has been firmly established -- particularly for the real-time detection of transient objects, which allows new classes of celestial objects to be discovered.    The Catalina Real-Time Transient Survey \citep{Drake09} was utilized as a large-scale test-bed for the use of ML to aid the optical detection and monitoring of variable and transient objects \citep[e.g.][]{Mahabal09,Djorgovski2014,Djorgovski2016}. These early successes paved the way for the Zwicky Transient Factory \citep{Bellm19}. 
The multi-filter optical survey with the ZTF was designed to provide a rich exploration of the transient sky, generating hundreds of thousands of real-time candidate alerts for each night of operation.  Machine learning is fundamental to accelerating and enabling the data analysis and candidate identification (and rejection) workflows of the ZTF.  ML is being used to perform time critical tasks such as morphological star/galaxy classification \citep{Tachibana18}, binary real/bogus classification of candidates, and asteroid detection \citep{Mahabal19}, and can play a role in the brokering of alerts with application to the LSST Alert Stream \citep{Narayan18}.   In a radio-based transient object project, \citet{Farah18} used a random forest as part of the UTMOST real-time detection pipeline, leading to the discovery of Fast Radio Burst FRB170827, however, \citet{Connor18} determined that CNNs were sub-optimal for some radio transient tasks, such as reducing the need for GPU-acclerated, brute-force dedispersion of time series signals.

\item {\em Galaxies}. One of the major areas of ML application has been in the classification of galaxies from optical and radio imaging surveys. Recent examples include: neural network-based Faranoff-Riley classifications of radio galaxies \citep{Aniyan17}; automated morphological annotation and assignment \citep{Beck18,Dominguez18,Kuminski18} and labelling \citep{Hocking18} of galaxy images, including detections from radio surveys \citep{Lukic18}.  The reference point for many of the automated classifiers is the work done by human volunteers for projects like Galaxy Zoo \citep{Lintott08}.  \citet{Cabrera18} uncovered human biases that existed in morphological classification, which could be reduced through supervised ML.  Other applications included predicting the H{\sc i} content of galaxies based on optical observations \citep{Rafieferantsoa18}, determining physical properties of galaxies from their emission-line spectra \citep{Ucci17}, point source detection from radio interferometry surveys \citep{Vafaei19}, and cross-identification of sources from the Radio Galaxy Zoo \citep{Alger18}.  
Training a CNN on mock images of rare ``blue nugget'' galaxies from cosmological simulations, such objects were successfully found in an observational sample from the CANDELS survey \citep{Huertas18}.

\item {\em Distance measures}.  Estimates of the distances to galaxies, quasars and other remote celestial objects has benefited greatly from the adoption of ML.   Redshifts can be accurately inferred from photometric measurements of galaxies \citep[e.g.][]{Koo85,Koo1999,Bolzonella00}, by training on samples where spectroscopic redshifts are also available.  In general, it is more challenging to make the required spectroscopic measurements for large samples of galaxies \citep{Ball2010}, whereas many surveys are able to provide a wealth of features for ML algorithms to learn from.     Recent work has included: comparisons between Gaussian Processes and other machine learning methods -- including ANNz \citep{collister04} -- from SDSS Data Release 12 \citet{Almosallam16}; removal of anomalies from training data \citet{Hoyle15};  application of deep neural networks \citet{Hoyle16} and SVM \citet{Jones17}; and the use of k-means clustering to identify features for input to photometric redshift estimation from SDSS datasets \citep{Stensbo-Smidt17}.    Multiple machine learning methods have been utilized for determining photometric redshifts for the Dark Energy Survey Science Verification shear catalogue (DES SV)\citet{Bonnett16}.  See also \citet{Morrison17,Cavuoti17b,Leistedt17} and the review by \citet{Salvato19}. \citet{Cavuoti17} compared multiple machine learning methods with Bayesian and spectral energy template fitting, showing that ML achieved the best accuracy at prediction when there was appropriate coverage by spectroscopic templates.  \citet{Beck17} reported similar outcomes when comparing machine learning with template-fitting approaches, highlighting the ``expected bad results'' for machine learning methods when no suitable spectroscopic templates were available.  Comparison between probability density functions obtained with ANNz2 \citep{Sadeh16} and METAPHOR (Machine-learning Estimation Tool for Accurate PHOtometric Redshifts) in \citet{Amaro19}.

\item {\em Gravitational lensing}. Concentrations of matter on galactic and cosmological scales bend and deflect the path of light rays from more distance sources.  Lensing provides unique probes of dark matter distributions, tests of cosmological models, and magnified views of otherwise faint objects.  However, finding lensed systems is observationally challenging.  
ML has helped in the discovery of previously unknown lensed quasars, e.g. \citet[][Gaussian mixture models]{Ostrovski17} and \citet[][RFs/k-NN]{Timlin18}. A major challenge for deep learning methods in lens finding is paucity of training data; this has been solved using simulated lenses at galaxy scale. Deep neural nets trained on simulations have resulted in lens discoveries in survey data including
the Kilo-Degree Survey \citep[][]{dejongFirstSecondData2015}
by \citet{petrilloFindingStrongGravitational2017, petrilloLinKSDiscoveringGalaxyscale2019};
and the Dark Energy Survey \citep{darkenergysurveycollaborationDarkEnergySurvey2016}
by \citet{jacobsFindingHighredshiftStrong2019, jacobsExtendedCatalogGalaxy2019}. A strong lens finding challenge was recently conducted using simulated data \citep{metcalfStrongGravitationalLens2019}, and deep learning-based methods outperformed all other methodologies including examination by human experts. 

\item {\em Gravitational wave astronomy}
The recent detection of gravitational wave signals from coalescing black hole binaries \citep{Abbott18}, and other related compact systems, has relied on real-time computation and analysis of streams of data from the Advanced Laser Interferometer Gravitational-Wave Observatory (LIGO) detectors \citep{Harry10}.   By incorporating machine learning, \citet{Powell17} improved performance in distinguishing between sources and noise signals, along with reducing the latency of the detection pipeline.   \citet{Zevin17} used crowd-sourced categorization of common ``glitch'' signals in order to train a ML system for real-time glitch classification. \citet{George18b,George18a} developed Deep Filtering, which utilizes two CNNs for detecting signals (classification) and performing parameter estimation (regression) in real time.  Testing first on mock data, they successfully recovered events from LIGO observations.   Theoretical insight into binary black hole mergers has also been achieved through ML, training on outputs from numerical relativity simulations \citep{Huerta18}.

\end{itemize}

\section{Concluding remarks}
\label{sct:conclusions}
Every week, new astronomical applications of machine learning and artificial intelligence are added to a growing corpus of work.  Random forests, support vector machines, neural networks (artificial, deep, and convolutional), and generative adversarial networks are now having a genuine impact across all domains of astronomy.   ML and AI simplify the processes of classification and regression, determination of clustering relationships, forecasting of time-based events, and the generation or reconstruction of missing information.   As methods become more sophisticated, the volume of training data grows, and classifications become more robust, ML and AI allow for new objects to be discovered, and for new scientific insight to be gathered.   

The adoption of ML and AI is emerging in planetary studies, investigations of the non-stellar components of the Milky Way and of stellar clusters, and in real-time monitoring of instruments.   Elsewhere, rapid progress is occurring in the use of ML and AI for classification and detection of Solar System objects, and the discovery of rare types of active galactic nuclei and quasars.  ML, in particular through early experimentation with GANs, offer an intriguing alternative to generating and understanding complex structures in cosmological simulations.  ML and AI are now firmly established in: solar astronomy (particularly forecasting of solar activity); the discovery of extra-solar planets and transient objects; and classification, discovery and gaining insights into the properties of all types of stars, variable stars, and stellar evolutionary products (neutron stars, pulsars, and black holes).  ML and AI offer new ways to find and understand galaxies,  gravitationally-lensed sources, and gravitational wave candidates.

As astronomy moves ever closer to the Exascale data era of the Square Kilometre Array, an increasing number of human-centred tasks and processes are being replaced by faster, automated processing.  The adoption of ML and AI techniques is driving a fundamental change in the way future astronomers will approach the process of ``discovery''.  To date, in the vast majority of cases, discoveries have occurred when astronomers look directly at their data: qualitative inspection supported by quantitative analysis (e.g. model fitting, simulation, etc.).   The volume (e.g. number of sources or quantity of data recorded per source) and the complexity (e.g. dimensionality) of data has not vastly exceeded available computing or visual resources. It has been possible to look at the majority of potential sources and false detections by eye, and to conduct visualization and analysis using a desktop-bound workspace. This is no longer the case. Continuous human monitoring of data streams from the SKA is likely to be a tedious task, and discoveries will be missed.  Computers excel at such repetitive actions [see, for example, \citet{Yeakel18}, who investigated the role of AI as a means to reduce tedium and detect anomalies in spacecraft systems, through the Cassini-Huygens mission's study of Saturn's magnetosphere], and allow astronomers to focus their attention on interpreting and explaining new types of astronomical phenomena and their connection to fundamental physics.

As the use of deep neural networks increases in astronomy (and a great many other fields), the question arises: what is going on inside the networks? Where AI systems are making medical diagnoses and driving autonomous vehicles this may be an urgent question; but an understanding of the errors, biases and limitations is also of growing importance in a scientific context. In the computer vision realm several attempts have been made to develop techniques for visualizing and interpreting deep neural network outputs \citep[see][]{montavonMethodsInterpretingUnderstanding2018}, for instance visualizing the feature detectors or building ``saliency maps'' of the important input pixels \citep{selvarajuGradCAMVisualExplanations2017}.  However, the utility and adequacy of these methods in astronomy, where precisely quantified errors are often required, is far from apparent. Several advancements have been made, for example the use of Bayesian neural networks \citep{denkerTransformingNeuralNetOutput1991} -- where the outputs are probability distributions -- in estimating gravitational lens model errors \citep{perreaultlevasseurUncertaintiesParametersEstimated2017} or the uncertainties in neutron capture mass models \citep{utamaRefiningMassFormulas2017}. Further work in understanding the internals of deep networks will be needed if the promise of deep neural networks for astronomical discovery is to be fully realized.

As progress in artificial intelligence and machine learning accelerates, particularly through advances in deep learning, the gap between human and automated pattern recognition capabilities is narrowing.   However, it is still not always obvious why and how classifications or discoveries are made by ever more complex neural networks.  There is still scope for more studies that consider the strengths and weaknesses of different ML and AI approaches when applied to a specific dataset -- particularly as new, experimental techniques continue to appear. Learning which types of objects are harder to detect or classify also provides insight, along with highlighting potential biases in human input.  As \citet{Ball2010} stated, and as still holds true, ``there is no simple method to select the optimal algorithm to use''.  For the time being, traditional statistical methods or visualization are still highly productive first steps, providing astronomers with a detailed understanding of their data. In the future, there is no doubt that the reach and maturity of machine learning and artificial intelligence in astronomy will continue to grow.

\section*{acknowledgements}
CJF and CJ thank WIREs for the invitation and opportunity to write this advanced review, and the two anonymous referees for their insightful comments.   CJF thanks Wael Farah for helpful discussions. This research has made extensive use of NASA’s Astrophysics Data System.

\bibliography{main}{}

\begin{thebibliography}{285}
\providecommand{\natexlab}[1]{#1}
\providecommand{\url}[1]{\texttt{#1}}
\expandafter\ifx\csname urlstyle\endcsname\relax
  \providecommand{\doi}[1]{doi: #1}\else
  \providecommand{\doi}{doi: \begingroup \urlstyle{rm}\Url}\fi

\bibitem[{Abazajian} and {et al.}(2009)]{Abazajian09}
Kevork~N. {Abazajian} and {et al.}
\newblock {The Seventh Data Release of the Sloan Digital Sky Survey}.
\newblock \emph{ApJSS}, 182:\penalty0 543--558, Jun 2009.
\newblock \doi{10.1088/0067-0049/182/2/543}.

\bibitem[{Abbott} et~al.(2018){Abbott}, {Abdalla}, {Alarcon}, {Allam},
  {Andrade-Oliveira}, {Annis}, {Avila}, {Banerji}, {Banik}, {Bechtol},
  {Bernstein}, {Bernstein}, {Bertin}, {Brooks}, {Buckley-Geer}, {Burke},
  {Camacho}, {Carnero Rosell}, {Carrasco Kind}, {Carretero}, {Castander},
  {Cawthon}, {Chan}, {Crocce}, {Cunha}, {D'Andrea}, {da Costa}, {Davis}, {De
  Vicente}, {DePoy}, {Desai}, {Diehl}, {Doel}, {Drlica-Wagner}, {Eifler},
  {Elvin-Poole}, {Estrada}, {Evrard}, {Flaugher}, {Fosalba}, {Frieman},
  {Garc{\'\i}a-Bellido}, {Gaztanaga}, {Gerdes}, {Giannantonio}, {Gruen},
  {Gruendl}, {Gschwend}, {Gutierrez}, {Hartley}, {Hollowood}, {Honscheid},
  {Hoyle}, {Jain}, {James}, {Jeltema}, {Johnson}, {Kent}, {Kokron}, {Krause},
  {Kuehn}, {Kuhlmann}, {Kuropatkin}, {Lacasa}, {Lahav}, {Lima}, {Lin}, {Maia},
  {Manera}, {Marriner}, {Marshall}, {Martini}, {Melchior}, {Menanteau},
  {Miller}, {Miquel}, {Mohr}, {Neilsen}, {Percival}, {Plazas}, {Porredon},
  {Romer}, {Roodman}, {Rosenfeld}, {Ross}, {Rozo}, {Rykoff}, {Sako}, {Sanchez},
  {Santiago}, {Scarpine}, {Schindler}, {Schubnell}, {Serrano},
  {Sevilla-Noarbe}, {Sheldon}, {Smith}, {Smith}, {Sobreira}, {Suchyta},
  {Swanson}, {Tarle}, {Thomas}, {Troxel}, {Tucker}, {Vikram}, {Walker},
  {Wechsler}, {Weller}, {Yanny}, and {Zhang}]{Abbott18}
T.~M.~C. {Abbott}, F.~B. {Abdalla}, A.~{Alarcon}, S.~{Allam},
  F.~{Andrade-Oliveira}, J.~{Annis}, S.~{Avila}, M.~{Banerji}, N.~{Banik},
  K.~{Bechtol}, R.~A. {Bernstein}, G.~M. {Bernstein}, E.~{Bertin}, D.~{Brooks},
  E.~{Buckley-Geer}, D.~L. {Burke}, H.~{Camacho}, A.~{Carnero Rosell},
  M.~{Carrasco Kind}, J.~{Carretero}, F.~J. {Castander}, R.~{Cawthon}, K.~C.
  {Chan}, M.~{Crocce}, C.~E. {Cunha}, C.~B. {D'Andrea}, L.~N. {da Costa},
  C.~{Davis}, J.~{De Vicente}, D.~L. {DePoy}, S.~{Desai}, H.~T. {Diehl},
  P.~{Doel}, A.~{Drlica-Wagner}, T.~F. {Eifler}, J.~{Elvin-Poole},
  J.~{Estrada}, A.~E. {Evrard}, B.~{Flaugher}, P.~{Fosalba}, J.~{Frieman},
  J.~{Garc{\'\i}a-Bellido}, E.~{Gaztanaga}, D.~W. {Gerdes}, T.~{Giannantonio},
  D.~{Gruen}, R.~A. {Gruendl}, J.~{Gschwend}, G.~{Gutierrez}, W.~G. {Hartley},
  D.~{Hollowood}, K.~{Honscheid}, B.~{Hoyle}, B.~{Jain}, D.~J. {James},
  T.~{Jeltema}, M.~D. {Johnson}, S.~{Kent}, N.~{Kokron}, E.~{Krause},
  K.~{Kuehn}, S.~{Kuhlmann}, N.~{Kuropatkin}, F.~{Lacasa}, O.~{Lahav},
  M.~{Lima}, H.~{Lin}, M.~A.~G. {Maia}, M.~{Manera}, J.~{Marriner}, J.~L.
  {Marshall}, P.~{Martini}, P.~{Melchior}, F.~{Menanteau}, C.~J. {Miller},
  R.~{Miquel}, J.~J. {Mohr}, E.~{Neilsen}, W.~J. {Percival}, A.~A. {Plazas},
  A.~{Porredon}, A.~K. {Romer}, A.~{Roodman}, R.~{Rosenfeld}, A.~J. {Ross},
  E.~{Rozo}, E.~S. {Rykoff}, M.~{Sako}, E.~{Sanchez}, B.~{Santiago},
  V.~{Scarpine}, R.~{Schindler}, M.~{Schubnell}, S.~{Serrano},
  I.~{Sevilla-Noarbe}, E.~{Sheldon}, R.~C. {Smith}, M.~{Smith}, F.~{Sobreira},
  E.~{Suchyta}, M.~E.~C. {Swanson}, G.~{Tarle}, D.~{Thomas}, M.~A. {Troxel},
  D.~L. {Tucker}, V.~{Vikram}, A.~R. {Walker}, R.~H. {Wechsler}, J.~{Weller},
  B.~{Yanny}, and Y.~{Zhang}.
\newblock {Dark Energy Survey Year 1 Results: Measurement of the Baryon
  Acoustic Oscillation scale in the distribution of galaxies to redshift 1}.
\newblock \emph{MNRAS}, 483:\penalty0 4866--4883, Dec 2018.
\newblock \doi{10.1093/mnras/sty3351}.

\bibitem[{Adelman-McCarthy} et~al.(2006){Adelman-McCarthy}, {Ag{\"u}eros},
  {Allam}, {Anderson}, {Anderson}, {Annis}, {Bahcall}, {Baldry}, {Barentine},
  {Berlind}, {Bernardi}, {Blanton}, {Boroski}, {Brewington}, {Brinchmann},
  {Brinkmann}, {Brunner}, {Budav{\'a}ri}, {Carey}, {Carr}, {Castander},
  {Connolly}, {Csabai}, {Czarapata}, {Dalcanton}, {Doi}, {Dong}, {Eisenstein},
  {Evans}, {Fan}, {Finkbeiner}, {Friedman}, {Frieman}, {Fukugita}, {Gillespie},
  {Glazebrook}, {Gray}, {Grebel}, {Gunn}, {Gurbani}, {de Haas}, {Hall},
  {Harris}, {Harvanek}, {Hawley}, {Hayes}, {Hendry}, {Hennessy}, {Hindsley},
  {Hirata}, {Hogan}, {Hogg}, {Holmgren}, {Holtzman}, {Ichikawa}, {Ivezi{\'c}},
  {Jester}, {Johnston}, {Jorgensen}, {Juri{\'c}}, {Kent}, {Kleinman}, {Knapp},
  {Kniazev}, {Kron}, {Krzesinski}, {Kuropatkin}, {Lamb}, {Lampeitl}, {Lee},
  {Leger}, {Lin}, {Long}, {Loveday}, {Lupton}, {Margon},
  {Mart{\'\i}nez-Delgado}, {Mand elbaum}, {Matsubara}, {McGehee}, {McKay},
  {Meiksin}, {Munn}, {Nakajima}, {Nash}, {Neilsen}, {Newberg}, {Newman},
  {Nichol}, {Nicinski}, {Nieto-Santisteban}, {Nitta}, {O'Mullane}, {Okamura},
  {Owen}, {Padmanabhan}, {Pauls}, {Peoples}, {Pier}, {Pope}, {Pourbaix},
  {Quinn}, {Richards}, {Richmond}, {Rockosi}, {Schlegel}, {Schneider},
  {Schroeder}, {Scranton}, {Seljak}, {Sheldon}, {Shimasaku}, {Smith},
  {Smol{\v{c}}i{\'c}}, {Snedden}, {Stoughton}, {Strauss}, {SubbaRao}, {Szalay},
  {Szapudi}, {Szkody}, {Tegmark}, {Thakar}, {Tucker}, {Uomoto}, {Vanden Berk},
  {Vandenberg}, {Vogeley}, {Voges}, {Vogt}, {Walkowicz}, {Weinberg}, {West},
  {White}, {Xu}, {Yanny}, {Yocum}, {York}, {Zehavi}, {Zibetti}, and
  {Zucker}]{Adelman06}
Jennifer~K. {Adelman-McCarthy}, Marcel~A. {Ag{\"u}eros}, Sahar~S. {Allam}, Kurt
  S.~J. {Anderson}, Scott~F. {Anderson}, James {Annis}, Neta~A. {Bahcall},
  Ivan~K. {Baldry}, J.~C. {Barentine}, Andreas {Berlind}, Mariangela
  {Bernardi}, Michael~R. {Blanton}, William~N. {Boroski}, Howard~J.
  {Brewington}, Jarle {Brinchmann}, J.~{Brinkmann}, Robert~J. {Brunner},
  Tam{\'a}s {Budav{\'a}ri}, Larry~N. {Carey}, Michael~A. {Carr}, Francisco~J.
  {Castander}, A.~J. {Connolly}, Istv{\'a}n {Csabai}, Paul~C. {Czarapata},
  Julianne~J. {Dalcanton}, Mamoru {Doi}, Feng {Dong}, Daniel~J. {Eisenstein},
  Michael~L. {Evans}, Xiaohui {Fan}, Douglas~P. {Finkbeiner}, Scott~D.
  {Friedman}, Joshua~A. {Frieman}, Masataka {Fukugita}, Bruce {Gillespie}, Karl
  {Glazebrook}, Jim {Gray}, Eva~K. {Grebel}, James~E. {Gunn}, Vijay~K.
  {Gurbani}, Ernst {de Haas}, Patrick~B. {Hall}, Frederick~H. {Harris}, Michael
  {Harvanek}, Suzanne~L. {Hawley}, Jeffrey {Hayes}, John~S. {Hendry},
  Gregory~S. {Hennessy}, Robert~B. {Hindsley}, Christopher~M. {Hirata},
  Craig~J. {Hogan}, David~W. {Hogg}, Donald~J. {Holmgren}, Jon~A. {Holtzman},
  Shin-ichi {Ichikawa}, {\v{Z}}eljko {Ivezi{\'c}}, Sebastian {Jester}, David~E.
  {Johnston}, Anders~M. {Jorgensen}, Mario {Juri{\'c}}, Stephen~M. {Kent},
  S.~J. {Kleinman}, G.~R. {Knapp}, Alexei~Yu. {Kniazev}, Richard~G. {Kron},
  Jurek {Krzesinski}, Nikolay {Kuropatkin}, Donald~Q. {Lamb}, Hubert
  {Lampeitl}, Brian~C. {Lee}, R.~French {Leger}, Huan {Lin}, Daniel~C. {Long},
  Jon {Loveday}, Robert~H. {Lupton}, Bruce {Margon}, David
  {Mart{\'\i}nez-Delgado}, Rachel {Mand elbaum}, Takahiko {Matsubara},
  Peregrine~M. {McGehee}, Timothy~A. {McKay}, Avery {Meiksin}, Jeffrey~A.
  {Munn}, Reiko {Nakajima}, Thomas {Nash}, Jr. {Neilsen}, Eric~H., Heidi~Jo
  {Newberg}, Peter~R. {Newman}, Robert~C. {Nichol}, Tom {Nicinski}, Maria
  {Nieto-Santisteban}, Atsuko {Nitta}, William {O'Mullane}, Sadanori {Okamura},
  Russell {Owen}, Nikhil {Padmanabhan}, George {Pauls}, Jr. {Peoples}, John,
  Jeffrey~R. {Pier}, Adrian~C. {Pope}, Dimitri {Pourbaix}, Thomas~R. {Quinn},
  Gordon~T. {Richards}, Michael~W. {Richmond}, Constance~M. {Rockosi}, David~J.
  {Schlegel}, Donald~P. {Schneider}, Joshua {Schroeder}, Ryan {Scranton},
  Uro{\v{s}} {Seljak}, Erin {Sheldon}, Kazu {Shimasaku}, J.~Allyn {Smith},
  Vernesa {Smol{\v{c}}i{\'c}}, Stephanie~A. {Snedden}, Chris {Stoughton},
  Michael~A. {Strauss}, Mark {SubbaRao}, Alexander~S. {Szalay}, Istv{\'a}n
  {Szapudi}, Paula {Szkody}, Max {Tegmark}, Aniruddha~R. {Thakar}, Douglas~L.
  {Tucker}, Alan {Uomoto}, Daniel~E. {Vanden Berk}, Jan {Vandenberg},
  Michael~S. {Vogeley}, Wolfgang {Voges}, Nicole~P. {Vogt}, Lucianne~M.
  {Walkowicz}, David~H. {Weinberg}, Andrew~A. {West}, Simon D.~M. {White},
  Yongzhong {Xu}, Brian {Yanny}, D.~R. {Yocum}, Donald~G. {York}, Idit
  {Zehavi}, Stefano {Zibetti}, and Daniel~B. {Zucker}.
\newblock {The Fourth Data Release of the Sloan Digital Sky Survey}.
\newblock \emph{ApJSS}, 162\penalty0 (1):\penalty0 38--48, Jan 2006.
\newblock \doi{10.1086/497917}.

\bibitem[{Agarwal} et~al.(2018){Agarwal}, {Dav{\'e}}, and {Bassett}]{Agarwal18}
Shankar {Agarwal}, Romeel {Dav{\'e}}, and Bruce~A. {Bassett}.
\newblock {Painting galaxies into dark matter haloes using machine learning}.
\newblock \emph{MNRAS}, 478:\penalty0 3410--3422, Aug 2018.
\newblock \doi{10.1093/mnras/sty1169}.

\bibitem[Agnello et~al.(2015)Agnello, Kelly, Treu, and
  Marshall]{agnelloDataMiningGravitationally2015a}
Adriano Agnello, Brandon~C. Kelly, Tommaso Treu, and Philip~J. Marshall.
\newblock Data mining for gravitationally lensed quasars.
\newblock \emph{MNRAS}, 448\penalty0 (2):\penalty0 1446--1462, April 2015.
\newblock ISSN 0035-8711.
\newblock \doi{10.1093/mnras/stv037}.

\bibitem[{Alger} et~al.(2018){Alger}, {Banfield}, {Ong}, {Rudnick}, {Wong},
  {Wolf}, {Andernach}, {Norris}, and {Shabala}]{Alger18}
M.~J. {Alger}, J.~K. {Banfield}, C.~S. {Ong}, L.~{Rudnick}, O.~I. {Wong},
  C.~{Wolf}, H.~{Andernach}, R.~P. {Norris}, and S.~S. {Shabala}.
\newblock {Radio Galaxy Zoo: machine learning for radio source host galaxy
  cross-identification}.
\newblock \emph{MNRAS}, 478:\penalty0 5547--5563, Aug 2018.
\newblock \doi{10.1093/mnras/sty1308}.

\bibitem[{Almosallam} et~al.(2016){Almosallam}, {Jarvis}, and
  {Roberts}]{Almosallam16}
Ibrahim~A. {Almosallam}, Matt~J. {Jarvis}, and Stephen~J. {Roberts}.
\newblock {GPZ: non-stationary sparse Gaussian processes for heteroscedastic
  uncertainty estimation in photometric redshifts}.
\newblock \emph{MNRAS}, 462\penalty0 (1):\penalty0 726--739, Oct 2016.
\newblock \doi{10.1093/mnras/stw1618}.

\bibitem[{Amaro} et~al.(2019){Amaro}, {Cavuoti}, {Brescia}, {Vellucci},
  {Longo}, {Bilicki}, {de Jong}, {Tortora}, {Radovich}, {Napolitano}, and
  {Buddelmeijer}]{Amaro19}
V.~{Amaro}, S.~{Cavuoti}, M.~{Brescia}, C.~{Vellucci}, G.~{Longo},
  M.~{Bilicki}, J.~T.~A. {de Jong}, C.~{Tortora}, M.~{Radovich}, N.~R.
  {Napolitano}, and H.~{Buddelmeijer}.
\newblock {Statistical analysis of probability density functions for
  photometric redshifts through the KiDS-ESO-DR3 galaxies}.
\newblock \emph{MNRAS}, 482\penalty0 (3):\penalty0 3116--3134, Jan 2019.
\newblock \doi{10.1093/mnras/sty2922}.

\bibitem[{Amendola} et~al.(2013){Amendola}, {Appleby}, {Bacon}, {Baker},
  {Baldi}, {Bartolo}, {Blanchard}, {Bonvin}, {Borgani}, {Branchini}, {Burrage},
  {Camera}, {Carbone}, {Casarini}, {Cropper}, {de Rham}, {Di Porto}, {Ealet},
  {Ferreira}, {Finelli}, {Garc{\'\i}a-Bellido}, {Giannantonio}, {Guzzo},
  {Heavens}, {Heisenberg}, {Heymans}, {Hoekstra}, {Hollenstein}, {Holmes},
  {Horst}, {Jahnke}, {Kitching}, {Koivisto}, {Kunz}, {La Vacca}, {March},
  {Majerotto}, {Markovic}, {Marsh}, {Marulli}, {Massey}, {Mellier}, {Mota},
  {Nunes}, {Percival}, {Pettorino}, {Porciani}, {Quercellini}, {Read},
  {Rinaldi}, {Sapone}, {Scaramella}, {Skordis}, {Simpson}, {Taylor}, {Thomas},
  {Trotta}, {Verde}, {Vernizzi}, {Vollmer}, {Wang}, {Weller}, and
  {Zlosnik}]{Amendola18}
Luca {Amendola}, Stephen {Appleby}, David {Bacon}, Tessa {Baker}, Marco
  {Baldi}, Nicola {Bartolo}, Alain {Blanchard}, Camille {Bonvin}, Stefano
  {Borgani}, Enzo {Branchini}, Clare {Burrage}, Stefano {Camera}, Carmelita
  {Carbone}, Luciano {Casarini}, Mark {Cropper}, Claudia {de Rham}, Cinzia {Di
  Porto}, Anne {Ealet}, Pedro~G. {Ferreira}, Fabio {Finelli}, Juan
  {Garc{\'\i}a-Bellido}, Tommaso {Giannantonio}, Luigi {Guzzo}, Alan {Heavens},
  Lavinia {Heisenberg}, Catherine {Heymans}, Henk {Hoekstra}, Lukas
  {Hollenstein}, Rory {Holmes}, Ole {Horst}, Knud {Jahnke}, Thomas~D.
  {Kitching}, Tomi {Koivisto}, Martin {Kunz}, Giuseppe {La Vacca}, Marisa
  {March}, Elisabetta {Majerotto}, Katarina {Markovic}, David {Marsh}, Federico
  {Marulli}, Richard {Massey}, Yannick {Mellier}, David~F. {Mota}, Nelson~J.
  {Nunes}, Will {Percival}, Valeria {Pettorino}, Cristiano {Porciani}, Claudia
  {Quercellini}, Justin {Read}, Massimiliano {Rinaldi}, Domenico {Sapone},
  Roberto {Scaramella}, Constantinos {Skordis}, Fergus {Simpson}, Andy
  {Taylor}, Shaun {Thomas}, Roberto {Trotta}, Licia {Verde}, Filippo
  {Vernizzi}, Adrian {Vollmer}, Yun {Wang}, Jochen {Weller}, and Tom {Zlosnik}.
\newblock {Cosmology and Fundamental Physics with the Euclid Satellite}.
\newblock \emph{Living Reviews in Relativity}, 16:\penalty0 6, Sep 2013.
\newblock \doi{10.12942/lrr-2013-6}.

\bibitem[Angel et~al.(1990)Angel, Wizinowich, {Lloyd-Hart}, and
  Sandler]{angelAdaptiveOpticsArray1990}
J.~R.~P. Angel, P.~Wizinowich, M.~{Lloyd-Hart}, and D.~Sandler.
\newblock Adaptive optics for array telescopes using neural-network techniques.
\newblock \emph{Nature}, 348\penalty0 (6298):\penalty0 221--224, November 1990.
\newblock ISSN 1476-4687.
\newblock \doi{10.1038/348221a0}.

\bibitem[{Aniyan} and {Thorat}(2017)]{Aniyan17}
A.~K. {Aniyan} and K.~{Thorat}.
\newblock {Classifying Radio Galaxies with the Convolutional Neural Network}.
\newblock \emph{ApJSS}, 230:\penalty0 20, Jun 2017.
\newblock \doi{10.3847/1538-4365/aa7333}.

\bibitem[{Araya} et~al.(2018){Araya}, {Mendoza}, {Solar}, {Mardones}, and
  {Bayo}]{Araya18}
M.~{Araya}, M.~{Mendoza}, M.~{Solar}, D.~{Mardones}, and A.~{Bayo}.
\newblock {Unsupervised learning of structure in spectroscopic cubes}.
\newblock \emph{A\&C}, 24:\penalty0 25--35, Jul 2018.
\newblock \doi{10.1016/j.ascom.2018.06.001}.

\bibitem[Armstrong et~al.(2016)Armstrong, Kirk, Lam, McCormac, Osborn, Spake,
  Walker, Brown, Kristiansen, Pollacco, West, and
  Wheatley]{armstrongK2VariableCatalogue2016}
D.~J. Armstrong, J.~Kirk, K.~W.~F. Lam, J.~McCormac, H.~P. Osborn, J.~Spake,
  S.~Walker, D.~J.~A. Brown, M.~H. Kristiansen, D.~Pollacco, R.~West, and P.~J.
  Wheatley.
\newblock K2 variable catalogue \textendash{} {{II}}. {{Machine}} learning
  classification of variable stars and eclipsing binaries in {{K2}} fields
  0\textendash{}4.
\newblock \emph{MNRAS}, 456\penalty0 (2):\penalty0 2260--2272, February 2016.
\newblock ISSN 0035-8711.
\newblock \doi{10.1093/mnras/stv2836}.

\bibitem[{Armstrong} et~al.(2017){Armstrong}, {Pollacco}, and
  {Santerne}]{Armstrong17}
D.~J. {Armstrong}, D.~{Pollacco}, and A.~{Santerne}.
\newblock {Transit shapes and self-organizing maps as a tool for ranking
  planetary candidates: application to Kepler and K2}.
\newblock \emph{MNRAS}, 465:\penalty0 2634--2642, Mar 2017.
\newblock \doi{10.1093/mnras/stw2881}.

\bibitem[{Armstrong} et~al.(2018){Armstrong}, {G{\"u}nther}, {McCormac},
  {Smith}, {Bayliss}, {Bouchy}, {Burleigh}, {Casewell}, {Eigm{\"u}ller},
  {Gillen}, {Goad}, {Hodgkin}, {Jenkins}, {Louden}, {Metrailler}, {Pollacco},
  {Poppenhaeger}, {Queloz}, {Raynard}, {Rauer}, {Udry}, {Walker}, {Watson},
  {West}, and {Wheatley}]{Armstrong18}
David~J. {Armstrong}, Maximilian~N. {G{\"u}nther}, James {McCormac}, Alexis
  M.~S. {Smith}, Daniel {Bayliss}, Fran{\c{c}}ois {Bouchy}, Matthew~R.
  {Burleigh}, Sarah {Casewell}, Philipp {Eigm{\"u}ller}, Edward {Gillen},
  Michael~R. {Goad}, Simon~T. {Hodgkin}, James~S. {Jenkins}, Tom {Louden},
  Lionel {Metrailler}, Don {Pollacco}, Katja {Poppenhaeger}, Didier {Queloz},
  Liam {Raynard}, Heike {Rauer}, St{\'e}phane {Udry}, Simon~R. {Walker},
  Christopher~A. {Watson}, Richard~G. {West}, and Peter~J. {Wheatley}.
\newblock {Automatic vetting of planet candidates from ground-based surveys:
  machine learning with NGTS}.
\newblock \emph{MNRAS}, 478:\penalty0 4225--4237, Aug 2018.
\newblock \doi{10.1093/mnras/sty1313}.

\bibitem[Ball et~al.(2004)Ball, Loveday, Fukugita, Nakamura, Okamura,
  Brinkmann, and Brunner]{ballGalaxyTypesSloan2004}
N.~M. Ball, J.~Loveday, M.~Fukugita, O.~Nakamura, S.~Okamura, J.~Brinkmann, and
  R.~J. Brunner.
\newblock Galaxy types in the {{Sloan Digital Sky Survey}} using supervised
  artificial neural networks.
\newblock \emph{MNRAS}, 348\penalty0 (3):\penalty0 1038--1046, March 2004.
\newblock ISSN 0035-8711.
\newblock \doi{10.1111/j.1365-2966.2004.07429.x}.

\bibitem[{Ball} and {Brunner}(2010)]{Ball2010}
Nicholas~M. {Ball} and Robert~J. {Brunner}.
\newblock {Data Mining and Machine Learning in Astronomy}.
\newblock \emph{International Journal of Modern Physics D}, 19:\penalty0
  1049--1106, January 2010.
\newblock \doi{10.1142/S0218271810017160}.

\bibitem[Baron and Poznanski(2016)]{baronWeirdestSDSSGalaxies2016}
Dalya Baron and Dovi Poznanski.
\newblock The weirdest {{SDSS}} galaxies: Results from an outlier detection
  algorithm.
\newblock \emph{arXiv:1611.07526}, November 2016.

\bibitem[{Barsdell} et~al.(2010){Barsdell}, {Barnes}, and
  {Fluke}]{Barsdell2010}
B.~R. {Barsdell}, D.~G. {Barnes}, and C.~J. {Fluke}.
\newblock {Analysing astronomy algorithms for graphics processing units and
  beyond}.
\newblock \emph{MNRAS}, 408:\penalty0 1936--1944, Nov 2010.
\newblock \doi{10.1111/j.1365-2966.2010.17257.x}.

\bibitem[{Beck} et~al.(2018){Beck}, {Scarlata}, {Fortson}, {Lintott},
  {Simmons}, {Galloway}, {Willett}, {Dickinson}, {Masters}, {Marshall}, and
  {Wright}]{Beck18}
Melanie~R. {Beck}, Claudia {Scarlata}, Lucy~F. {Fortson}, Chris~J. {Lintott},
  B.~D. {Simmons}, Melanie~A. {Galloway}, Kyle~W. {Willett}, Hugh {Dickinson},
  Karen~L. {Masters}, Philip~J. {Marshall}, and Darryl {Wright}.
\newblock {Integrating human and machine intelligence in galaxy morphology
  classification tasks}.
\newblock \emph{MNRAS}, 476:\penalty0 5516--5534, Jun 2018.
\newblock \doi{10.1093/mnras/sty503}.

\bibitem[{Beck} et~al.(2017){Beck}, {Lin}, {Ishida}, {Gieseke}, {de Souza},
  {Costa-Duarte}, {Hattab}, and {Krone-Martins}]{Beck17}
R.~{Beck}, C.~A. {Lin}, E.~E.~O. {Ishida}, F.~{Gieseke}, R.~S. {de Souza},
  M.~V. {Costa-Duarte}, M.~W. {Hattab}, and A.~{Krone-Martins}.
\newblock {On the realistic validation of photometric redshifts}.
\newblock \emph{MNRAS}, 468:\penalty0 4323--4339, Jul 2017.
\newblock \doi{10.1093/mnras/stx687}.

\bibitem[Bell et~al.(2009)Bell, Hey, and Szalay]{Bell2009}
Gordon Bell, Tony Hey, and Alex Szalay.
\newblock Beyond the data deluge.
\newblock \emph{Science}, 323\penalty0 (5919):\penalty0 1297--1298, 2009.
\newblock ISSN 0036-8075.
\newblock \doi{10.1126/science.1170411}.

\bibitem[{Bellm} et~al.(2019){Bellm}, {Kulkarni}, {Graham}, {Dekany}, {Smith},
  {Riddle}, {Masci}, {Helou}, {Prince}, {Adams}, {Barbarino}, {Barlow},
  {Bauer}, {Beck}, {Belicki}, {Biswas}, {Blagorodnova}, {Bodewits}, {Bolin},
  {Brinnel}, {Brooke}, {Bue}, {Bulla}, {Burruss}, {Cenko}, {Chang}, {Connolly},
  {Coughlin}, {Cromer}, {Cunningham}, {De}, {Delacroix}, {Desai}, {Duev},
  {Eadie}, {Farnham}, {Feeney}, {Feindt}, {Flynn}, {Franckowiak}, {Frederick},
  {Fremling}, {Gal-Yam}, {Gezari}, {Giomi}, {Goldstein}, {Golkhou}, {Goobar},
  {Groom}, {Hacopians}, {Hale}, {Henning}, {Ho}, {Hover}, {Howell}, {Hung},
  {Huppenkothen}, {Imel}, {Ip}, {Ivezi{\'c}}, {Jackson}, {Jones}, {Juric},
  {Kasliwal}, {Kaspi}, {Kaye}, {Kelley}, {Kowalski}, {Kramer}, {Kupfer},
  {Landry}, {Laher}, {Lee}, {Lin}, {Lin}, {Lunnan}, {Giomi}, {Mahabal}, {Mao},
  {Miller}, {Monkewitz}, {Murphy}, {Ngeow}, {Nordin}, {Nugent}, {Ofek},
  {Patterson}, {Penprase}, {Porter}, {Rauch}, {Rebbapragada}, {Reiley},
  {Rigault}, {Rodriguez}, {van Roestel}, {Rusholme}, {van Santen}, {Schulze},
  {Shupe}, {Singer}, {Soumagnac}, {Stein}, {Surace}, {Sollerman}, {Szkody},
  {Taddia}, {Terek}, {Van Sistine}, {van Velzen}, {Vestrand}, {Walters},
  {Ward}, {Ye}, {Yu}, {Yan}, and {Zolkower}]{Bellm19}
Eric~C. {Bellm}, Shrinivas~R. {Kulkarni}, Matthew~J. {Graham}, Richard
  {Dekany}, Roger~M. {Smith}, Reed {Riddle}, Frank~J. {Masci}, George {Helou},
  Thomas~A. {Prince}, Scott~M. {Adams}, C.~{Barbarino}, Tom {Barlow}, James
  {Bauer}, Ron {Beck}, Justin {Belicki}, Rahul {Biswas}, Nadejda
  {Blagorodnova}, Dennis {Bodewits}, Bryce {Bolin}, Valery {Brinnel}, Tim
  {Brooke}, Brian {Bue}, Mattia {Bulla}, Rick {Burruss}, S.~Bradley {Cenko},
  Chan-Kao {Chang}, Andrew {Connolly}, Michael {Coughlin}, John {Cromer},
  Virginia {Cunningham}, Kishalay {De}, Alex {Delacroix}, Vandana {Desai},
  Dmitry~A. {Duev}, Gwendolyn {Eadie}, Tony~L. {Farnham}, Michael {Feeney},
  Ulrich {Feindt}, David {Flynn}, Anna {Franckowiak}, S.~{Frederick},
  C.~{Fremling}, Avishay {Gal-Yam}, Suvi {Gezari}, Matteo {Giomi}, Daniel~A.
  {Goldstein}, V.~Zach {Golkhou}, Ariel {Goobar}, Steven {Groom}, Eugean
  {Hacopians}, David {Hale}, John {Henning}, Anna Y.~Q. {Ho}, David {Hover},
  Justin {Howell}, Tiara {Hung}, Daniela {Huppenkothen}, David {Imel},
  Wing-Huen {Ip}, {\v{Z}}eljko {Ivezi{\'c}}, Edward {Jackson}, Lynne {Jones},
  Mario {Juric}, Mansi~M. {Kasliwal}, S.~{Kaspi}, Stephen {Kaye}, Michael S.~P.
  {Kelley}, Marek {Kowalski}, Emily {Kramer}, Thomas {Kupfer}, Walter {Landry},
  Russ~R. {Laher}, Chien-De {Lee}, Hsing~Wen {Lin}, Zhong-Yi {Lin}, Ragnhild
  {Lunnan}, Matteo {Giomi}, Ashish {Mahabal}, Peter {Mao}, Adam~A. {Miller},
  Serge {Monkewitz}, Patrick {Murphy}, Chow-Choong {Ngeow}, Jakob {Nordin},
  Peter {Nugent}, Eran {Ofek}, Maria~T. {Patterson}, Bryan {Penprase}, Michael
  {Porter}, Ludwig {Rauch}, Umaa {Rebbapragada}, Dan {Reiley}, Mickael
  {Rigault}, Hector {Rodriguez}, Jan {van Roestel}, Ben {Rusholme}, Jakob {van
  Santen}, S.~{Schulze}, David~L. {Shupe}, Leo~P. {Singer}, Maayane~T.
  {Soumagnac}, Robert {Stein}, Jason {Surace}, Jesper {Sollerman}, Paula
  {Szkody}, F.~{Taddia}, Scott {Terek}, Angela {Van Sistine}, Sjoert {van
  Velzen}, W.~Thomas {Vestrand}, Richard {Walters}, Charlotte {Ward}, Quan-Zhi
  {Ye}, Po-Chieh {Yu}, Lin {Yan}, and Jeffry {Zolkower}.
\newblock {The Zwicky Transient Facility: System Overview, Performance, and
  First Results}.
\newblock \emph{PASP}, 131\penalty0 (995):\penalty0 018002, Jan 2019.
\newblock \doi{10.1088/1538-3873/aaecbe}.

\bibitem[{Benavente} et~al.(2017){Benavente}, {Protopapas}, and
  {Pichara}]{Benavente17}
Patricio {Benavente}, Pavlos {Protopapas}, and Karim {Pichara}.
\newblock {Automatic Survey-invariant Classification of Variable Stars}.
\newblock \emph{ApJ}, 845:\penalty0 147, Aug 2017.
\newblock \doi{10.3847/1538-4357/aa7f2d}.

\bibitem[{Benvenuto} et~al.(2018){Benvenuto}, {Piana}, {Campi}, and
  {Massone}]{Benvenuto18}
Federico {Benvenuto}, Michele {Piana}, Cristina {Campi}, and Anna~Maria
  {Massone}.
\newblock {A Hybrid Supervised/Unsupervised Machine Learning Approach to Solar
  Flare Prediction}.
\newblock \emph{ApJ}, 853:\penalty0 90, Jan 2018.
\newblock \doi{10.3847/1538-4357/aaa23c}.

\bibitem[Bertin and Arnouts(1996)]{bertinSExtractorSoftwareSource1996a}
E.~Bertin and S.~Arnouts.
\newblock {{SExtractor}}: {{Software}} for source extraction.
\newblock \emph{A\&ASS}, 117:\penalty0 393, June 1996.
\newblock \doi{10.1051/aas:1996164}.

\bibitem[{Bethapudi} and {Desai}(2018)]{Bethapudi18}
S.~{Bethapudi} and S.~{Desai}.
\newblock {Separation of pulsar signals from noise using supervised machine
  learning algorithms}.
\newblock \emph{A\&C}, 23:\penalty0 15--26, Apr 2018.
\newblock \doi{10.1016/j.ascom.2018.02.002}.

\bibitem[{Bilicki} et~al.(2018){Bilicki}, {Hoekstra}, {Brown}, {Amaro},
  {Blake}, {Cavuoti}, {de Jong}, {Georgiou}, {Hildebrandt}, {Wolf}, {Amon},
  {Brescia}, {Brough}, {Costa-Duarte}, {Erben}, {Glazebrook}, {Grado},
  {Heymans}, {Jarrett}, {Joudaki}, {Kuijken}, {Longo}, {Napolitano},
  {Parkinson}, {Vellucci}, {Kleijn}, and {Wang}]{Bilicki2018}
M.~{Bilicki}, H.~{Hoekstra}, M.~J.~I. {Brown}, V.~{Amaro}, C.~{Blake},
  S.~{Cavuoti}, J.~T.~A. {de Jong}, C.~{Georgiou}, H.~{Hildebrandt}, C.~{Wolf},
  A.~{Amon}, M.~{Brescia}, S.~{Brough}, M.~V. {Costa-Duarte}, T.~{Erben},
  K.~{Glazebrook}, A.~{Grado}, C.~{Heymans}, T.~{Jarrett}, S.~{Joudaki},
  K.~{Kuijken}, G.~{Longo}, N.~{Napolitano}, D.~{Parkinson}, C.~{Vellucci},
  G.~A.~Verdoes {Kleijn}, and L.~{Wang}.
\newblock {Photometric redshifts for the Kilo-Degree Survey. Machine-learning
  analysis with artificial neural networks}.
\newblock \emph{A\&A}, 616:\penalty0 A69, August 2018.
\newblock \doi{10.1051/0004-6361/201731942}.

\bibitem[Biswas et~al.(2013)Biswas, Blackburn, Cao, Essick, Hodge,
  Katsavounidis, Kim, Kim, Le~Bigot, Lee, Oh, Oh, Son, Tao, Vaulin, and
  Wang]{biswasApplicationMachineLearning2013}
Rahul Biswas, Lindy Blackburn, Junwei Cao, Reed Essick, Kari~Alison Hodge,
  Erotokritos Katsavounidis, Kyungmin Kim, Young-Min Kim, Eric-Olivier
  Le~Bigot, Chang-Hwan Lee, John~J. Oh, Sang~Hoon Oh, Edwin~J. Son, Ye~Tao,
  Ruslan Vaulin, and Xiaoge Wang.
\newblock Application of machine learning algorithms to the study of noise
  artifacts in gravitational-wave data.
\newblock \emph{Phys. Rev. D}, 88\penalty0 (6):\penalty0 062003, September
  2013.
\newblock \doi{10.1103/PhysRevD.88.062003}.

\bibitem[{Bolzonella} et~al.(2000){Bolzonella}, {Miralles}, and
  {Pell{\'o}}]{Bolzonella00}
M.~{Bolzonella}, J.~M. {Miralles}, and R.~{Pell{\'o}}.
\newblock {Photometric redshifts based on standard SED fitting procedures}.
\newblock \emph{A\&A}, 363:\penalty0 476--492, Nov 2000.

\bibitem[{Bonnett} et~al.(2016){Bonnett}, {Troxel}, {Hartley}, {Amara},
  {Leistedt}, {Becker}, {Bernstein}, {Bridle}, {Bruderer}, {Busha}, {Carrasco
  Kind}, {Childress}, {Castander}, {Chang}, {Crocce}, {Davis}, {Eifler},
  {Frieman}, {Gangkofner}, {Gaztanaga}, {Glazebrook}, {Gruen}, {Kacprzak},
  {King}, {Kwan}, {Lahav}, {Lewis}, {Lidman}, {Lin}, {MacCrann}, {Miquel},
  {O'Neill}, {Palmese}, {Peiris}, {Refregier}, {Rozo}, {Rykoff}, {Sadeh},
  {S{\'a}nchez}, {Sheldon}, {Uddin}, {Wechsler}, {Zuntz}, {Abbott}, {Abdalla},
  {Allam}, {Armstrong}, {Banerji}, {Bauer}, {Benoit-L{\'e}vy}, {Bertin},
  {Brooks}, {Buckley-Geer}, {Burke}, {Capozzi}, {Carnero Rosell}, {Carretero},
  {Cunha}, {D'Andrea}, {da Costa}, {DePoy}, {Desai}, {Diehl}, {Dietrich},
  {Doel}, {Fausti Neto}, {Fernandez}, {Flaugher}, {Fosalba}, {Gerdes},
  {Gruendl}, {Honscheid}, {Jain}, {James}, {Jarvis}, {Kim}, {Kuehn},
  {Kuropatkin}, {Li}, {Lima}, {Maia}, {March}, {Marshall}, {Martini},
  {Melchior}, {Miller}, {Neilsen}, {Nichol}, {Nord}, {Ogando}, {Plazas},
  {Reil}, {Romer}, {Roodman}, {Sako}, {Sanchez}, {Santiago}, {Smith},
  {Soares-Santos}, {Sobreira}, {Suchyta}, {Swanson}, {Tarle}, {Thaler},
  {Thomas}, {Vikram}, {Walker}, and {Dark Energy Survey
  Collaboration}]{Bonnett16}
C.~{Bonnett}, M.~A. {Troxel}, W.~{Hartley}, A.~{Amara}, B.~{Leistedt}, M.~R.
  {Becker}, G.~M. {Bernstein}, S.~L. {Bridle}, C.~{Bruderer}, M.~T. {Busha},
  M.~{Carrasco Kind}, M.~J. {Childress}, F.~J. {Castander}, C.~{Chang},
  M.~{Crocce}, T.~M. {Davis}, T.~F. {Eifler}, J.~{Frieman}, C.~{Gangkofner},
  E.~{Gaztanaga}, K.~{Glazebrook}, D.~{Gruen}, T.~{Kacprzak}, A.~{King},
  J.~{Kwan}, O.~{Lahav}, G.~{Lewis}, C.~{Lidman}, H.~{Lin}, N.~{MacCrann},
  R.~{Miquel}, C.~R. {O'Neill}, A.~{Palmese}, H.~V. {Peiris}, A.~{Refregier},
  E.~{Rozo}, E.~S. {Rykoff}, I.~{Sadeh}, C.~{S{\'a}nchez}, E.~{Sheldon},
  S.~{Uddin}, R.~H. {Wechsler}, J.~{Zuntz}, T.~{Abbott}, F.~B. {Abdalla},
  S.~{Allam}, R.~{Armstrong}, M.~{Banerji}, A.~H. {Bauer},
  A.~{Benoit-L{\'e}vy}, E.~{Bertin}, D.~{Brooks}, E.~{Buckley-Geer}, D.~L.
  {Burke}, D.~{Capozzi}, A.~{Carnero Rosell}, J.~{Carretero}, C.~E. {Cunha},
  C.~B. {D'Andrea}, L.~N. {da Costa}, D.~L. {DePoy}, S.~{Desai}, H.~T. {Diehl},
  J.~P. {Dietrich}, P.~{Doel}, A.~{Fausti Neto}, E.~{Fernandez}, B.~{Flaugher},
  P.~{Fosalba}, D.~W. {Gerdes}, R.~A. {Gruendl}, K.~{Honscheid}, B.~{Jain},
  D.~J. {James}, M.~{Jarvis}, A.~G. {Kim}, K.~{Kuehn}, N.~{Kuropatkin}, T.~S.
  {Li}, M.~{Lima}, M.~A.~G. {Maia}, M.~{March}, J.~L. {Marshall}, P.~{Martini},
  P.~{Melchior}, C.~J. {Miller}, E.~{Neilsen}, R.~C. {Nichol}, B.~{Nord},
  R.~{Ogando}, A.~A. {Plazas}, K.~{Reil}, A.~K. {Romer}, A.~{Roodman},
  M.~{Sako}, E.~{Sanchez}, B.~{Santiago}, R.~C. {Smith}, M.~{Soares-Santos},
  F.~{Sobreira}, E.~{Suchyta}, M.~E.~C. {Swanson}, G.~{Tarle}, J.~{Thaler},
  D.~{Thomas}, V.~{Vikram}, A.~R. {Walker}, and {Dark Energy Survey
  Collaboration}.
\newblock {Redshift distributions of galaxies in the Dark Energy Survey Science
  Verification shear catalogue and implications for weak lensing}.
\newblock \emph{PRD}, 94\penalty0 (4):\penalty0 042005, Aug 2016.
\newblock \doi{10.1103/PhysRevD.94.042005}.

\bibitem[{Booth} et~al.(2009){Booth}, {de Blok}, {Jonas}, and
  {Fanaroff}]{Booth09}
R.~S. {Booth}, W.~J.~G. {de Blok}, J.~L. {Jonas}, and B.~{Fanaroff}.
\newblock {MeerKAT Key Project Science, Specifications, and Proposals}.
\newblock art. arXiv:0910.2935, Oct 2009.

\bibitem[Borne(2009)]{borne_scientific_2009}
Kirk Borne.
\newblock Scientific {Data} {Mining} in {Astronomy}.
\newblock In H.~{Kargupta}, J.~{Han}, P.S. {Yu}, R.~{Motwani}, and V.~{Kumar},
  editors, \emph{Next Generation of Data Mining}, pages 91--114, 2009.

\bibitem[{Borucki} and {et al.}(2010)]{Borucki10}
William~J. {Borucki} and {et al.}
\newblock {Kepler Planet-Detection Mission: Introduction and First Results}.
\newblock \emph{Science}, 327:\penalty0 977, Feb 2010.
\newblock \doi{10.1126/science.1185402}.

\bibitem[Breiman(1996)]{Breiman1996}
Leo Breiman.
\newblock Bagging predictors.
\newblock \emph{Mach Learn}, 24\penalty0 (2):\penalty0 123--140, August 1996.
\newblock ISSN 1573-0565.
\newblock \doi{10.1007/BF00058655}.

\bibitem[Breiman(2001)]{Breiman2001}
Leo Breiman.
\newblock Random {{Forests}}.
\newblock \emph{Machine Learning}, 45\penalty0 (1):\penalty0 5--32, October
  2001.
\newblock ISSN 1573-0565.
\newblock \doi{10.1023/A:1010933404324}.

\bibitem[{Brescia} et~al.(2013){Brescia}, {Cavuoti}, {D'Abrusco}, {Longo}, and
  {Mercurio}]{Brescia13}
M.~{Brescia}, S.~{Cavuoti}, R.~{D'Abrusco}, G.~{Longo}, and A.~{Mercurio}.
\newblock {Photometric Redshifts for Quasars in Multi-band Surveys}.
\newblock \emph{ApJ}, 772\penalty0 (2):\penalty0 140, Aug 2013.
\newblock \doi{10.1088/0004-637X/772/2/140}.

\bibitem[{Brescia} et~al.(2014){Brescia}, {Cavuoti}, {Longo}, {Nocella},
  {Garofalo}, {Manna}, {Esposito}, {Albano}, {Guglielmo}, {D'Angelo}, {Di
  Guido}, {Djorgovski}, {Donalek}, {Mahabal}, {Graham}, {Fiore}, and
  {D'Abrusco}]{Brescia14}
Massimo {Brescia}, Stefano {Cavuoti}, Giuseppe {Longo}, Alfonso {Nocella},
  Mauro {Garofalo}, Francesco {Manna}, Francesco {Esposito}, Giovanni {Albano},
  Marisa {Guglielmo}, Giovanni {D'Angelo}, Alessandro {Di Guido}, S.~George
  {Djorgovski}, Ciro {Donalek}, Ashish~A. {Mahabal}, Matthew~J. {Graham},
  Michelangelo {Fiore}, and Raffaele {D'Abrusco}.
\newblock {DAMEWARE: A Web Cyberinfrastructure for Astrophysical Data Mining}.
\newblock \emph{PASP}, 126\penalty0 (942):\penalty0 783, Aug 2014.
\newblock \doi{10.1086/677725}.

\bibitem[{Brescia} et~al.(2016){Brescia}, {Cavuoti}, {Esposito}, {Fiore},
  {Garofalo}, {Guglielmo}, {Longo}, {Manna}, {Nocella}, and
  {Vellucci}]{Brescia16}
Massimo {Brescia}, Stefano {Cavuoti}, Francesco {Esposito}, Michelangelo
  {Fiore}, Mauro {Garofalo}, Marisa {Guglielmo}, Giuseppe {Longo}, Francesco
  {Manna}, Alfonso {Nocella}, and Civita {Vellucci}.
\newblock {DAMEWARE - Data Mining \& Exploration Web Application Resource}.
\newblock art. arXiv:1603.00720, Mar 2016.

\bibitem[{Bron} et~al.(2018){Bron}, {Daudon}, {Pety}, {Levrier}, {Gerin},
  {Gratier}, {Orkisz}, {Guzman}, {Bardeau}, {Goicoechea}, {Liszt}, {{\"O}berg},
  {Peretto}, {Sievers}, and {Tremblin}]{Bron18}
Emeric {Bron}, Chlo{\'e} {Daudon}, J{\'e}r{\^o}me {Pety}, Fran{\c{c}}ois
  {Levrier}, Maryvonne {Gerin}, Pierre {Gratier}, Jan~H. {Orkisz}, Viviana
  {Guzman}, S{\'e}bastien {Bardeau}, Javier~R. {Goicoechea}, Harvey {Liszt},
  Karin {{\"O}berg}, Nicolas {Peretto}, Albrecht {Sievers}, and Pascal
  {Tremblin}.
\newblock {Clustering the Orion B giant molecular cloud based on its molecular
  emission}.
\newblock \emph{A\&A}, 610:\penalty0 A12, Feb 2018.
\newblock \doi{10.1051/0004-6361/201731833}.

\bibitem[Brunner et~al.(2002)Brunner, Djorgovski, Prince, and
  Szalay]{Brunner2002}
Robert~J. Brunner, S.~George Djorgovski, Thomas~A. Prince, and Alex~S. Szalay.
\newblock \emph{Massive Datasets in Astronomy}, pages 931--979.
\newblock Springer US, Boston, MA, 2002.
\newblock ISBN 978-1-4615-0005-6.

\bibitem[{Bu} et~al.(2017){Bu}, {Lei}, {Zhao}, {Bu}, and {Pan}]{Bu17}
Yude {Bu}, Zhenxin {Lei}, Gang {Zhao}, Jingde {Bu}, and Jingchang {Pan}.
\newblock {Searching for Hot Subdwarf Stars from the LAMOST Spectra. I.
  Method}.
\newblock \emph{ApJSS}, 233:\penalty0 2, Nov 2017.
\newblock \doi{10.3847/1538-4365/aa91cd}.

\bibitem[{Cabrera-Vives} et~al.(2018){Cabrera-Vives}, {Miller}, and
  {Schneider}]{Cabrera18}
Guillermo {Cabrera-Vives}, Christopher~J. {Miller}, and Jeff {Schneider}.
\newblock {Systematic Labeling Bias in Galaxy Morphologies}.
\newblock \emph{AJ}, 156:\penalty0 284, Dec 2018.
\newblock \doi{10.3847/1538-3881/aae9f4}.

\bibitem[Carrasco~Kind and
  Brunner(2013)]{carrascokindTPZPhotometricRedshift2013}
Matias Carrasco~Kind and Robert~J. Brunner.
\newblock {{TPZ}}: Photometric redshift {{PDFs}} and ancillary information by
  using prediction trees and random forests.
\newblock \emph{MNRAS}, 432\penalty0 (2):\penalty0 1483--1501, June 2013.
\newblock ISSN 0035-8711.
\newblock \doi{10.1093/mnras/stt574}.

\bibitem[Carrasco~Kind and
  Brunner(2014)]{carrascokindSOMzPhotometricRedshift2014}
Matias Carrasco~Kind and Robert~J. Brunner.
\newblock {{SOMz}}: Photometric redshift {{PDFs}} with self-organizing maps and
  random atlas.
\newblock \emph{MNRAS}, 438\penalty0 (4):\penalty0 3409--3421, March 2014.
\newblock ISSN 0035-8711.
\newblock \doi{10.1093/mnras/stt2456}.

\bibitem[{Castro-Ginard} et~al.(2018){Castro-Ginard}, {Jordi}, {Luri}, {Julbe},
  {Morvan}, {Balaguer-N{\'u}{\~n}ez}, and {Cantat- Gaudin}]{Castro-Ginard18}
A.~{Castro-Ginard}, C.~{Jordi}, X.~{Luri}, F.~{Julbe}, M.~{Morvan},
  L.~{Balaguer-N{\'u}{\~n}ez}, and T.~{Cantat- Gaudin}.
\newblock {A new method for unveiling open clusters in Gaia. New nearby open
  clusters confirmed by DR2}.
\newblock \emph{A\&A}, 618:\penalty0 A59, October 2018.
\newblock \doi{10.1051/0004-6361/201833390}.

\bibitem[{Cavuoti} et~al.(2012){Cavuoti}, {Brescia}, {Longo}, and
  {Mercurio}]{Cavuoti12}
S.~{Cavuoti}, M.~{Brescia}, G.~{Longo}, and A.~{Mercurio}.
\newblock {Photometric redshifts with the quasi Newton algorithm (MLPQNA)
  Results in the PHAT1 contest}.
\newblock \emph{A\&A}, 546:\penalty0 A13, Oct 2012.
\newblock \doi{10.1051/0004-6361/201219755}.

\bibitem[{Cavuoti} et~al.(2017{\natexlab{a}}){Cavuoti}, {Amaro}, {Brescia},
  {Vellucci}, {Tortora}, and {Longo}]{Cavuoti17b}
S.~{Cavuoti}, V.~{Amaro}, M.~{Brescia}, C.~{Vellucci}, C.~{Tortora}, and
  G.~{Longo}.
\newblock {METAPHOR: a machine-learning-based method for the probability
  density estimation of photometric redshifts}.
\newblock \emph{MNRAS}, 465:\penalty0 1959--1973, Feb 2017{\natexlab{a}}.
\newblock \doi{10.1093/mnras/stw2930}.

\bibitem[{Cavuoti} et~al.(2017{\natexlab{b}}){Cavuoti}, {Tortora}, {Brescia},
  {Longo}, {Radovich}, {Napolitano}, {Amaro}, {Vellucci}, {La Barbera},
  {Getman}, and {Grado}]{Cavuoti17}
S.~{Cavuoti}, C.~{Tortora}, M.~{Brescia}, G.~{Longo}, M.~{Radovich}, N.~R.
  {Napolitano}, V.~{Amaro}, C.~{Vellucci}, F.~{La Barbera}, F.~{Getman}, and
  A.~{Grado}.
\newblock {A cooperative approach among methods for photometric redshifts
  estimation: an application to KiDS data}.
\newblock \emph{MNRAS}, 466:\penalty0 2039--2053, Apr 2017{\natexlab{b}}.
\newblock \doi{10.1093/mnras/stw3208}.

\bibitem[{Chen} et~al.(2019){Chen}, {Huang}, {Yuan}, {Wang}, {Fan}, {Xiang},
  {Zhang}, {Tian}, and {Liu}]{Chen19}
B.~Q. {Chen}, Y.~{Huang}, H.~B. {Yuan}, C.~{Wang}, D.~W. {Fan}, M.~S. {Xiang},
  H.~W. {Zhang}, Z.~J. {Tian}, and X.~W. {Liu}.
\newblock {Three-dimensional interstellar dust reddening maps of the Galactic
  plane}.
\newblock \emph{MNRAS}, 483:\penalty0 4277--4289, Mar 2019.
\newblock \doi{10.1093/mnras/sty3341}.

\bibitem[{Chen} et~al.(2018){Chen}, {Lin}, {Alexandersen}, {Lehner}, {Wang},
  {Wang}, {Yoshida}, {Komiyama}, and {Miyazaki}]{Chen18}
Ying-Tung {Chen}, Hsing-Wen {Lin}, Mike {Alexandersen}, Matthew~J. {Lehner},
  Shiang-Yu {Wang}, Jen-Hung {Wang}, Fumi {Yoshida}, Yutaka {Komiyama}, and
  Satoshi {Miyazaki}.
\newblock {Searching for moving objects in HSC-SSP: Pipeline and preliminary
  results}.
\newblock \emph{PASJ}, 70:\penalty0 S38, Jan 2018.
\newblock \doi{10.1093/pasj/psx145}.

\bibitem[{Chilingarian} et~al.(2009){Chilingarian}, {Cayatte}, {Revaz},
  {Dodonov}, {Durand}, {Durret}, {Micol}, and {Slezak}]{Chilingarian09}
Igor {Chilingarian}, V{\'e}ronique {Cayatte}, Yves {Revaz}, Serguei {Dodonov},
  Daniel {Durand}, Florence {Durret}, Alberto {Micol}, and Eric {Slezak}.
\newblock {A Population of Compact Elliptical Galaxies Detected with the
  Virtual Observatory}.
\newblock \emph{Science}, 326\penalty0 (5958):\penalty0 1379, Dec 2009.
\newblock \doi{10.1126/science.1175930}.

\bibitem[{Chyla} et~al.(2015){Chyla}, {Accomazzi}, {Holachek}, {Grant},
  {Elliott}, {Henneken}, {Thompson}, {Kurtz}, {Murray}, and
  {Sudilovsky}]{Chyla15}
R.~{Chyla}, A.~{Accomazzi}, A.~{Holachek}, C.~S. {Grant}, J.~{Elliott}, E.~A.
  {Henneken}, D.~M. {Thompson}, M.~J. {Kurtz}, S.~S. {Murray}, and
  V.~{Sudilovsky}.
\newblock {ADS 2.0: New Architecture, API and Services}.
\newblock In A.~R. {Taylor} and E.~{Rosolowsky}, editors, \emph{Astronomical
  Data Analysis Software an Systems XXIV (ADASS XXIV)}, volume 495 of
  \emph{Astronomical Society of the Pacific Conference Series}, page 401, Sep
  2015.

\bibitem[{Ciuca} and {Hern{\'a}ndez}(2017)]{Ciuca17}
Razvan {Ciuca} and Oscar~F. {Hern{\'a}ndez}.
\newblock {A Bayesian framework for cosmic string searches in CMB maps}.
\newblock \emph{Journal of Cosmology and Astro-Particle Physics},
  2017:\penalty0 028, Aug 2017.
\newblock \doi{10.1088/1475-7516/2017/08/028}.

\bibitem[{Cohen} et~al.(2017){Cohen}, {Sesar}, {Bahnolzer}, {He}, {Kulkarni},
  {Prince}, {Bellm}, and {Laher}]{Cohen17}
Judith~G. {Cohen}, Branimir {Sesar}, Sophianna {Bahnolzer}, Kevin {He},
  Shrinivas~R. {Kulkarni}, Thomas~A. {Prince}, Eric {Bellm}, and Russ~R.
  {Laher}.
\newblock {The Outer Halo of the Milky Way as Probed by RR Lyr Variables from
  the Palomar Transient Facility}.
\newblock \emph{ApJ}, 849:\penalty0 150, Nov 2017.
\newblock \doi{10.3847/1538-4357/aa9120}.

\bibitem[{Collister} and {Lahav}(2004)]{collister04}
Adrian~A. {Collister} and Ofer {Lahav}.
\newblock {ANNz: Estimating Photometric Redshifts Using Artificial Neural
  Networks}.
\newblock \emph{PASP}, 116\penalty0 (818):\penalty0 345--351, Apr 2004.
\newblock \doi{10.1086/383254}.

\bibitem[{Connor} and {van Leeuwen}(2018)]{Connor18}
Liam {Connor} and Joeri {van Leeuwen}.
\newblock {Applying Deep Learning to Fast Radio Burst Classification}.
\newblock \emph{AJ}, 156:\penalty0 256, Dec 2018.
\newblock \doi{10.3847/1538-3881/aae649}.

\bibitem[Conselice(2006)]{conseliceFundamentalPropertiesGalaxies2006}
Christopher~J. Conselice.
\newblock The fundamental properties of galaxies and a new galaxy
  classification system.
\newblock \emph{MNRAS}, 373\penalty0 (4):\penalty0 1389--1408, December 2006.
\newblock ISSN 0035-8711.
\newblock \doi{10.1111/j.1365-2966.2006.11114.x}.

\bibitem[Cortes and Vapnik(1995)]{Cortes1995}
Corinna Cortes and Vladimir Vapnik.
\newblock Support-vector networks.
\newblock \emph{Mach Learn}, 20\penalty0 (3):\penalty0 273--297, September
  1995.
\newblock ISSN 0885-6125, 1573-0565.
\newblock \doi{10.1007/BF00994018}.

\bibitem[{Dark Energy Survey Collaboration} et~al.(2016){Dark Energy Survey
  Collaboration}, Abbott, Abdalla, Aleksi{\'c}, Allam, Amara, Bacon, Balbinot,
  Banerji, Bechtol, {Benoit-L{\'e}vy}, Bernstein, Bertin, Blazek, Bonnett,
  Bridle, Brooks, Brunner, {Buckley-Geer}, Burke, Caminha, Capozzi, Carlsen,
  {Carnero-Rosell}, Carollo, {Carrasco-Kind}, Carretero, Castander, Clerkin,
  Collett, Conselice, Crocce, Cunha, D'Andrea, {da Costa}, Davis, Desai, Diehl,
  Dietrich, Dodelson, Doel, {Drlica-Wagner}, Estrada, Etherington, Evrard,
  Fabbri, Finley, Flaugher, Foley, Fosalba, Frieman, {Garc{\'i}a-Bellido},
  Gaztanaga, Gerdes, Giannantonio, Goldstein, Gruen, Gruendl, Guarnieri,
  Gutierrez, Hartley, Honscheid, Jain, James, Jeltema, Jouvel, Kessler, King,
  Kirk, Kron, Kuehn, Kuropatkin, Lahav, Li, Lima, Lin, Maia, Makler, Manera,
  Maraston, Marshall, Martini, McMahon, Melchior, Merson, Miller, Miquel, Mohr,
  {Morice-Atkinson}, Naidoo, Neilsen, Nichol, Nord, Ogando, Ostrovski, Palmese,
  Papadopoulos, Peiris, Peoples, Percival, Plazas, Reed, Refregier, Romer,
  Roodman, Ross, Rozo, Rykoff, Sadeh, Sako, S{\'a}nchez, Sanchez, Santiago,
  Scarpine, Schubnell, {Sevilla-Noarbe}, Sheldon, Smith, Smith,
  {Soares-Santos}, Sobreira, Soumagnac, Suchyta, Sullivan, Swanson, Tarle,
  Thaler, Thomas, Thomas, Tucker, Vieira, Vikram, Walker, Wechsler, Weller,
  Wester, Whiteway, Wilcox, Yanny, Zhang, and
  Zuntz]{darkenergysurveycollaborationDarkEnergySurvey2016}
{Dark Energy Survey Collaboration}, T.~Abbott, F.~B. Abdalla, J.~Aleksi{\'c},
  S.~Allam, A.~Amara, D.~Bacon, E.~Balbinot, M.~Banerji, K.~Bechtol,
  A.~{Benoit-L{\'e}vy}, G.~M. Bernstein, E.~Bertin, J.~Blazek, C.~Bonnett,
  S.~Bridle, D.~Brooks, R.~J. Brunner, E.~{Buckley-Geer}, D.~L. Burke, G.~B.
  Caminha, D.~Capozzi, J.~Carlsen, A.~{Carnero-Rosell}, M.~Carollo,
  M.~{Carrasco-Kind}, J.~Carretero, F.~J. Castander, L.~Clerkin, T.~Collett,
  C.~Conselice, M.~Crocce, C.~E. Cunha, C.~B. D'Andrea, L.~N. {da Costa}, T.~M.
  Davis, S.~Desai, H.~T. Diehl, J.~P. Dietrich, S.~Dodelson, P.~Doel,
  A.~{Drlica-Wagner}, J.~Estrada, J.~Etherington, A.~E. Evrard, J.~Fabbri,
  D.~A. Finley, B.~Flaugher, R.~J. Foley, P.~Fosalba, J.~Frieman,
  J.~{Garc{\'i}a-Bellido}, E.~Gaztanaga, D.~W. Gerdes, T.~Giannantonio, D.~A.
  Goldstein, D.~Gruen, R.~A. Gruendl, P.~Guarnieri, G.~Gutierrez, W.~Hartley,
  K.~Honscheid, B.~Jain, D.~J. James, T.~Jeltema, S.~Jouvel, R.~Kessler,
  A.~King, D.~Kirk, R.~Kron, K.~Kuehn, N.~Kuropatkin, O.~Lahav, T.~S. Li,
  M.~Lima, H.~Lin, M.~A.~G. Maia, M.~Makler, M.~Manera, C.~Maraston, J.~L.
  Marshall, P.~Martini, R.~G. McMahon, P.~Melchior, A.~Merson, C.~J. Miller,
  R.~Miquel, J.~J. Mohr, X.~{Morice-Atkinson}, K.~Naidoo, E.~Neilsen, R.~C.
  Nichol, B.~Nord, R.~Ogando, F.~Ostrovski, A.~Palmese, A.~Papadopoulos, H.~V.
  Peiris, J.~Peoples, W.~J. Percival, A.~A. Plazas, S.~L. Reed, A.~Refregier,
  A.~K. Romer, A.~Roodman, A.~Ross, E.~Rozo, E.~S. Rykoff, I.~Sadeh, M.~Sako,
  C.~S{\'a}nchez, E.~Sanchez, B.~Santiago, V.~Scarpine, M.~Schubnell,
  I.~{Sevilla-Noarbe}, E.~Sheldon, M.~Smith, R.~C. Smith, M.~{Soares-Santos},
  F.~Sobreira, M.~Soumagnac, E.~Suchyta, M.~Sullivan, M.~Swanson, G.~Tarle,
  J.~Thaler, D.~Thomas, R.~C. Thomas, D.~Tucker, J.~D. Vieira, V.~Vikram, A.~R.
  Walker, R.~H. Wechsler, J.~Weller, W.~Wester, L.~Whiteway, H.~Wilcox,
  B.~Yanny, Y.~Zhang, and J.~Zuntz.
\newblock The {{Dark Energy Survey}}: More than dark energy - an overview.
\newblock \emph{MNRAS}, 460:\penalty0 1270--1299, August 2016.
\newblock ISSN 0035-8711.
\newblock \doi{10.1093/mnras/stw641}.

\bibitem[{de Jong} et~al.(2013){de Jong}, {Verdoes Kleijn}, {Kuijken}, and
  {Valentijn}]{dejong13}
Jelte T.~A. {de Jong}, Gijs~A. {Verdoes Kleijn}, Konrad~H. {Kuijken}, and
  Edwin~A. {Valentijn}.
\newblock {The Kilo-Degree Survey}.
\newblock \emph{Experimental Astronomy}, 35\penalty0 (1-2):\penalty0 25--44,
  Jan 2013.
\newblock \doi{10.1007/s10686-012-9306-1}.

\bibitem[{de Jong} et~al.(2015){de Jong}, Verdoes~Kleijn, Boxhoorn,
  Buddelmeijer, Capaccioli, Getman, Grado, Helmich, Huang, Irisarri, Kuijken,
  La~Barbera, McFarland, Napolitano, Radovich, Sikkema, Valentijn, Begeman,
  Brescia, Cavuoti, Choi, Cordes, Covone, Dall'Ora, Hildebrandt, Longo,
  Nakajima, Paolillo, Puddu, Rifatto, Tortora, {van Uitert}, Buddendiek,
  {Harnois-D{\'e}raps}, Erben, Eriksen, Heymans, Hoekstra, Joachimi, Kitching,
  Klaes, Koopmans, K{\"o}hlinger, Roy, Sif{\'o}n, Schneider, Sutherland, Viola,
  and Vriend]{dejongFirstSecondData2015}
Jelte T.~A. {de Jong}, Gijs~A. Verdoes~Kleijn, Danny~R. Boxhoorn, Hugo
  Buddelmeijer, Massimo Capaccioli, Fedor Getman, Aniello Grado, Ewout Helmich,
  Zhuoyi Huang, Nancy Irisarri, Konrad Kuijken, Francesco La~Barbera, John~P.
  McFarland, Nicola~R. Napolitano, Mario Radovich, Gert Sikkema, Edwin~A.
  Valentijn, Kor~G. Begeman, Massimo Brescia, Stefano Cavuoti, Ami Choi,
  Oliver-Mark Cordes, Giovanni Covone, Massimo Dall'Ora, Hendrik Hildebrandt,
  Giuseppe Longo, Reiko Nakajima, Maurizio Paolillo, Emanuella Puddu, Agatino
  Rifatto, Crescenzo Tortora, Edo {van Uitert}, Axel Buddendiek, Joachim
  {Harnois-D{\'e}raps}, Thomas Erben, Martin~B. Eriksen, Catherine Heymans,
  Henk Hoekstra, Benjamin Joachimi, Thomas~D. Kitching, Dominik Klaes, L{\'e}on
  V.~E. Koopmans, Fabian K{\"o}hlinger, Nivya Roy, Crist{\'o}bal Sif{\'o}n,
  Peter Schneider, Will~J. Sutherland, Massimo Viola, and Willem-Jan Vriend.
\newblock The first and second data releases of the {{Kilo}}-{{Degree Survey}}.
\newblock \emph{A\&A}, 582:\penalty0 A62, October 2015.
\newblock \doi{10.1051/0004-6361/201526601}.

\bibitem[de~Le{\'o}n et~al.(2010)de~Le{\'o}n, Licandro, {Serra-Ricart},
  {Pinilla-Alonso}, and Campins]{leonObservationsCompositionalPhysical2010}
J.~de~Le{\'o}n, J.~Licandro, M.~{Serra-Ricart}, N.~{Pinilla-Alonso}, and
  H.~Campins.
\newblock Observations, compositional, and physical characterization of
  near-{{Earth}} and {{Mars}}-crosser asteroids from a spectroscopic survey.
\newblock \emph{A\&A}, 517:\penalty0 A23, July 2010.
\newblock ISSN 0004-6361, 1432-0746.
\newblock \doi{10.1051/0004-6361/200913852}.

\bibitem[{D{\'e}nes} et~al.(2018){D{\'e}nes}, {McClure-Griffiths}, {Dickey},
  {Dawson}, and {Murray}]{Denes18}
H.~{D{\'e}nes}, N.~M. {McClure-Griffiths}, J.~M. {Dickey}, J.~R. {Dawson}, and
  C.~E. {Murray}.
\newblock {Calibrating the HISA temperature: Measuring the temperature of the
  Riegel-Crutcher cloud}.
\newblock \emph{MNRAS}, 479:\penalty0 1465--1490, Sep 2018.
\newblock \doi{10.1093/mnras/sty1384}.

\bibitem[Denker and Lecun(1991)]{denkerTransformingNeuralNetOutput1991}
John Denker and Yann Lecun.
\newblock Transforming {{Neural}}-{{Net Output Levels}} to {{Probability
  Distributions}}.
\newblock In \emph{Advances in {{Neural Information Processing Systems}} 3},
  pages 853--859. {Morgan Kaufmann}, 1991.

\bibitem[{Dewdney} et~al.(2009){Dewdney}, {Hall}, {Schilizzi}, and
  {Lazio}]{Dewdney09}
P.~E. {Dewdney}, P.~J. {Hall}, R.~T. {Schilizzi}, and T.~J.~L.~W. {Lazio}.
\newblock {The Square Kilometre Array}.
\newblock \emph{IEEE Proceedings}, 97:\penalty0 1482--1496, Aug 2009.
\newblock \doi{10.1109/JPROC.2009.2021005}.

\bibitem[{Diakogiannis} et~al.(2019){Diakogiannis}, {Lewis}, {Ibata},
  {Guglielmo}, {Wilkinson}, and {Power}]{Diakogiannis19}
Foivos~I. {Diakogiannis}, Geraint~F. {Lewis}, Rodrigo~A. {Ibata}, Magda
  {Guglielmo}, Mark~I. {Wilkinson}, and Chris {Power}.
\newblock {Reliable mass calculation in spherical gravitating systems}.
\newblock \emph{MNRAS}, 482:\penalty0 3356--3372, Jan 2019.
\newblock \doi{10.1093/mnras/sty2931}.

\bibitem[Dieleman et~al.(2015)Dieleman, Willett, and
  Dambre]{dieleman_rotation-invariant_2015}
Sander Dieleman, Kyle~W. Willett, and Joni Dambre.
\newblock Rotation-invariant convolutional neural networks for galaxy
  morphology prediction.
\newblock \emph{MNRAS}, 450:\penalty0 1441--1459, June 2015.
\newblock ISSN 0035-8711.
\newblock \doi{10.1093/mnras/stv632}.

\bibitem[{Djorgovski}(2014)]{Djorgovski2014}
G.~{Djorgovski}.
\newblock {Astrophysics in the Era of Massive Time-Domain Surveys}.
\newblock In P.~R. {Wozniak}, M.~J. {Graham}, A.~A. {Mahabal}, and R.~{Seaman},
  editors, \emph{The Third Hot-wiring the Transient Universe Workshop}, pages
  215--215, Jan 2014.

\bibitem[Djorgovski et~al.(2016)Djorgovski, Graham, Donalek, Mahabal, Drake,
  Turmon, and Fuchs]{Djorgovski2016}
S.G. Djorgovski, M.J. Graham, C.~Donalek, A.A. Mahabal, A.J. Drake, M.~Turmon,
  and T.~Fuchs.
\newblock Real-time data mining of massive data streams from synoptic sky
  surveys.
\newblock \emph{Future Gener. Comput. Syst.}, 59\penalty0 (C):\penalty0
  95--104, June 2016.
\newblock ISSN 0167-739X.
\newblock \doi{10.1016/j.future.2015.10.013}.

\bibitem[{Dom{\'\i}nguez S{\'a}nchez} et~al.(2018){Dom{\'\i}nguez S{\'a}nchez},
  {Huertas-Company}, {Bernardi}, {Tuccillo}, and {Fischer}]{Dominguez18}
H.~{Dom{\'\i}nguez S{\'a}nchez}, M.~{Huertas-Company}, M.~{Bernardi},
  D.~{Tuccillo}, and J.~L. {Fischer}.
\newblock {Improving galaxy morphologies for SDSS with Deep Learning}.
\newblock \emph{MNRAS}, 476:\penalty0 3661--3676, May 2018.
\newblock \doi{10.1093/mnras/sty338}.

\bibitem[{Drake} et~al.(2009){Drake}, {Djorgovski}, {Mahabal}, {Beshore},
  {Larson}, {Graham}, {Williams}, {Christensen}, {Catelan}, {Boattini},
  {Gibbs}, {Hill}, and {Kowalski}]{Drake09}
A.~J. {Drake}, S.~G. {Djorgovski}, A.~{Mahabal}, E.~{Beshore}, S.~{Larson},
  M.~J. {Graham}, R.~{Williams}, E.~{Christensen}, M.~{Catelan}, A.~{Boattini},
  A.~{Gibbs}, R.~{Hill}, and R.~{Kowalski}.
\newblock {First Results from the Catalina Real-Time Transient Survey}.
\newblock \emph{ApJ}, 696\penalty0 (1):\penalty0 870--884, May 2009.
\newblock \doi{10.1088/0004-637X/696/1/870}.

\bibitem[{Duev} et~al.(2019){Duev}, {Mahabal}, {Ye}, {Tirumala}, {Belicki},
  {Dekany}, {Frederick}, {Graham}, {Laher}, {Masci}, {Prince}, {Riddle},
  {Rosnet}, and {Soumagnac}]{Duev19}
Dmitry~A. {Duev}, Ashish {Mahabal}, Quanzhi {Ye}, Kushal {Tirumala}, Justin
  {Belicki}, Richard {Dekany}, Sara {Frederick}, Matthew~J. {Graham}, Russ~R.
  {Laher}, Frank~J. {Masci}, Thomas~A. {Prince}, Reed {Riddle}, Philippe
  {Rosnet}, and Maayane~T. {Soumagnac}.
\newblock {DeepStreaks: identifying fast-moving objects in the Zwicky Transient
  Facility data with deep learning}.
\newblock \emph{MNRAS}, 486\penalty0 (3):\penalty0 4158--4165, Jul 2019.
\newblock \doi{10.1093/mnras/stz1096}.

\bibitem[Eatough et~al.(2010)Eatough, Molkenthin, Kramer, Noutsos, Keith,
  Stappers, and Lyne]{eatoughSelectionRadioPulsar2010}
R.~P. Eatough, N.~Molkenthin, M.~Kramer, A.~Noutsos, M.~J. Keith, B.~W.
  Stappers, and A.~G. Lyne.
\newblock Selection of radio pulsar candidates using artificial neural
  networks.
\newblock \emph{MNRAS}, 407\penalty0 (4):\penalty0 2443--2450, October 2010.
\newblock ISSN 0035-8711.
\newblock \doi{10.1111/j.1365-2966.2010.17082.x}.

\bibitem[{Erasmus} et~al.(2017){Erasmus}, {Mommert}, {Trilling}, {Sickafoose},
  {van Gend}, and {Hora}]{Erasmus17}
N.~{Erasmus}, M.~{Mommert}, D.~E. {Trilling}, A.~A. {Sickafoose}, C.~{van
  Gend}, and J.~L. {Hora}.
\newblock {Characterization of Near-Earth Asteroids Using KMTNET-SAAO}.
\newblock \emph{AJ}, 154:\penalty0 162, Oct 2017.
\newblock \doi{10.3847/1538-3881/aa88be}.

\bibitem[{Erasmus} et~al.(2018){Erasmus}, {McNeill}, {Mommert}, {Trilling},
  {Sickafoose}, and {van Gend}]{Erasmus18}
N.~{Erasmus}, A.~{McNeill}, M.~{Mommert}, D.~E. {Trilling}, A.~A. {Sickafoose},
  and C.~{van Gend}.
\newblock {Taxonomy and Light-curve Data of 1000 Serendipitously Observed
  Main-belt Asteroids}.
\newblock \emph{ApJSS}, 237:\penalty0 19, Jul 2018.
\newblock \doi{10.3847/1538-4365/aac38f}.

\bibitem[Fadely et~al.(2012)Fadely, Hogg, and
  Willman]{fadelySTARGALAXYCLASSIFICATIONMULTIBAND2012}
Ross Fadely, David~W. Hogg, and Beth Willman.
\newblock {{STAR}}-{{GALAXY CLASSIFICATION IN MULTI}}-{{BAND OPTICAL IMAGING}}.
\newblock \emph{ApJ}, 760\penalty0 (1):\penalty0 15, October 2012.
\newblock ISSN 0004-637X.
\newblock \doi{10.1088/0004-637X/760/1/15}.

\bibitem[{Farah} et~al.(2018){Farah}, {Flynn}, {Bailes}, {Jameson},
  {Bannister}, {Barr}, {Bateman}, {Bhandari}, {Caleb}, {Campbell-Wilson},
  {Chang}, {Deller}, {Green}, {Hunstead}, {Jankowski}, {Keane}, {Macquart},
  {M{\"o}ller}, {Onken}, {Os{\l}owski}, {Parthasarathy}, {Plant}, {Ravi},
  {Shannon}, {Tucker}, {Venkatraman Krishnan}, and {Wolf}]{Farah18}
W.~{Farah}, C.~{Flynn}, M.~{Bailes}, A.~{Jameson}, K.~W. {Bannister}, E.~D.
  {Barr}, T.~{Bateman}, S.~{Bhandari}, M.~{Caleb}, D.~{Campbell-Wilson}, S.~W.
  {Chang}, A.~{Deller}, A.~J. {Green}, R.~{Hunstead}, F.~{Jankowski},
  E.~{Keane}, J.~P. {Macquart}, A.~{M{\"o}ller}, C.~A. {Onken},
  S.~{Os{\l}owski}, A.~{Parthasarathy}, K.~{Plant}, V.~{Ravi}, R.~M. {Shannon},
  B.~E. {Tucker}, V.~{Venkatraman Krishnan}, and C.~{Wolf}.
\newblock {FRB microstructure revealed by the real-time detection of
  FRB170827}.
\newblock \emph{MNRAS}, 478:\penalty0 1209--1217, July 2018.
\newblock \doi{10.1093/mnras/sty1122}.

\bibitem[Fayyad et~al.(1996)Fayyad, Piatetsky-Shapiro, and Smyth]{Fayyad1996}
U.~Fayyad, G.~Piatetsky-Shapiro, and P.~Smyth.
\newblock From data mining to knowledge discovery in databases.
\newblock \emph{AI Magazine}, 17\penalty0 (3):\penalty0 37--53, 1996.

\bibitem[Feigelson and Babu(2012)]{Feigelson2012}
E.D. Feigelson and G.J. Babu.
\newblock Big data in astronomy.
\newblock \emph{Significance}, 9\penalty0 (4):\penalty0 22--25, 2012.
\newblock \doi{10.1111/j.1740-9713.2012.00587.x}.

\bibitem[Firth et~al.(2003)Firth, Lahav, and
  Somerville]{firthEstimatingPhotometricRedshifts2003}
Andrew~E. Firth, Ofer Lahav, and Rachel~S. Somerville.
\newblock Estimating photometric redshifts with artificial neural networks.
\newblock \emph{MNRAS}, 339\penalty0 (4):\penalty0 1195--1202, March 2003.
\newblock ISSN 0035-8711.
\newblock \doi{10.1046/j.1365-8711.2003.06271.x}.

\bibitem[{Florios} et~al.(2018){Florios}, {Kontogiannis}, {Park}, {Guerra},
  {Benvenuto}, {Bloomfield}, and {Georgoulis}]{Florios18}
Kostas {Florios}, Ioannis {Kontogiannis}, Sung-Hong {Park}, Jordan~A. {Guerra},
  Federico {Benvenuto}, D.~Shaun {Bloomfield}, and Manolis~K. {Georgoulis}.
\newblock {Forecasting Solar Flares Using Magnetogram-based Predictors and
  Machine Learning}.
\newblock \emph{SoPh}, 293:\penalty0 28, Feb 2018.
\newblock \doi{10.1007/s11207-018-1250-4}.

\bibitem[{Fluke} et~al.(2011){Fluke}, {Barnes}, {Barsdell}, and
  {Hassan}]{Fluke2011}
Christopher~J. {Fluke}, David~G. {Barnes}, Benjamin~R. {Barsdell}, and Amr~H.
  {Hassan}.
\newblock {Astrophysical Supercomputing with GPUs: Critical Decisions for Early
  Adopters}.
\newblock \emph{PASA}, 28:\penalty0 15--27, Jan 2011.
\newblock \doi{10.1071/AS10019}.

\bibitem[Francis et~al.(1992)Francis, Hewett, Foltz, and
  Chaffee]{francisObjectiveClassificationScheme1992}
Paul~J. Francis, Paul~C. Hewett, Craig~B. Foltz, and Frederic~H. Chaffee.
\newblock An {{Objective Classification Scheme}} for {{QSO Spectra}}.
\newblock \emph{ApJ}, 398:\penalty0 476, October 1992.
\newblock ISSN 0004-637X.
\newblock \doi{10.1086/171870}.

\bibitem[{French} and {Zabludoff}(2018)]{French18}
K.~Decker {French} and Ann~I. {Zabludoff}.
\newblock {Identifying Tidal Disruption Events via Prior Photometric Selection
  of Their Preferred Hosts}.
\newblock \emph{ApJ}, 868:\penalty0 99, Dec 2018.
\newblock \doi{10.3847/1538-4357/aaea64}.

\bibitem[Freund and Schapire(1995)]{Freund1995}
Yoav Freund and Robert~E. Schapire.
\newblock A desicion-theoretic generalization of on-line learning and an
  application to boosting.
\newblock In Paul Vit{\'a}nyi, editor, \emph{Computational {{Learning
  Theory}}}, Lecture {{Notes}} in {{Computer Science}}, pages 23--37. {Springer
  Berlin Heidelberg}, 1995.
\newblock ISBN 978-3-540-49195-8.

\bibitem[Friedman(2001)]{Friedman2001}
Jerome~H. Friedman.
\newblock Greedy {{Function Approximation}}: {{A Gradient Boosting Machine}}.
\newblock \emph{The Annals of Statistics}, 29\penalty0 (5):\penalty0
  1189--1232, 2001.
\newblock ISSN 0090-5364.

\bibitem[{Frontera-Pons} et~al.(2017){Frontera-Pons}, {Sureau}, {Bobin}, and
  {Le Floc'h}]{Frontera17}
J.~{Frontera-Pons}, F.~{Sureau}, J.~{Bobin}, and E.~{Le Floc'h}.
\newblock {Unsupervised feature-learning for galaxy SEDs with denoising
  autoencoders}.
\newblock \emph{A\&A}, 603:\penalty0 A60, Jul 2017.
\newblock \doi{10.1051/0004-6361/201630240}.

\bibitem[{Fujimoto} et~al.(2018){Fujimoto}, {Fukushima}, and
  {Murase}]{Fujimoto18}
Yuki {Fujimoto}, Kenji {Fukushima}, and Koichi {Murase}.
\newblock {Methodology study of machine learning for the neutron star equation
  of state}.
\newblock \emph{PhysRevD}, 98:\penalty0 023019, Jul 2018.
\newblock \doi{10.1103/PhysRevD.98.023019}.

\bibitem[Fukushima(1980)]{Fukushima1980}
Kunihiko Fukushima.
\newblock Neocognitron: {{A}} self-organizing neural network model for a
  mechanism of pattern recognition unaffected by shift in position.
\newblock \emph{Biol. Cybernetics}, 36\penalty0 (4):\penalty0 193--202, April
  1980.
\newblock ISSN 0340-1200, 1432-0770.
\newblock \doi{10.1007/BF00344251}.

\bibitem[{Fussell} and {Moews}(2019)]{Fussell19}
Levi {Fussell} and Ben {Moews}.
\newblock {Forging new worlds: high-resolution synthetic galaxies with chained
  generative adversarial networks}.
\newblock \emph{MNRAS}, 485:\penalty0 3203--3214, Mar 2019.
\newblock \doi{10.1093/mnras/stz602}.

\bibitem[{Gaia Collaboration} and {et al.}(2016{\natexlab{a}})]{GAIA16a}
{Gaia Collaboration} and {et al.}
\newblock {The Gaia mission}.
\newblock \emph{A\&A}, 595:\penalty0 A1, Nov 2016{\natexlab{a}}.
\newblock \doi{10.1051/0004-6361/201629272}.

\bibitem[{Gaia Collaboration} and {et al.}(2016{\natexlab{b}})]{GAIA16b}
{Gaia Collaboration} and {et al.}
\newblock {Gaia Data Release 1. Summary of the astrometric, photometric, and
  survey properties}.
\newblock \emph{A\&A}, 595:\penalty0 A2, Nov 2016{\natexlab{b}}.
\newblock \doi{10.1051/0004-6361/201629512}.

\bibitem[{Gaia Collaboration} and {et al.}(2018)]{GAIA18}
{Gaia Collaboration} and {et al.}
\newblock {Gaia Data Release 2. Summary of the contents and survey properties}.
\newblock \emph{A\&A}, 616:\penalty0 A1, Aug 2018.
\newblock \doi{10.1051/0004-6361/201833051}.

\bibitem[{Gao}(2018b)]{Gao18b}
Xin-Hua {Gao}.
\newblock {Memberships, distance and proper-motion of the open cluster NGC 188
  based on a machine learning method}.
\newblock \emph{Ap\&SS}, 363:\penalty0 232, Nov 2018b.
\newblock \doi{10.1007/s10509-018-3453-4}.

\bibitem[{Gao}(2018a)]{Gao18a}
Xinhua {Gao}.
\newblock {A Machine-learning-based Investigation of the Open Cluster M67}.
\newblock \emph{ApJ}, 869:\penalty0 9, Dec 2018a.
\newblock \doi{10.3847/1538-4357/aae8dd}.

\bibitem[{Garcia-Dias} et~al.(2018){Garcia-Dias}, {Allende Prieto},
  {S{\'a}nchez Almeida}, and {Ordov{\'a}s-Pascual}]{Garcia18}
Rafael {Garcia-Dias}, Carlos {Allende Prieto}, Jorge {S{\'a}nchez Almeida}, and
  Ignacio {Ordov{\'a}s-Pascual}.
\newblock {Machine learning in APOGEE. Unsupervised spectral classification
  with K-means}.
\newblock \emph{A\&A}, 612:\penalty0 A98, May 2018.
\newblock \doi{10.1051/0004-6361/201732134}.

\bibitem[{George} and {Huerta}(2018{\natexlab{a}})]{George18a}
Daniel {George} and E.~A. {Huerta}.
\newblock {Deep neural networks to enable real-time multimessenger
  astrophysics}.
\newblock \emph{PRD}, 97:\penalty0 044039, February 2018{\natexlab{a}}.
\newblock \doi{10.1103/PhysRevD.97.044039}.

\bibitem[{George} and {Huerta}(2018{\natexlab{b}})]{George18b}
Daniel {George} and E.~A. {Huerta}.
\newblock {Deep Learning for real-time gravitational wave detection and
  parameter estimation: Results with Advanced LIGO data}.
\newblock \emph{Physics Letters B}, 778:\penalty0 64--70, March
  2018{\natexlab{b}}.
\newblock \doi{10.1016/j.physletb.2017.12.053}.

\bibitem[{Gichu} and {Ogohara}(2019)]{Gichu19}
Ryusei {Gichu} and Kazunori {Ogohara}.
\newblock {Segmentation of dust storm areas on Mars images using principal
  component analysis and neural network}.
\newblock \emph{Progress in Earth and Planetary Science}, 6:\penalty0 19, Feb
  2019.
\newblock \doi{10.1186/s40645-019-0266-1}.

\bibitem[{Gieseke} et~al.(2017){Gieseke}, {Bloemen}, {van den Bogaard},
  {Heskes}, {Kindler}, {Scalzo}, {Ribeiro}, {van Roestel}, {Groot}, {Yuan},
  {M{\"o}ller}, and {Tucker}]{Gieseke17}
Fabian {Gieseke}, Steven {Bloemen}, Cas {van den Bogaard}, Tom {Heskes}, Jonas
  {Kindler}, Richard~A. {Scalzo}, Val{\'e}rio A.~R.~M. {Ribeiro}, Jan {van
  Roestel}, Paul~J. {Groot}, Fang {Yuan}, Anais {M{\"o}ller}, and Brad~E.
  {Tucker}.
\newblock {Convolutional neural networks for transient candidate vetting in
  large-scale surveys}.
\newblock \emph{MNRAS}, 472:\penalty0 3101--3114, Dec 2017.
\newblock \doi{10.1093/mnras/stx2161}.

\bibitem[{Giles} and {Walkowicz}(2019)]{Giles19}
Daniel {Giles} and Lucianne {Walkowicz}.
\newblock {Systematic serendipity: a test of unsupervised machine learning as a
  method for anomaly detection}.
\newblock \emph{MNRAS}, 484:\penalty0 834--849, Mar 2019.
\newblock \doi{10.1093/mnras/sty3461}.

\bibitem[{Gomez Gonzalez} et~al.(2018){Gomez Gonzalez}, {Absil}, and {Van
  Droogenbroeck}]{Gomez18}
C.~A. {Gomez Gonzalez}, O.~{Absil}, and M.~{Van Droogenbroeck}.
\newblock {Supervised detection of exoplanets in high-contrast imaging
  sequences}.
\newblock \emph{A\&A}, 613:\penalty0 A71, Jun 2018.
\newblock \doi{10.1051/0004-6361/201731961}.

\bibitem[{Gonz{\'a}lez} and {Guzm{\'a}n}(2018)]{Gonzalez18}
J.~A. {Gonz{\'a}lez} and F.~S. {Guzm{\'a}n}.
\newblock {Characterizing the velocity of a wandering black hole and properties
  of the surrounding medium using convolutional neural networks}.
\newblock \emph{PhysRevD}, 97:\penalty0 063001, Mar 2018.
\newblock \doi{10.1103/PhysRevD.97.063001}.

\bibitem[{Gonz{\'a}lez-Solares} et~al.(2008){Gonz{\'a}lez-Solares}, {Walton},
  {Greimel}, {Drew}, {Irwin}, {Sale}, {Andrews}, {Aungwerojwit}, {Barlow}, {van
  den Besselaar}, {Corradi}, {G{\"a}nsicke}, {Groot}, {Hales}, {Hopewell},
  {Hu}, {Irwin}, {Knigge}, {Lagadec}, {Leisy}, {Lewis}, {Mampaso}, {Matsuura},
  {Moont}, {Morales-Rueda}, {Morris}, {Naylor}, {Parker}, {Prema}, {Pyrzas},
  {Rixon}, {Rodr{\'\i}guez-Gil}, {Roelofs}, {Sabin}, {Skillen}, {Suso}, {Tata},
  {Viironen}, {Vink}, {Witham}, {Wright}, {Zijlstra}, {Zurita}, {Drake},
  {Fabregat}, {Lennon}, {Lucas}, {Mart{\'\i}n}, {Phillipps}, {Steeghs}, and
  {Unruh}]{Gonzalez08}
E.~A. {Gonz{\'a}lez-Solares}, N.~A. {Walton}, R.~{Greimel}, J.~E. {Drew}, M.~J.
  {Irwin}, S.~E. {Sale}, K.~{Andrews}, A.~{Aungwerojwit}, M.~J. {Barlow},
  E.~{van den Besselaar}, R.~L.~M. {Corradi}, B.~T. {G{\"a}nsicke}, P.~J.
  {Groot}, A.~S. {Hales}, E.~C. {Hopewell}, Haili {Hu}, J.~{Irwin},
  C.~{Knigge}, E.~{Lagadec}, P.~{Leisy}, J.~R. {Lewis}, A.~{Mampaso},
  M.~{Matsuura}, B.~{Moont}, L.~{Morales-Rueda}, R.~A.~H. {Morris},
  T.~{Naylor}, Q.~A. {Parker}, P.~{Prema}, S.~{Pyrzas}, G.~T. {Rixon},
  P.~{Rodr{\'\i}guez-Gil}, G.~{Roelofs}, L.~{Sabin}, I.~{Skillen}, J.~{Suso},
  R.~{Tata}, K.~{Viironen}, J.~S. {Vink}, A.~{Witham}, N.~J. {Wright}, A.~A.
  {Zijlstra}, A.~{Zurita}, J.~{Drake}, J.~{Fabregat}, D.~J. {Lennon}, P.~W.
  {Lucas}, E.~L. {Mart{\'\i}n}, S.~{Phillipps}, D.~{Steeghs}, and Y.~C.
  {Unruh}.
\newblock {Initial data release from the INT Photometric H{\ensuremath{\alpha}}
  Survey of the Northern Galactic Plane (IPHAS)}.
\newblock \emph{MNRAS}, 388\penalty0 (1):\penalty0 89--104, Jul 2008.
\newblock \doi{10.1111/j.1365-2966.2008.13399.x}.

\bibitem[{Goulding} et~al.(2018){Goulding}, {Greene}, {Bezanson}, {Greco},
  {Johnson}, {Leauthaud}, {Matsuoka}, {Medezinski}, and
  {Price-Whelan}]{Goulding18}
Andy~D. {Goulding}, Jenny~E. {Greene}, Rachel {Bezanson}, Johnny {Greco}, Sean
  {Johnson}, Alexie {Leauthaud}, Yoshiki {Matsuoka}, Elinor {Medezinski}, and
  Adrian~M. {Price-Whelan}.
\newblock {Galaxy interactions trigger rapid black hole growth: An
  unprecedented view from the Hyper Suprime-Cam survey}.
\newblock \emph{PASJ}, 70:\penalty0 S37, January 2018.
\newblock \doi{10.1093/pasj/psx135}.

\bibitem[{Graham} et~al.(2013){Graham}, {Djorgovski}, {Mahabal}, {Donalek}, and
  {Drake}]{Graham13}
Matthew~J. {Graham}, S.~G. {Djorgovski}, Ashish~A. {Mahabal}, Ciro {Donalek},
  and Andrew~J. {Drake}.
\newblock {Machine-assisted discovery of relationships in astronomy}.
\newblock \emph{MNRAS}, 431\penalty0 (3):\penalty0 2371--2384, May 2013.
\newblock \doi{10.1093/mnras/stt329}.

\bibitem[{Harry} and {LIGO Scientific Collaboration}(2010)]{Harry10}
Gregory~M. {Harry} and {LIGO Scientific Collaboration}.
\newblock {Advanced LIGO: the next generation of gravitational wave detectors}.
\newblock \emph{Classical and Quantum Gravity}, 27\penalty0 (8):\penalty0
  084006, Apr 2010.
\newblock \doi{10.1088/0264-9381/27/8/084006}.

\bibitem[{Hartley} et~al.(2017){Hartley}, {Flamary}, {Jackson}, {Tagore}, and
  {Metcalf}]{Hartley17}
P.~{Hartley}, R.~{Flamary}, N.~{Jackson}, A.~S. {Tagore}, and R.~B. {Metcalf}.
\newblock {Support vector machine classification of strong gravitational
  lenses}.
\newblock \emph{MNRAS}, 471:\penalty0 3378--3397, Nov 2017.
\newblock \doi{10.1093/mnras/stx1733}.

\bibitem[{Heck} et~al.(1985){Heck}, {Murtagh}, and {Ponz}]{Heck1985}
A.~{Heck}, F.~{Murtagh}, and D.~{Ponz}.
\newblock {The Increasing Importance of Statistical Methods in Astronomy}.
\newblock \emph{The Messenger}, 41:\penalty0 22--25, September 1985.

\bibitem[{Hedges} et~al.(2018){Hedges}, {Hodgkin}, and {Kennedy}]{Hedges18}
Christina {Hedges}, Simon {Hodgkin}, and Grant {Kennedy}.
\newblock {Discovery of new dipper stars with K2: a window into the inner disc
  region of T Tauri stars}.
\newblock \emph{MNRAS}, 476:\penalty0 2968--2998, May 2018.
\newblock \doi{10.1093/mnras/sty328}.

\bibitem[{Heinze} et~al.(2018){Heinze}, {Tonry}, {Denneau}, {Flewelling},
  {Stalder}, {Rest}, {Smith}, {Smartt}, and {Weiland}]{Heinze18}
A.~N. {Heinze}, J.~L. {Tonry}, L.~{Denneau}, H.~{Flewelling}, B.~{Stalder},
  A.~{Rest}, K.~W. {Smith}, S.~J. {Smartt}, and H.~{Weiland}.
\newblock {A First Catalog of Variable Stars Measured by the Asteroid
  Terrestrial-impact Last Alert System (ATLAS)}.
\newblock \emph{AJ}, 156:\penalty0 241, November 2018.
\newblock \doi{10.3847/1538-3881/aae47f}.

\bibitem[{Hildebrandt} et~al.(2010){Hildebrandt}, {Arnouts}, {Capak},
  {Moustakas}, {Wolf}, {Abdalla}, {Assef}, {Banerji}, {Ben{\'\i}tez},
  {Brammer}, {Budav{\'a}ri}, {Carliles}, {Coe}, {Dahlen}, {Feldmann}, {Gerdes},
  {Gillis}, {Ilbert}, {Kotulla}, {Lahav}, {Li}, {Miralles}, {Purger},
  {Schmidt}, and {Singal}]{Hildebrandt10}
H.~{Hildebrandt}, S.~{Arnouts}, P.~{Capak}, L.~A. {Moustakas}, C.~{Wolf}, F.~B.
  {Abdalla}, R.~J. {Assef}, M.~{Banerji}, N.~{Ben{\'\i}tez}, G.~B. {Brammer},
  T.~{Budav{\'a}ri}, S.~{Carliles}, D.~{Coe}, T.~{Dahlen}, R.~{Feldmann},
  D.~{Gerdes}, B.~{Gillis}, O.~{Ilbert}, R.~{Kotulla}, O.~{Lahav}, I.~H. {Li},
  J.~M. {Miralles}, N.~{Purger}, S.~{Schmidt}, and J.~{Singal}.
\newblock {PHAT: PHoto-z Accuracy Testing}.
\newblock \emph{A\&A}, 523:\penalty0 A31, Nov 2010.
\newblock \doi{10.1051/0004-6361/201014885}.

\bibitem[{Ho}(2019)]{Ho19}
I.~Ting {Ho}.
\newblock {A Machine Learning Artificial Neural Network Calibration of the
  Strong-Line Oxygen Abundance}.
\newblock \emph{MNRAS}, 485:\penalty0 3569--3579, Mar 2019.
\newblock \doi{10.1093/mnras/stz649}.

\bibitem[{Ho, T.K.}(1995)]{Ho1995}
{Ho, T.K.}
\newblock Random decision forests.
\newblock In \emph{Proceedings of 3rd {{International Conference}} on
  {{Document Analysis}} and {{Recognition}}}, volume~1, pages 278--282, August
  1995.
\newblock \doi{10.1109/ICDAR.1995.598994}.

\bibitem[{Hocking} et~al.(2018){Hocking}, {Geach}, {Sun}, and
  {Davey}]{Hocking18}
Alex {Hocking}, James~E. {Geach}, Yi~{Sun}, and Neil {Davey}.
\newblock {An automatic taxonomy of galaxy morphology using unsupervised
  machine learning}.
\newblock \emph{MNRAS}, 473:\penalty0 1108--1129, Jan 2018.
\newblock \doi{10.1093/mnras/stx2351}.

\bibitem[{Hoyle}(2016)]{Hoyle16}
B.~{Hoyle}.
\newblock {Measuring photometric redshifts using galaxy images and Deep Neural
  Networks}.
\newblock \emph{A\&C}, 16:\penalty0 34--40, Jul 2016.
\newblock \doi{10.1016/j.ascom.2016.03.006}.

\bibitem[Hoyle(2016)]{hoyleMeasuringPhotometricRedshifts2016a}
B.~Hoyle.
\newblock Measuring photometric redshifts using galaxy images and {{Deep Neural
  Networks}}.
\newblock \emph{A\&C}, 16:\penalty0 34--40, July 2016.
\newblock ISSN 2213-1337.
\newblock \doi{10.1016/j.ascom.2016.03.006}.

\bibitem[{Hoyle} et~al.(2015){Hoyle}, {Rau}, {Paech}, {Bonnett}, {Seitz}, and
  {Weller}]{Hoyle15}
Ben {Hoyle}, Markus~Michael {Rau}, Kerstin {Paech}, Christopher {Bonnett},
  Stella {Seitz}, and Jochen {Weller}.
\newblock {Anomaly detection for machine learning redshifts applied to SDSS
  galaxies}.
\newblock \emph{MNRAS}, 452\penalty0 (4):\penalty0 4183--4194, Oct 2015.
\newblock \doi{10.1093/mnras/stv1551}.

\bibitem[Hoyle et~al.(2015{\natexlab{a}})Hoyle, Rau, Zitlau, Seitz, and
  Weller]{hoyleFeatureImportanceMachine2015}
Ben Hoyle, Markus~Michael Rau, Roman Zitlau, Stella Seitz, and Jochen Weller.
\newblock Feature importance for machine learning redshifts applied to {{SDSS}}
  galaxies.
\newblock \emph{MNRAS}, 449\penalty0 (2):\penalty0 1275--1283, May
  2015{\natexlab{a}}.
\newblock ISSN 0035-8711.
\newblock \doi{10.1093/mnras/stv373}.

\bibitem[Hoyle et~al.(2015{\natexlab{b}})Hoyle, Rau, Zitlau, Seitz, and
  Weller]{hoyleFeatureImportanceMachine2015a}
Ben Hoyle, Markus~Michael Rau, Roman Zitlau, Stella Seitz, and Jochen Weller.
\newblock Feature importance for machine learning redshifts applied to {{SDSS}}
  galaxies.
\newblock \emph{MNRAS}, 449\penalty0 (2):\penalty0 1275--1283, May
  2015{\natexlab{b}}.
\newblock ISSN 0035-8711.
\newblock \doi{10.1093/mnras/stv373}.

\bibitem[{Huerta} et~al.(2018){Huerta}, {Moore}, {Kumar}, {George}, {Chua},
  {Haas}, {Wessel}, {Johnson}, {Glennon}, {Rebei}, {Holgado}, {Gair}, and
  {Pfeiffer}]{Huerta18}
E.~A. {Huerta}, C.~J. {Moore}, Prayush {Kumar}, Daniel {George}, Alvin J.~K.
  {Chua}, Roland {Haas}, Erik {Wessel}, Daniel {Johnson}, Derek {Glennon}, Adam
  {Rebei}, A.~Miguel {Holgado}, Jonathan~R. {Gair}, and Harald~P. {Pfeiffer}.
\newblock {Eccentric, nonspinning, inspiral, Gaussian-process merger
  approximant for the detection and characterization of eccentric binary black
  hole mergers}.
\newblock \emph{PhysRevD}, 97:\penalty0 024031, Jan 2018.
\newblock \doi{10.1103/PhysRevD.97.024031}.

\bibitem[{Huertas-Company} et~al.(2015){Huertas-Company}, Gravet,
  {Cabrera-Vives}, {P{\'e}rez-Gonz{\'a}lez}, Kartaltepe, Barro, Bernardi, Mei,
  Shankar, Dimauro, Bell, Kocevski, Koo, Faber, and
  Mcintosh]{huertas-companyCATALOGVISUALLIKEMORPHOLOGIES2015}
M.~{Huertas-Company}, R.~Gravet, G.~{Cabrera-Vives}, P.~G.
  {P{\'e}rez-Gonz{\'a}lez}, J.~S. Kartaltepe, G.~Barro, M.~Bernardi, S.~Mei,
  F.~Shankar, P.~Dimauro, E.~F. Bell, D.~Kocevski, D.~C. Koo, S.~M. Faber, and
  D.~H. Mcintosh.
\newblock A {{CATALOG OF VISUAL}}-{{LIKE MORPHOLOGIES IN THE}} 5 {{CANDELS
  FIELDS USING DEEP LEARNING}}.
\newblock \emph{ApJS}, 221\penalty0 (1):\penalty0 8, October 2015.
\newblock ISSN 0067-0049.
\newblock \doi{10.1088/0067-0049/221/1/8}.

\bibitem[{Huertas-Company} et~al.(2018){Huertas-Company}, {Primack}, {Dekel},
  {Koo}, {Lapiner}, {Ceverino}, {Simons}, {Snyder}, {Bernardi}, {Chen},
  {Dom{\'\i}nguez-S{\'a}nchez}, {Lee}, {Margalef-Bentabol}, and
  {Tuccillo}]{Huertas18}
M.~{Huertas-Company}, J.~R. {Primack}, A.~{Dekel}, D.~C. {Koo}, S.~{Lapiner},
  D.~{Ceverino}, R.~C. {Simons}, G.~F. {Snyder}, M.~{Bernardi}, Z.~{Chen},
  H.~{Dom{\'\i}nguez-S{\'a}nchez}, C.~T. {Lee}, B.~{Margalef-Bentabol}, and
  D.~{Tuccillo}.
\newblock {Deep Learning Identifies High-z Galaxies in a Central Blue Nugget
  Phase in a Characteristic Mass Range}.
\newblock \emph{ApJ}, 858:\penalty0 114, May 2018.
\newblock \doi{10.3847/1538-4357/aabfed}.

\bibitem[{Hui} et~al.(2018){Hui}, {Aragon}, {Cui}, and {Flegal}]{Hui18}
Jianan {Hui}, Miguel {Aragon}, Xinping {Cui}, and James~M. {Flegal}.
\newblock {A machine learning approach to galaxy-LSS classification - I.
  Imprints on halo merger trees}.
\newblock \emph{MNRAS}, 475:\penalty0 4494--4503, Apr 2018.
\newblock \doi{10.1093/mnras/stx3235}.

\bibitem[{Inceoglu} et~al.(2018){Inceoglu}, {Jeppesen}, {Kongstad},
  {Hern{\'a}ndez Marcano}, {Jacobsen}, and {Karoff}]{Inceoglu18}
Fadil {Inceoglu}, Jacob~H. {Jeppesen}, Peter {Kongstad}, N{\'e}stor~J.
  {Hern{\'a}ndez Marcano}, Rune~H. {Jacobsen}, and Christoffer {Karoff}.
\newblock {Using Machine Learning Methods to Forecast if Solar Flares Will Be
  Associated with CMEs and SEPs}.
\newblock \emph{ApJ}, 861:\penalty0 128, Jul 2018.
\newblock \doi{10.3847/1538-4357/aac81e}.

\bibitem[{Iodice} et~al.(2016){Iodice}, {Capaccioli}, {Grado}, {Limatola},
  {Spavone}, {Napolitano}, {Paolillo}, {Peletier}, {Cantiello}, {Lisker},
  {Wittmann}, {Venhola}, {Hilker}, {D'Abrusco}, {Pota}, and
  {Schipani}]{Iodice16}
E.~{Iodice}, M.~{Capaccioli}, A.~{Grado}, L.~{Limatola}, M.~{Spavone}, N.~R.
  {Napolitano}, M.~{Paolillo}, R.~F. {Peletier}, M.~{Cantiello}, T.~{Lisker},
  C.~{Wittmann}, A.~{Venhola}, M.~{Hilker}, R.~{D'Abrusco}, V.~{Pota}, and
  P.~{Schipani}.
\newblock {The Fornax Deep Survey with VST. I. The Extended and Diffuse Stellar
  Halo of NGC 1399 out to 192 kpc}.
\newblock \emph{ApJ}, 820\penalty0 (1):\penalty0 42, Mar 2016.
\newblock \doi{10.3847/0004-637X/820/1/42}.

\bibitem[{Ivezi{\'c}} and {et al.}(2019)]{Ivezic19}
{\v{Z}}eljko {Ivezi{\'c}} and {et al.}
\newblock {LSST: From Science Drivers to Reference Design and Anticipated Data
  Products}.
\newblock \emph{ApJ}, 873:\penalty0 111, Mar 2019.
\newblock \doi{10.3847/1538-4357/ab042c}.

\bibitem[{Ivezi{\'c}} et~al.(2014){Ivezi{\'c}}, {Connelly}, {Vand erPlas}, and
  {Gray}]{Ivezic2014}
{\v{Z}}eljko {Ivezi{\'c}}, Andrew~J. {Connelly}, Jacob~T. {Vand erPlas}, and
  Alexander {Gray}.
\newblock \emph{{Statistics, Data Mining, and Machine Learning in Astronomy}}.
\newblock 2014.

\bibitem[{Jacobs} et~al.(2017){Jacobs}, {Glazebrook}, {Collett}, {More}, and
  {McCarthy}]{Jacobs17}
C.~{Jacobs}, K.~{Glazebrook}, T.~{Collett}, A.~{More}, and C.~{McCarthy}.
\newblock {Finding strong lenses in CFHTLS using convolutional neural
  networks}.
\newblock \emph{MNRAS}, 471:\penalty0 167--181, Oct 2017.
\newblock \doi{10.1093/mnras/stx1492}.

\bibitem[Jacobs et~al.(2019{\natexlab{a}})Jacobs, Collett, Glazebrook,
  {Buckley-Geer}, Diehl, Lin, McCarthy, Qin, Odden, Escudero, Dial, Yung,
  Gaitsch, Pellico, Lindgren, Abbott, Annis, Avila, Brooks, Burke, Rosell,
  Kind, Carretero, da~Costa, Vicente, Fosalba, Frieman, {Garc{\'i}a-Bellido},
  Gaztanaga, Goldstein, Gruen, Gruendl, Gschwend, Hollowood, Honscheid, Hoyle,
  James, Krause, Kuropatkin, Lahav, Lima, Maia, Marshall, Miquel, Plazas,
  Roodman, Sanchez, Scarpine, Serrano, {Sevilla-Noarbe}, Smith, Sobreira,
  Suchyta, Swanson, Tarle, Vikram, Walker, and
  {and}]{jacobsExtendedCatalogGalaxy2019}
C.~Jacobs, T.~Collett, K.~Glazebrook, E.~{Buckley-Geer}, H.~T. Diehl, H.~Lin,
  C.~McCarthy, A.~K. Qin, C.~Odden, M.~Caso Escudero, P.~Dial, V.~J. Yung,
  S.~Gaitsch, A.~Pellico, K.~A. Lindgren, T.~M.~C. Abbott, J.~Annis, S.~Avila,
  D.~Brooks, D.~L. Burke, A.~Carnero Rosell, M.~Carrasco Kind, J.~Carretero,
  L.~N. da~Costa, J.~De Vicente, P.~Fosalba, J.~Frieman,
  J.~{Garc{\'i}a-Bellido}, E.~Gaztanaga, D.~A. Goldstein, D.~Gruen, R.~A.
  Gruendl, J.~Gschwend, D.~L. Hollowood, K.~Honscheid, B.~Hoyle, D.~J. James,
  E.~Krause, N.~Kuropatkin, O.~Lahav, M.~Lima, M.~A.~G. Maia, J.~L. Marshall,
  R.~Miquel, A.~A. Plazas, A.~Roodman, E.~Sanchez, V.~Scarpine, S.~Serrano,
  I.~{Sevilla-Noarbe}, M.~Smith, F.~Sobreira, E.~Suchyta, M.~E.~C. Swanson,
  G.~Tarle, V.~Vikram, A.~R. Walker, and Y.~Zhang {and}.
\newblock An {{Extended Catalog}} of {{Galaxy}}\textendash{{Galaxy Strong
  Gravitational Lenses Discovered}} in {{DES Using Convolutional Neural
  Networks}}.
\newblock \emph{ApJS}, 243\penalty0 (1):\penalty0 17, July 2019{\natexlab{a}}.
\newblock ISSN 0067-0049.
\newblock \doi{10.3847/1538-4365/ab26b6}.

\bibitem[Jacobs et~al.(2019{\natexlab{b}})Jacobs, Collett, Glazebrook,
  McCarthy, Qin, Abbott, Abdalla, Annis, Avila, Bechtol, Bertin, Brooks,
  {Buckley-Geer}, Burke, Carnero~Rosell, Carrasco~Kind, Carretero, {da Costa},
  Davis, De~Vicente, Desai, Diehl, Doel, Eifler, Flaugher, Frieman,
  {Garc{\'i}a-Bellido}, Gaztanaga, Gerdes, Goldstein, Gruen, Gruendl, Gschwend,
  Gutierrez, Hartley, Hollowood, Honscheid, Hoyle, James, Kuehn, Kuropatkin,
  Lahav, Li, Lima, Lin, Maia, Martini, Miller, Miquel, Nord, Plazas, Sanchez,
  Scarpine, Schubnell, Serrano, {Sevilla-Noarbe}, Smith, {Soares-Santos},
  Sobreira, Suchyta, Swanson, Tarle, Vikram, Walker, Zhang, and
  Zuntz]{jacobsFindingHighredshiftStrong2019}
C.~Jacobs, T.~Collett, K.~Glazebrook, C.~McCarthy, A.~K. Qin, T.~M.~C. Abbott,
  F.~B. Abdalla, J.~Annis, S.~Avila, K.~Bechtol, E.~Bertin, D.~Brooks,
  E.~{Buckley-Geer}, D.~L. Burke, A.~Carnero~Rosell, M.~Carrasco~Kind,
  J.~Carretero, L.~N. {da Costa}, C.~Davis, J.~De~Vicente, S.~Desai, H.~T.
  Diehl, P.~Doel, T.~F. Eifler, B.~Flaugher, J.~Frieman,
  J.~{Garc{\'i}a-Bellido}, E.~Gaztanaga, D.~W. Gerdes, D.~A. Goldstein,
  D.~Gruen, R.~A. Gruendl, J.~Gschwend, G.~Gutierrez, W.~G. Hartley, D.~L.
  Hollowood, K.~Honscheid, B.~Hoyle, D.~J. James, K.~Kuehn, N.~Kuropatkin,
  O.~Lahav, T.~S. Li, M.~Lima, H.~Lin, M.~a.~G. Maia, P.~Martini, C.~J. Miller,
  R.~Miquel, B.~Nord, A.~A. Plazas, E.~Sanchez, V.~Scarpine, M.~Schubnell,
  S.~Serrano, I.~{Sevilla-Noarbe}, M.~Smith, M.~{Soares-Santos}, F.~Sobreira,
  E.~Suchyta, M.~E.~C. Swanson, G.~Tarle, V.~Vikram, A.~R. Walker, Y.~Zhang,
  and J.~Zuntz.
\newblock Finding high-redshift strong lenses in {{DES}} using convolutional
  neural networks.
\newblock \emph{MNRAS}, 484\penalty0 (4):\penalty0 5330--5349, April
  2019{\natexlab{b}}.
\newblock ISSN 0035-8711.
\newblock \doi{10.1093/mnras/stz272}.

\bibitem[{Jaschek}(1978)]{Jaschek78}
C.~{Jaschek}.
\newblock {Data Growth in Astronomy}.
\newblock \emph{QJRAS}, 19:\penalty0 269, September 1978.

\bibitem[Jaschek(1968)]{Jaschek68}
Carlos Jaschek.
\newblock Information problems in astrophysics.
\newblock \emph{PASP}, 80:\penalty0 654, dec 1968.
\newblock \doi{10.1086/128707}.

\bibitem[{Johnston} et~al.(2007){Johnston}, {Bailes}, {Bartel}, {Baugh},
  {Bietenholz}, {Blake}, {Braun}, {Brown}, {Chatterjee}, {Darling}, {Deller},
  {Dodson}, {Edwards}, {Ekers}, {Ellingsen}, {Feain}, {Gaensler}, {Haverkorn},
  {Hobbs}, {Hopkins}, {Jackson}, {James}, {Joncas}, {Kaspi}, {Kilborn},
  {Koribalski}, {Kothes}, {Landecker}, {Lenc}, {Lovell}, {Macquart},
  {Manchester}, {Matthews}, {McClure-Griffiths}, {Norris}, {Pen}, {Phillips},
  {Power}, {Protheroe}, {Sadler}, {Schmidt}, {Stairs}, {Staveley-Smith},
  {Stil}, {Taylor}, {Tingay}, {Tzioumis}, {Walker}, {Wall}, and
  {Wolleben}]{Johnston07}
S.~{Johnston}, M.~{Bailes}, N.~{Bartel}, C.~{Baugh}, M.~{Bietenholz},
  C.~{Blake}, R.~{Braun}, J.~{Brown}, S.~{Chatterjee}, J.~{Darling},
  A.~{Deller}, R.~{Dodson}, P.~G. {Edwards}, R.~{Ekers}, S.~{Ellingsen},
  I.~{Feain}, B.~M. {Gaensler}, M.~{Haverkorn}, G.~{Hobbs}, A.~{Hopkins},
  C.~{Jackson}, C.~{James}, G.~{Joncas}, V.~{Kaspi}, V.~{Kilborn},
  B.~{Koribalski}, R.~{Kothes}, T.~L. {Landecker}, E.~{Lenc}, J.~{Lovell},
  J.~P. {Macquart}, R.~{Manchester}, D.~{Matthews}, N.~M. {McClure-Griffiths},
  R.~{Norris}, U.~L. {Pen}, C.~{Phillips}, C.~{Power}, R.~{Protheroe},
  E.~{Sadler}, B.~{Schmidt}, I.~{Stairs}, L.~{Staveley-Smith}, J.~{Stil},
  R.~{Taylor}, S.~{Tingay}, A.~{Tzioumis}, M.~{Walker}, J.~{Wall}, and
  M.~{Wolleben}.
\newblock {Science with the Australian Square Kilometre Array Pathfinder}.
\newblock \emph{PASA}, 24:\penalty0 174--188, Dec 2007.
\newblock \doi{10.1071/AS07033}.

\bibitem[{Johnston} et~al.(2008){Johnston}, {Taylor}, {Bailes}, {Bartel},
  {Baugh}, {Bietenholz}, {Blake}, {Braun}, {Brown}, {Chatterjee}, {Darling},
  {Deller}, {Dodson}, {Edwards}, {Ekers}, {Ellingsen}, {Feain}, {Gaensler},
  {Haverkorn}, {Hobbs}, {Hopkins}, {Jackson}, {James}, {Joncas}, {Kaspi},
  {Kilborn}, {Koribalski}, {Kothes}, {Landecker}, {Lenc}, {Lovell}, {Macquart},
  {Manchester}, {Matthews}, {McClure-Griffiths}, {Norris}, {Pen}, {Phillips},
  {Power}, {Protheroe}, {Sadler}, {Schmidt}, {Stairs}, {Staveley-Smith},
  {Stil}, {Tingay}, {Tzioumis}, {Walker}, {Wall}, and {Wolleben}]{Johnston08}
S.~{Johnston}, R.~{Taylor}, M.~{Bailes}, N.~{Bartel}, C.~{Baugh},
  M.~{Bietenholz}, C.~{Blake}, R.~{Braun}, J.~{Brown}, S.~{Chatterjee},
  J.~{Darling}, A.~{Deller}, R.~{Dodson}, P.~{Edwards}, R.~{Ekers},
  S.~{Ellingsen}, I.~{Feain}, B.~{Gaensler}, M.~{Haverkorn}, G.~{Hobbs},
  A.~{Hopkins}, C.~{Jackson}, C.~{James}, G.~{Joncas}, V.~{Kaspi},
  V.~{Kilborn}, B.~{Koribalski}, R.~{Kothes}, T.~{Landecker}, E.~{Lenc},
  J.~{Lovell}, J.~P. {Macquart}, R.~{Manchester}, D.~{Matthews},
  N.~{McClure-Griffiths}, R.~{Norris}, U.~L. {Pen}, C.~{Phillips}, C.~{Power},
  R.~{Protheroe}, E.~{Sadler}, B.~{Schmidt}, I.~{Stairs}, L.~{Staveley-Smith},
  J.~{Stil}, S.~{Tingay}, A.~{Tzioumis}, M.~{Walker}, J.~{Wall}, and
  M.~{Wolleben}.
\newblock {Science with ASKAP. The Australian square-kilometre-array
  pathfinder}.
\newblock \emph{ExA}, 22:\penalty0 151--273, Dec 2008.
\newblock \doi{10.1007/s10686-008-9124-7}.

\bibitem[{Jones} and {Singal}(2017)]{Jones17}
E.~{Jones} and J.~{Singal}.
\newblock {Analysis of a custom support vector machine for photometric redshift
  estimation and the inclusion of galaxy shape information}.
\newblock \emph{A\&A}, 600:\penalty0 A113, Apr 2017.
\newblock \doi{10.1051/0004-6361/201629558}.

\bibitem[{Kaiser}(2004)]{Kaiser04}
Nicholas {Kaiser}.
\newblock {Pan-STARRS: a wide-field optical survey telescope array}.
\newblock In Jr. {Oschmann}, Jacobus~M., editor, \emph{Ground-based
  Telescopes}, volume 5489 of \emph{Society of Photo-Optical Instrumentation
  Engineers (SPIE) Conference Series}, pages 11--22, Oct 2004.
\newblock \doi{10.1117/12.552472}.

\bibitem[{Kang} et~al.(2019){Kang}, {Fan}, {Mao}, {Wu}, {Feng}, and
  {Yin}]{Kang19}
Shi-Ju {Kang}, Jun-Hui {Fan}, Weiming {Mao}, Qingwen {Wu}, Jianchao {Feng}, and
  Yue {Yin}.
\newblock {Evaluating the Optical Classification of Fermi BCUs Using Machine
  Learning}.
\newblock \emph{ApJ}, 872:\penalty0 189, Feb 2019.
\newblock \doi{10.3847/1538-4357/ab0383}.

\bibitem[{Keller} et~al.(2007){Keller}, {Schmidt}, {Bessell}, {Conroy},
  {Francis}, {Granlund}, {Kowald}, {Oates}, {Martin-Jones}, {Preston},
  {Tisserand }, {Vaccarella}, and {Waterson}]{Keller07}
S.~C. {Keller}, B.~P. {Schmidt}, M.~S. {Bessell}, P.~G. {Conroy}, P.~{Francis},
  A.~{Granlund}, E.~{Kowald}, A.~P. {Oates}, T.~{Martin-Jones}, T.~{Preston},
  P.~{Tisserand }, A.~{Vaccarella}, and M.~F. {Waterson}.
\newblock {The SkyMapper Telescope and The Southern Sky Survey}.
\newblock \emph{PASA}, 24\penalty0 (1):\penalty0 1--12, May 2007.
\newblock \doi{10.1071/AS07001}.

\bibitem[{Kim} and {Brunner}(2017)]{Kim17}
Edward~J. {Kim} and Robert~J. {Brunner}.
\newblock {Star-galaxy classification using deep convolutional neural
  networks}.
\newblock \emph{MNRAS}, 464:\penalty0 4463--4475, Feb 2017.
\newblock \doi{10.1093/mnras/stw2672}.

\bibitem[Kohonen(1990)]{kohonenSelforganizingMap1990}
T.~Kohonen.
\newblock The self-organizing map.
\newblock \emph{Proceedings of the IEEE}, 78\penalty0 (9):\penalty0 1464--1480,
  September 1990.
\newblock ISSN 0018-9219.
\newblock \doi{10.1109/5.58325}.

\bibitem[{Kong} et~al.(2018){Kong}, {Luo}, {Li}, {Wang}, {Li}, and
  {Zhao}]{Kong18}
Xiao {Kong}, A.~Li {Luo}, Xiang-Ru {Li}, You-Fen {Wang}, Yin-Bi {Li}, and
  Jing-Kun {Zhao}.
\newblock {Spectral Feature Extraction for DB White Dwarfs Through Machine
  Learning Applied to New Discoveries in the Sdss DR12 and DR14}.
\newblock \emph{PASP}, 130:\penalty0 084203, Aug 2018.
\newblock \doi{10.1088/1538-3873/aac7a8}.

\bibitem[{Koo}(1985)]{Koo85}
D.~C. {Koo}.
\newblock {Optical multicolors : a poor person's Z machine for galaxies.}
\newblock \emph{AJ}, 90:\penalty0 418--440, Mar 1985.
\newblock \doi{10.1086/113748}.

\bibitem[{Koo}(1999)]{Koo1999}
D.~C. {Koo}.
\newblock {Overview - Photometric Redshifts: A Perspective from an Old-Timer[!]
  on their Past, Present, and Potential}.
\newblock In Ray {Weymann}, Lisa {Storrie-Lombardi}, Marcin {Sawicki}, and
  Robert {Brunner}, editors, \emph{Photometric Redshifts and the Detection of
  High Redshift Galaxies}, volume 191 of \emph{Astronomical Society of the
  Pacific Conference Series}, page~3, Jan 1999.

\bibitem[{Krause} et~al.(2017){Krause}, {Pueschel}, and {Maier}]{Krause17}
Maria {Krause}, Elisa {Pueschel}, and Gernot {Maier}.
\newblock {Improved {\ensuremath{\gamma}}/hadron separation for the detection
  of faint {\ensuremath{\gamma}}-ray sources using boosted decision trees}.
\newblock \emph{Astroparticle Physics}, 89:\penalty0 1--9, Mar 2017.
\newblock \doi{10.1016/j.astropartphys.2017.01.004}.

\bibitem[{Kriessler} et~al.(1998){Kriessler}, {Han}, {Odewahn}, and
  {Beers}]{AutomatedMorphologicalClassification}
J.~R. {Kriessler}, E.~H. {Han}, S.~C. {Odewahn}, and T.~C. {Beers}.
\newblock {Automated Morphological Classification of Galaxies and the
  Morphology-Density Relation}.
\newblock In \emph{American Astronomical Society Meeting Abstracts}, volume
  193, page 38.20, Dec 1998.

\bibitem[Krizhevsky et~al.(2012)Krizhevsky, Sutskever, and
  Hinton]{krizhevskyImageNetClassificationDeep2012}
Alex Krizhevsky, Ilya Sutskever, and Geoffrey~E. Hinton.
\newblock {{ImageNet Classification}} with {{Deep Convolutional Neural
  Networks}}.
\newblock In F.~Pereira, C.~J.~C. Burges, L.~Bottou, and K.~Q. Weinberger,
  editors, \emph{Advances in {{Neural Information Processing Systems}} 25},
  pages 1097--1105. {Curran Associates, Inc.}, 2012.

\bibitem[{Ksoll} et~al.(2018){Ksoll}, {Gouliermis}, {Klessen}, {Grebel},
  {Sabbi}, {Anderson}, {Lennon}, {Cignoni}, {de Marchi}, {Smith}, {Tosi}, and
  {van der Marel}]{Ksoll18}
Victor~F. {Ksoll}, Dimitrios~A. {Gouliermis}, Ralf~S. {Klessen}, Eva~K.
  {Grebel}, Elena {Sabbi}, Jay {Anderson}, Daniel~J. {Lennon}, Michele
  {Cignoni}, Guido {de Marchi}, Linda~J. {Smith}, Monica {Tosi}, and Roeland~P.
  {van der Marel}.
\newblock {Hubble Tarantula Treasury Project - VI. Identification of
  pre-main-sequence stars using machine-learning techniques}.
\newblock \emph{MNRAS}, 479:\penalty0 2389--2414, Sep 2018.
\newblock \doi{10.1093/mnras/sty1317}.

\bibitem[{Kuminski} and {Shamir}(2018)]{Kuminski18}
E.~{Kuminski} and L.~{Shamir}.
\newblock {A hybrid approach to machine learning annotation of large galaxy
  image databases}.
\newblock \emph{A\&C}, 25:\penalty0 257--269, Oct 2018.
\newblock \doi{10.1016/j.ascom.2018.10.008}.

\bibitem[{Kuntzer} and {Courbin}(2017)]{Kuntzer17}
T.~{Kuntzer} and F.~{Courbin}.
\newblock {Detecting unresolved binary stars in Euclid VIS images}.
\newblock \emph{A\&A}, 606:\penalty0 A119, Oct 2017.
\newblock \doi{10.1051/0004-6361/201730792}.

\bibitem[{Kurtz} et~al.(2000){Kurtz}, {Eichhorn}, {Accomazzi}, {Grant},
  {Murray}, and {Watson}]{Kurtz00}
Michael~J. {Kurtz}, Guenther {Eichhorn}, Alberto {Accomazzi}, Carolyn~S.
  {Grant}, Stephen~S. {Murray}, and Joyce~M. {Watson}.
\newblock {The NASA Astrophysics Data System: Overview}.
\newblock \emph{A\&ASS}, 143:\penalty0 41--59, Apr 2000.
\newblock \doi{10.1051/aas:2000170}.

\bibitem[Lahav et~al.(1996)Lahav, Nairn, Sodr{\'e}, and
  {Storrie-Lombardi}]{lahavNeuralComputationTool1996}
O.~Lahav, A.~Nairn, L.~Sodr{\'e}, and M.~C. {Storrie-Lombardi}.
\newblock Neural computation as a tool for galaxy classification: Methods and
  examples.
\newblock \emph{MNRAS}, 283\penalty0 (1):\penalty0 207--221, October 1996.
\newblock ISSN 0035-8711.
\newblock \doi{10.1093/mnras/283.1.207}.

\bibitem[{Lanusse} et~al.(2018){Lanusse}, {Ma}, {Li}, {Collett}, {Li},
  {Ravanbakhsh}, {Mandelbaum}, and {P{\'o}czos}]{Lanusse18}
Fran{\c{c}}ois {Lanusse}, Quanbin {Ma}, Nan {Li}, Thomas~E. {Collett},
  Chun-Liang {Li}, Siamak {Ravanbakhsh}, Rachel {Mandelbaum}, and Barnab{\'a}s
  {P{\'o}czos}.
\newblock {CMU DeepLens: deep learning for automatic image-based galaxy-galaxy
  strong lens finding}.
\newblock \emph{MNRAS}, 473:\penalty0 3895--3906, January 2018.
\newblock \doi{10.1093/mnras/stx1665}.

\bibitem[{Laureijs} and {et al.}(2011)]{Laureijs11}
R.~{Laureijs} and {et al.}
\newblock {Euclid Definition Study Report}.
\newblock art. arXiv:1110.3193, Oct 2011.

\bibitem[LeCun et~al.(1989)LeCun, Boser, Denker, Henderson, Howard, Hubbard,
  and Jackel]{LeCun1989}
Y.~LeCun, B.~Boser, J.~S. Denker, D.~Henderson, R.~E. Howard, W.~Hubbard, and
  L.~D. Jackel.
\newblock Backpropagation {{Applied}} to {{Handwritten Zip Code Recognition}}.
\newblock \emph{Neural Comput.}, 1\penalty0 (4):\penalty0 541--551, December
  1989.
\newblock ISSN 0899-7667.
\newblock \doi{10.1162/neco.1989.1.4.541}.

\bibitem[{Leistedt} and {Hogg}(2017)]{Leistedt17}
Boris {Leistedt} and David~W. {Hogg}.
\newblock {Data-driven, Interpretable Photometric Redshifts Trained on
  Heterogeneous and Unrepresentative Data}.
\newblock \emph{ApJ}, 838:\penalty0 5, Mar 2017.
\newblock \doi{10.3847/1538-4357/aa6332}.

\bibitem[{Lemen} et~al.(2012){Lemen}, {Title}, {Akin}, {Boerner}, {Chou},
  {Drake}, {Duncan}, {Edwards}, {Friedlaender}, {Heyman}, {Hurlburt}, {Katz},
  {Kushner}, {Levay}, {Lindgren}, {Mathur}, {McFeaters}, {Mitchell}, {Rehse},
  {Schrijver}, {Springer}, {Stern}, {Tarbell}, {Wuelser}, {Wolfson}, {Yanari},
  {Bookbinder}, {Cheimets}, {Caldwell}, {Deluca}, {Gates}, {Golub}, {Park},
  {Podgorski}, {Bush}, {Scherrer}, {Gummin}, {Smith}, {Auker}, {Jerram},
  {Pool}, {Soufli}, {Windt}, {Beardsley}, {Clapp}, {Lang}, and
  {Waltham}]{Lemen12}
James~R. {Lemen}, Alan~M. {Title}, David~J. {Akin}, Paul~F. {Boerner},
  Catherine {Chou}, Jerry~F. {Drake}, Dexter~W. {Duncan}, Christopher~G.
  {Edwards}, Frank~M. {Friedlaender}, Gary~F. {Heyman}, Neal~E. {Hurlburt},
  Noah~L. {Katz}, Gary~D. {Kushner}, Michael {Levay}, Russell~W. {Lindgren},
  Dnyanesh~P. {Mathur}, Edward~L. {McFeaters}, Sarah {Mitchell}, Roger~A.
  {Rehse}, Carolus~J. {Schrijver}, Larry~A. {Springer}, Robert~A. {Stern},
  Theodore~D. {Tarbell}, Jean-Pierre {Wuelser}, C.~Jacob {Wolfson}, Carl
  {Yanari}, Jay~A. {Bookbinder}, Peter~N. {Cheimets}, David {Caldwell},
  Edward~E. {Deluca}, Richard {Gates}, Leon {Golub}, Sang {Park}, William~A.
  {Podgorski}, Rock~I. {Bush}, Philip~H. {Scherrer}, Mark~A. {Gummin}, Peter
  {Smith}, Gary {Auker}, Paul {Jerram}, Peter {Pool}, Regina {Soufli}, David~L.
  {Windt}, Sarah {Beardsley}, Matthew {Clapp}, James {Lang}, and Nicholas
  {Waltham}.
\newblock {The Atmospheric Imaging Assembly (AIA) on the Solar Dynamics
  Observatory (SDO)}.
\newblock \emph{Sol Phys}, 275:\penalty0 17--40, Jan 2012.
\newblock \doi{10.1007/s11207-011-9776-8}.

\bibitem[{L{\'e}pine} and {Shara}(2005)]{Lepine05}
S{\'e}bastien {L{\'e}pine} and Michael~M. {Shara}.
\newblock {A Catalog of Northern Stars with Annual Proper Motions Larger than
  0.15'' (LSPM-NORTH Catalog)}.
\newblock \emph{AJ}, 129\penalty0 (3):\penalty0 1483--1522, Mar 2005.
\newblock \doi{10.1086/427854}.

\bibitem[{Li} et~al.(2018){Li}, {Luo}, {Du}, {Zuo}, {Wang}, {Zhao}, {Jiang},
  {Zhang}, {Liu}, {Qin}, {Wang}, {Du}, {Guo}, {Wang}, {Han}, {Xiang}, {Huang},
  {Chen}, {Chen}, {Kong}, {Hou}, {Song}, {Wang}, {Wu}, {Zhang}, {Zhang},
  {Wang}, {Cao}, {Hou}, and {Zhao}]{Li18}
Yin-Bi {Li}, A.~Li {Luo}, Chang-De {Du}, Fang {Zuo}, Meng-Xin {Wang}, Gang
  {Zhao}, Bi-Wei {Jiang}, Hua-Wei {Zhang}, Chao {Liu}, Li~{Qin}, Rui {Wang},
  Bing {Du}, Yan-Xin {Guo}, Bo~{Wang}, Zhan-Wen {Han}, Mao-Sheng {Xiang}, Yang
  {Huang}, Bing-Qiu {Chen}, Jian-Jun {Chen}, Xiao {Kong}, Wen {Hou}, Yi-Han
  {Song}, You-Fen {Wang}, Ke-Fei {Wu}, Jian-Nan {Zhang}, Yong {Zhang}, Yue-Fei
  {Wang}, Zi-Huang {Cao}, Yong-Hui {Hou}, and Yong-Heng {Zhao}.
\newblock {Carbon Stars Identified from LAMOST DR4 Using Machine Learning}.
\newblock \emph{ApJSS}, 234:\penalty0 31, Feb 2018.
\newblock \doi{10.3847/1538-4365/aaa415}.

\bibitem[{Li} and {Yang}(2018)]{Li18b}
Yun {Li} and Shihai {Yang}.
\newblock {Research on the fault diagnosis and self-healing technology of
  unattended Antarctic telescope}.
\newblock In \emph{Society of Photo-Optical Instrumentation Engineers (SPIE)
  Conference Series}, volume 10700, page 107004W, Jul 2018.
\newblock \doi{10.1117/12.2311946}.

\bibitem[{Lin} et~al.(2018){Lin}, {Chen}, {Wang}, {Wang}, {Yoshida}, {Ip},
  {Miyazaki}, and {Terai}]{Lin18}
Hsing-Wen {Lin}, Ying-Tung {Chen}, Jen-Hung {Wang}, Shiang-Yu {Wang}, Fumi
  {Yoshida}, Wing-Huen {Ip}, Satoshi {Miyazaki}, and Tsuyoshi {Terai}.
\newblock {Machine-learning-based real-bogus system for the HSC-SSP moving
  object detection pipeline}.
\newblock \emph{PASJ}, 70:\penalty0 S39, Jan 2018.
\newblock \doi{10.1093/pasj/psx082}.

\bibitem[{Lintott} et~al.(2008){Lintott}, {Schawinski}, {Slosar}, {Land},
  {Bamford}, {Thomas}, {Raddick}, {Nichol}, {Szalay}, {Andreescu}, {Murray},
  and {Vandenberg}]{Lintott08}
Chris~J. {Lintott}, Kevin {Schawinski}, An{\v{z}}e {Slosar}, Kate {Land},
  Steven {Bamford}, Daniel {Thomas}, M.~Jordan {Raddick}, Robert~C. {Nichol},
  Alex {Szalay}, Dan {Andreescu}, Phil {Murray}, and Jan {Vandenberg}.
\newblock {Galaxy Zoo: morphologies derived from visual inspection of galaxies
  from the Sloan Digital Sky Survey}.
\newblock \emph{MNRAS}, 389\penalty0 (3):\penalty0 1179--1189, Sep 2008.
\newblock \doi{10.1111/j.1365-2966.2008.13689.x}.

\bibitem[{Liu} et~al.(2017){Liu}, {Deng}, {Wang}, and {Wang}]{Liu17b}
Chang {Liu}, Na~{Deng}, Jason T.~L. {Wang}, and Haimin {Wang}.
\newblock {Predicting Solar Flares Using SDO/HMI Vector Magnetic Data Products
  and the Random Forest Algorithm}.
\newblock \emph{ApJ}, 843:\penalty0 104, Jul 2017.
\newblock \doi{10.3847/1538-4357/aa789b}.

\bibitem[Liu et~al.(2008)Liu, Ting, and Zhou]{liuIsolationForest2008}
F.~T. Liu, K.~M. Ting, and Z.~Zhou.
\newblock Isolation {{Forest}}.
\newblock In \emph{2008 {{Eighth IEEE International Conference}} on {{Data
  Mining}}}, pages 413--422, December 2008.
\newblock \doi{10.1109/ICDM.2008.17}.

\bibitem[{Liu} et~al.(2018){Liu}, {Ye}, {Shen}, {Wang}, and
  {Erd{\'e}lyi}]{Liu18}
Jiajia {Liu}, Yudong {Ye}, Chenglong {Shen}, Yuming {Wang}, and Robert
  {Erd{\'e}lyi}.
\newblock {A New Tool for CME Arrival Time Prediction using Machine Learning
  Algorithms: CAT-PUMA}.
\newblock \emph{ApJ}, 855:\penalty0 109, Mar 2018.
\newblock \doi{10.3847/1538-4357/aaae69}.

\bibitem[{Liu}(2017)]{Liu17}
Junyu {Liu}.
\newblock {Artificial neural network in cosmic landscape}.
\newblock \emph{Journal of High Energy Physics}, 2017:\penalty0 149, Dec 2017.
\newblock \doi{10.1007/JHEP12(2017)149}.

\bibitem[{LSST Science Collaboration} and {et al.}(2009)]{LSST09}
{LSST Science Collaboration} and {et al.}
\newblock {LSST Science Book, Version 2.0}.
\newblock art. arXiv:0912.0201, Dec 2009.

\bibitem[{Lucie-Smith} et~al.(2018){Lucie-Smith}, {Peiris}, {Pontzen}, and
  {Lochner}]{Lucie18}
Luisa {Lucie-Smith}, Hiranya~V. {Peiris}, Andrew {Pontzen}, and Michelle
  {Lochner}.
\newblock {Machine learning cosmological structure formation}.
\newblock \emph{MNRAS}, 479:\penalty0 3405--3414, Sep 2018.
\newblock \doi{10.1093/mnras/sty1719}.

\bibitem[{Lukic} et~al.(2018){Lukic}, {Br{\"u}ggen}, {Banfield}, {Wong},
  {Rudnick}, {Norris}, and {Simmons}]{Lukic18}
V.~{Lukic}, M.~{Br{\"u}ggen}, J.~K. {Banfield}, O.~I. {Wong}, L.~{Rudnick},
  R.~P. {Norris}, and B.~{Simmons}.
\newblock {Radio Galaxy Zoo: compact and extended radio source classification
  with deep learning}.
\newblock \emph{MNRAS}, 476:\penalty0 246--260, May 2018.
\newblock \doi{10.1093/mnras/sty163}.

\bibitem[{Ma} et~al.(2019){Ma}, {Xu}, {Zhu}, {Hu}, {Li}, {Shan}, {Zhu}, {Gu},
  {Li}, {Liu}, and {Wu}]{Ma2019}
Zhixian {Ma}, Haiguang {Xu}, Jie {Zhu}, Dan {Hu}, Weitian {Li}, Chenxi {Shan},
  Zhenghao {Zhu}, Liyi {Gu}, Jinjin {Li}, Chengze {Liu}, and Xiangping {Wu}.
\newblock {A Machine Learning Based Morphological Classification of 14,245
  Radio AGNs Selected from the Best-Heckman Sample}.
\newblock \emph{ApJSS}, 240:\penalty0 34, Feb 2019.
\newblock \doi{10.3847/1538-4365/aaf9a2}.

\bibitem[Ma et~al.(2019)Ma, Xu, Zhu, Hu, Li, Shan, Zhu, Gu, Li, Liu, and
  Wu]{maMachineLearningBased2019}
Zhixian Ma, Haiguang Xu, Jie Zhu, Dan Hu, Weitian Li, Chenxi Shan, Zhenghao
  Zhu, Liyi Gu, Jinjin Li, Chengze Liu, and Xiangping Wu.
\newblock A {{Machine Learning Based Morphological Classification}} of 14,245
  {{Radio AGNs Selected}} from the {{Best}}\textendash{{Heckman Sample}}.
\newblock \emph{ApJS}, 240\penalty0 (2):\penalty0 34, February 2019.
\newblock ISSN 0067-0049.
\newblock \doi{10.3847/1538-4365/aaf9a2}.

\bibitem[{Mahabal} et~al.(2011){Mahabal}, {Djorgovski}, {Drake}, {Donalek},
  {Graham}, {Williams}, {Chen}, {Moghaddam}, {Turmon}, {Beshore}, and
  {Larson}]{Mahabal11}
A.~A. {Mahabal}, S.~G. {Djorgovski}, A.~J. {Drake}, C.~{Donalek}, M.~J.
  {Graham}, R.~D. {Williams}, Y.~{Chen}, B.~{Moghaddam}, M.~{Turmon},
  E.~{Beshore}, and S.~{Larson}.
\newblock {Discovery, classification, and scientific exploration of transient
  events from the Catalina Real-time Transient Survey}.
\newblock \emph{Bulletin of the Astronomical Society of India}, 39\penalty0
  (3):\penalty0 387--408, Sep 2011.

\bibitem[{Mahabal} et~al.(2019){Mahabal}, {Rebbapragada}, {Walters}, {Masci},
  {Blagorodnova}, {van Roestel}, {Ye}, {Biswas}, {Burdge}, {Chang}, {Duev},
  {Golkhou}, {Miller}, {Nordin}, {Ward}, {Adams}, {Bellm}, {Branton}, {Bue},
  {Cannella}, {Connolly}, {Dekany}, {Feindt}, {Hung}, {Fortson}, {Frederick},
  {Fremling}, {Gezari}, {Graham}, {Groom}, {Kasliwal}, {Kulkarni}, {Kupfer},
  {Lin}, {Lintott}, {Lunnan}, {Parejko}, {Prince}, {Riddle}, {Rusholme},
  {Saunders}, {Sedaghat}, {Shupe}, {Singer}, {Soumagnac}, {Szkody},
  {Tachibana}, {Tirumala}, {van Velzen}, and {Wright}]{Mahabal19}
Ashish {Mahabal}, Umaa {Rebbapragada}, Richard {Walters}, Frank~J. {Masci},
  Nadejda {Blagorodnova}, Jan {van Roestel}, Quan-Zhi {Ye}, Rahul {Biswas},
  Kevin {Burdge}, Chan-Kao {Chang}, Dmitry~A. {Duev}, V.~Zach {Golkhou},
  Adam~A. {Miller}, Jakob {Nordin}, Charlotte {Ward}, Scott {Adams}, Eric~C.
  {Bellm}, Doug {Branton}, Brian {Bue}, Chris {Cannella}, Andrew {Connolly},
  Richard {Dekany}, Ulrich {Feindt}, Tiara {Hung}, Lucy {Fortson}, Sara
  {Frederick}, C.~{Fremling}, Suvi {Gezari}, Matthew {Graham}, Steven {Groom},
  Mansi~M. {Kasliwal}, Shrinivas {Kulkarni}, Thomas {Kupfer}, Hsing~Wen {Lin},
  Chris {Lintott}, Ragnhild {Lunnan}, John {Parejko}, Thomas~A. {Prince}, Reed
  {Riddle}, Ben {Rusholme}, Nicholas {Saunders}, Nima {Sedaghat}, David~L.
  {Shupe}, Leo~P. {Singer}, Maayane~T. {Soumagnac}, Paula {Szkody}, Yutaro
  {Tachibana}, Kushal {Tirumala}, Sjoert {van Velzen}, and Darryl {Wright}.
\newblock {Machine Learning for the Zwicky Transient Facility}.
\newblock \emph{PASP}, 131:\penalty0 038002, Mar 2019.
\newblock \doi{10.1088/1538-3873/aaf3fa}.

\bibitem[{Mahabal} et~al.(2009){Mahabal}, {Djorgovski}, {Drake}, {Hensley},
  {Donalek}, {Graham}, {Williams}, {Glikman}, {Baltay}, {Rabinowitz}, and {PQ
  Survey Team}]{Mahabal09}
Ashish~A. {Mahabal}, S.~G. {Djorgovski}, A.~J. {Drake}, B.~{Hensley},
  C.~{Donalek}, M.~J. {Graham}, R.~D. {Williams}, E.~{Glikman}, C.~{Baltay},
  D.~{Rabinowitz}, and {PQ Survey Team}.
\newblock {Towards the Automated Classification of Variable Objects and
  Transients}.
\newblock \emph{{Bulletin of the American Astronomical Society}}, 42:\penalty0
  427.06, Jan 2009.

\bibitem[{Marchetti} et~al.(2017){Marchetti}, {Rossi}, {Kordopatis}, {Brown},
  {Rimoldi}, {Starkenburg}, {Youakim}, and {Ashley}]{Marchetti17}
T.~{Marchetti}, E.~M. {Rossi}, G.~{Kordopatis}, A.~G.~A. {Brown}, A.~{Rimoldi},
  E.~{Starkenburg}, K.~{Youakim}, and R.~{Ashley}.
\newblock {An artificial neural network to discover hypervelocity stars:
  candidates in Gaia DR1/TGAS}.
\newblock \emph{MNRAS}, 470:\penalty0 1388--1403, Sep 2017.
\newblock \doi{10.1093/mnras/stx1304}.

\bibitem[{M{\'a}rquez-Neila} et~al.(2018){M{\'a}rquez-Neila}, {Fisher},
  {Sznitman}, and {Heng}]{Marquez18}
Pablo {M{\'a}rquez-Neila}, Chloe {Fisher}, Raphael {Sznitman}, and Kevin
  {Heng}.
\newblock {Supervised machine learning for analysing spectra of exoplanetary
  atmospheres}.
\newblock \emph{Nature Astronomy}, 2:\penalty0 719--724, Jun 2018.
\newblock \doi{10.1038/s41550-018-0504-2}.

\bibitem[{Martin} et~al.(2005){Martin}, {Fanson}, {Schiminovich}, {Morrissey},
  {Friedman}, {Barlow}, {Conrow}, {Grange}, {Jelinsky}, {Milliard}, {Siegmund},
  {Bianchi}, {Byun}, {Donas}, {Forster}, {Heckman}, {Lee}, {Madore}, {Malina},
  {Neff}, {Rich}, {Small}, {Surber}, {Szalay}, {Welsh}, and {Wyder}]{Martin05}
D.~Christopher {Martin}, James {Fanson}, David {Schiminovich}, Patrick
  {Morrissey}, Peter~G. {Friedman}, Tom~A. {Barlow}, Tim {Conrow}, Robert
  {Grange}, Patrick~N. {Jelinsky}, Bruno {Milliard}, Oswald H.~W. {Siegmund},
  Luciana {Bianchi}, Yong-Ik {Byun}, Jose {Donas}, Karl {Forster}, Timothy~M.
  {Heckman}, Young-Wook {Lee}, Barry~F. {Madore}, Roger~F. {Malina}, Susan~G.
  {Neff}, R.~Michael {Rich}, Todd {Small}, Frank {Surber}, Alex~S. {Szalay},
  Barry {Welsh}, and Ted~K. {Wyder}.
\newblock {The Galaxy Evolution Explorer: A Space Ultraviolet Survey Mission}.
\newblock \emph{ApJL}, 619\penalty0 (1):\penalty0 L1--L6, Jan 2005.
\newblock \doi{10.1086/426387}.

\bibitem[Masters et~al.(2015)Masters, Capak, Stern, Ilbert, Salvato, Schmidt,
  Longo, Rhodes, Paltani, Mobasher, Hoekstra, Hildebrandt, Coupon, Steinhardt,
  Speagle, Faisst, Kalinich, Brodwin, Brescia, and
  Cavuoti]{mastersMAPPINGGALAXYCOLOR2015}
Daniel Masters, Peter Capak, Daniel Stern, Olivier Ilbert, Mara Salvato, Samuel
  Schmidt, Giuseppe Longo, Jason Rhodes, Stephane Paltani, Bahram Mobasher,
  Henk Hoekstra, Hendrik Hildebrandt, Jean Coupon, Charles Steinhardt, Josh
  Speagle, Andreas Faisst, Adam Kalinich, Mark Brodwin, Massimo Brescia, and
  Stefano Cavuoti.
\newblock {{MAPPING THE GALAXY COLOR}}\textendash{{REDSHIFT RELATION}}:
  {{OPTIMAL PHOTOMETRIC REDSHIFT CALIBRATION STRATEGIES FOR COSMOLOGY
  SURVEYS}}.
\newblock \emph{ApJ}, 813\penalty0 (1):\penalty0 53, October 2015.
\newblock ISSN 0004-637X.
\newblock \doi{10.1088/0004-637X/813/1/53}.

\bibitem[{McLeod} et~al.(2017){McLeod}, {Libeskind}, {Lahav}, and
  {Hoffman}]{McLeod17}
M.~{McLeod}, N.~{Libeskind}, O.~{Lahav}, and Y.~{Hoffman}.
\newblock {Estimating the mass of the Local Group using machine learning
  applied to numerical simulations}.
\newblock \emph{Journal of Cosmology and Astro-Particle Physics},
  2017:\penalty0 034, Dec 2017.
\newblock \doi{10.1088/1475-7516/2017/12/034}.

\bibitem[Metcalf et~al.(2019)Metcalf, Meneghetti, Avestruz, Bellagamba, Bom,
  Bertin, Cabanac, Courbin, Davies, Decenci{\`e}re, Flamary, Gavazzi, Geiger,
  Hartley, {Huertas-Company}, Jackson, Jacobs, Jullo, Kneib, Koopmans, Lanusse,
  Li, Ma, Makler, Li, Lightman, Petrillo, Serjeant, Sch{\"a}fer, Sonnenfeld,
  Tagore, Tortora, Tuccillo, Valent{\'i}n, {Velasco-Forero}, Kleijn, and
  Vernardos]{metcalfStrongGravitationalLens2019}
R.~B. Metcalf, M.~Meneghetti, C.~Avestruz, F.~Bellagamba, C.~R. Bom, E.~Bertin,
  R.~Cabanac, F.~Courbin, A.~Davies, E.~Decenci{\`e}re, R.~Flamary, R.~Gavazzi,
  M.~Geiger, P.~Hartley, M.~{Huertas-Company}, N.~Jackson, C.~Jacobs, E.~Jullo,
  J.-P. Kneib, L.~V.~E. Koopmans, F.~Lanusse, C.-L. Li, Q.~Ma, M.~Makler,
  N.~Li, M.~Lightman, C.~E. Petrillo, S.~Serjeant, C.~Sch{\"a}fer,
  A.~Sonnenfeld, A.~Tagore, C.~Tortora, D.~Tuccillo, M.~B. Valent{\'i}n,
  S.~{Velasco-Forero}, G.~A.~Verdoes Kleijn, and G.~Vernardos.
\newblock The strong gravitational lens finding challenge.
\newblock \emph{A\&A}, 625:\penalty0 A119, May 2019.
\newblock ISSN 0004-6361, 1432-0746.
\newblock \doi{10.1051/0004-6361/201832797}.

\bibitem[{Michilli} et~al.(2018){Michilli}, {Hessels}, {Lyon}, {Tan}, {Bassa},
  {Cooper}, {Kondratiev}, {Sanidas}, {Stappers}, and {van Leeuwen}]{Michilli18}
D.~{Michilli}, J.~W.~T. {Hessels}, R.~J. {Lyon}, C.~M. {Tan}, C.~{Bassa},
  S.~{Cooper}, V.~I. {Kondratiev}, S.~{Sanidas}, B.~W. {Stappers}, and J.~{van
  Leeuwen}.
\newblock {Single-pulse classifier for the LOFAR Tied-Array All-sky Survey}.
\newblock \emph{MNRAS}, 480:\penalty0 3457--3467, Nov 2018.
\newblock \doi{10.1093/mnras/sty2072}.

\bibitem[{Miettinen}(2018)]{Miettinen18}
O.~{Miettinen}.
\newblock {Protostellar classification using supervised machine learning
  algorithms}.
\newblock \emph{Ap\&SS}, 363:\penalty0 197, Sep 2018.
\newblock \doi{10.1007/s10509-018-3418-7}.

\bibitem[{Mislis} et~al.(2018){Mislis}, {Pyrzas}, and {Alsubai}]{Mislis18}
D.~{Mislis}, S.~{Pyrzas}, and K.~A. {Alsubai}.
\newblock {TSARDI: a Machine Learning data rejection algorithm for transiting
  exoplanet light curves}.
\newblock \emph{MNRAS}, 481:\penalty0 1624--1630, Dec 2018.
\newblock \doi{10.1093/mnras/sty2361}.

\bibitem[Montavon et~al.(2018)Montavon, Samek, and
  M{\"u}ller]{montavonMethodsInterpretingUnderstanding2018}
Gr{\'e}goire Montavon, Wojciech Samek, and Klaus-Robert M{\"u}ller.
\newblock Methods for interpreting and understanding deep neural networks.
\newblock \emph{Digital Signal Processing}, 73:\penalty0 1--15, February 2018.
\newblock ISSN 1051-2004.
\newblock \doi{10.1016/j.dsp.2017.10.011}.

\bibitem[{Morello} et~al.(2018){Morello}, {Morris}, {Van Dyk}, {Marston}, and
  {Mauerhan}]{Morello18}
Giuseppe {Morello}, P.~W. {Morris}, S.~D. {Van Dyk}, A.~P. {Marston}, and J.~C.
  {Mauerhan}.
\newblock {Applications of machine-learning algorithms for infrared colour
  selection of Galactic Wolf-Rayet stars}.
\newblock \emph{MNRAS}, 473:\penalty0 2565--2574, Jan 2018.
\newblock \doi{10.1093/mnras/stx2474}.

\bibitem[{Morello} et~al.(2019){Morello}, {Barr}, {Cooper}, {Bailes}, {Bates},
  {Bhat}, {Burgay}, {Burke-Spolaor}, {Cameron}, {Champion}, {Eatough}, {Flynn},
  {Jameson}, {Johnston}, {Keith}, {Keane}, {Kramer}, {Levin}, {Ng}, {Petroff},
  {Possenti}, {Stappers}, {van Straten}, and {Tiburzi}]{Morello19}
V.~{Morello}, E.~D. {Barr}, S.~{Cooper}, M.~{Bailes}, S.~{Bates}, N.~D.~R.
  {Bhat}, M.~{Burgay}, S.~{Burke-Spolaor}, A.~D. {Cameron}, D.~J. {Champion},
  R.~P. {Eatough}, C.~M.~L. {Flynn}, A.~{Jameson}, S.~{Johnston}, M.~J.
  {Keith}, E.~F. {Keane}, M.~{Kramer}, L.~{Levin}, C.~{Ng}, E.~{Petroff},
  A.~{Possenti}, B.~W. {Stappers}, W.~{van Straten}, and C.~{Tiburzi}.
\newblock {The High Time Resolution Universe survey - XIV. Discovery of 23
  pulsars through GPU-accelerated reprocessing}.
\newblock \emph{MNRAS}, 483:\penalty0 3673--3685, Mar 2019.
\newblock \doi{10.1093/mnras/sty3328}.

\bibitem[{Morrison} et~al.(2017){Morrison}, {Hildebrandt}, {Schmidt}, {Baldry},
  {Bilicki}, {Choi}, {Erben}, and {Schneider}]{Morrison17}
C.~B. {Morrison}, H.~{Hildebrandt}, S.~J. {Schmidt}, I.~K. {Baldry},
  M.~{Bilicki}, A.~{Choi}, T.~{Erben}, and P.~{Schneider}.
\newblock {the-wizz: clustering redshift estimation for everyone}.
\newblock \emph{MNRAS}, 467:\penalty0 3576--3589, May 2017.
\newblock \doi{10.1093/mnras/stx342}.

\bibitem[{Mukkavilli} et~al.(2018){Mukkavilli}, {Meger}, and
  {Dudek}]{Mukkavilli18}
S.~K. {Mukkavilli}, D.~{Meger}, and G.~{Dudek}.
\newblock {EnviRoNet - Planetary Science Applications}.
\newblock In \emph{AGU Fall Meeting Abstracts}, volume 2018, pages P43J--3875,
  Dec 2018.

\bibitem[{Mukund} et~al.(2018){Mukund}, {Thakur}, {Abraham}, {Aniyan}, {Mitra},
  {Sajeeth Philip}, {Vaghmare}, and {Acharjya}]{Mukund18}
Nikhil {Mukund}, Saurabh {Thakur}, Sheelu {Abraham}, A.~K. {Aniyan}, Sanjit
  {Mitra}, Ninan {Sajeeth Philip}, Kaustubh {Vaghmare}, and D.~P. {Acharjya}.
\newblock {An Information Retrieval and Recommendation System for Astronomical
  Observatories}.
\newblock \emph{ApJSS}, 235:\penalty0 22, Mar 2018.
\newblock \doi{10.3847/1538-4365/aaadb2}.

\bibitem[{M{\"u}ller} et~al.(2018){M{\"u}ller}, {Hackstein}, {Greiner},
  {Frank}, {Bomans}, {Dettmar}, and {En{\ss}lin}]{Muller18}
Ancla {M{\"u}ller}, Moritz {Hackstein}, Maksim {Greiner}, Philipp {Frank},
  Dominik~J. {Bomans}, Ralf-J{\"u}rgen {Dettmar}, and Torsten {En{\ss}lin}.
\newblock {Sharpening up Galactic all-sky maps with complementary data. A
  machine learning approach}.
\newblock \emph{A\&A}, 620:\penalty0 A64, Nov 2018.
\newblock \doi{10.1051/0004-6361/201833604}.

\bibitem[{Nadler} et~al.(2018){Nadler}, {Mao}, {Wechsler}, {Garrison-Kimmel},
  and {Wetzel}]{Nadler18}
Ethan~O. {Nadler}, Yao-Yuan {Mao}, Risa~H. {Wechsler}, Shea {Garrison-Kimmel},
  and Andrew {Wetzel}.
\newblock {Modeling the Impact of Baryons on Subhalo Populations with Machine
  Learning}.
\newblock \emph{ApJ}, 859:\penalty0 129, June 2018.
\newblock \doi{10.3847/1538-4357/aac266}.

\bibitem[Nakoneczny et~al.(2019)Nakoneczny, Bilicki, Solarz, Pollo, Maddox,
  Spiniello, Brescia, and Napolitano]{nakonecznyCatalogQuasarsKiloDegree2019}
S.~Nakoneczny, M.~Bilicki, A.~Solarz, A.~Pollo, N.~Maddox, C.~Spiniello,
  M.~Brescia, and N.~R. Napolitano.
\newblock Catalog of quasars from the {{Kilo}}-{{Degree Survey Data Release}}
  3.
\newblock \emph{A\&A}, 624:\penalty0 A13, April 2019.
\newblock ISSN 0004-6361, 1432-0746.
\newblock \doi{10.1051/0004-6361/201834794}.

\bibitem[{Narayan} et~al.(2018){Narayan}, {Zaidi}, {Soraisam}, {Wang},
  {Lochner}, {Matheson}, {Saha}, {Yang}, {Zhao}, {Kececioglu}, {Scheidegger},
  {Snodgrass}, {Axelrod}, {Jenness}, {Maier}, {Ridgway}, {Seaman}, {Evans},
  {Singh}, {Taylor}, {Toeniskoetter}, {Welch}, {Zhu}, and {ANTARES
  Collaboration}]{Narayan18}
Gautham {Narayan}, Tayeb {Zaidi}, Monika~D. {Soraisam}, Zhe {Wang}, Michelle
  {Lochner}, Thomas {Matheson}, Abhijit {Saha}, Shuo {Yang}, Zhenge {Zhao},
  John {Kececioglu}, Carlos {Scheidegger}, Richard~T. {Snodgrass}, Tim
  {Axelrod}, Tim {Jenness}, Robert~S. {Maier}, Stephen~T. {Ridgway}, Robert~L.
  {Seaman}, Eric~Michael {Evans}, Navdeep {Singh}, Clark {Taylor}, Jackson
  {Toeniskoetter}, Eric {Welch}, Songzhe {Zhu}, and {ANTARES Collaboration}.
\newblock {Machine-learning-based Brokers for Real-time Classification of the
  LSST Alert Stream}.
\newblock \emph{ApJSS}, 236:\penalty0 9, May 2018.
\newblock \doi{10.3847/1538-4365/aab781}.

\bibitem[{Naul} et~al.(2018){Naul}, {Bloom}, {P{\'e}rez}, and {van der
  Walt}]{Naul18}
Brett {Naul}, Joshua~S. {Bloom}, Fernando {P{\'e}rez}, and St{\'e}fan {van der
  Walt}.
\newblock {A recurrent neural network for classification of unevenly sampled
  variable stars}.
\newblock \emph{Nature Astronomy}, 2:\penalty0 151--155, Nov 2018.
\newblock \doi{10.1038/s41550-017-0321-z}.

\bibitem[{Nguyen} et~al.(2018){Nguyen}, {Pankratius}, {Eckman}, and
  {Seager}]{Nguyen18}
T.~{Nguyen}, V.~{Pankratius}, L.~{Eckman}, and S.~{Seager}.
\newblock {Computer-aided discovery of debris disk candidates: A case study
  using the Wide-Field Infrared Survey Explorer (WISE) catalog}.
\newblock \emph{A\&C}, 23:\penalty0 72--82, Apr 2018.
\newblock \doi{10.1016/j.ascom.2018.02.004}.

\bibitem[{Nishizuka} et~al.(2017){Nishizuka}, {Sugiura}, {Kubo}, {Den},
  {Watari}, and {Ishii}]{Nishizuka17}
N.~{Nishizuka}, K.~{Sugiura}, Y.~{Kubo}, M.~{Den}, S.~{Watari}, and M.~{Ishii}.
\newblock {Solar Flare Prediction Model with Three Machine-learning Algorithms
  using Ultraviolet Brightening and Vector Magnetograms}.
\newblock \emph{ApJ}, 835:\penalty0 156, Feb 2017.
\newblock \doi{10.3847/1538-4357/835/2/156}.

\bibitem[{Norris}(2017)]{Norris17}
Ray~P. {Norris}.
\newblock {Discovering the Unexpected in Astronomical Survey Data}.
\newblock \emph{PASA}, 34:\penalty0 e007, Jan 2017.
\newblock \doi{10.1017/pasa.2016.63}.

\bibitem[Odewahn et~al.(1992)Odewahn, Stockwell, Pennington, Humphreys, and
  Zumach]{odewahnAutomatedStarGalaxy1992}
S.~C. Odewahn, E.~B. Stockwell, R.~L. Pennington, R.~M. Humphreys, and W.~A.
  Zumach.
\newblock Automated {{Star}}/{{Galaxy Discrimination With Neural Networks}}.
\newblock \emph{The Astronomical Journal}, 103:\penalty0 318, January 1992.
\newblock ISSN 0004-6256.
\newblock \doi{10.1086/116063}.

\bibitem[{Ostrovski} et~al.(2017){Ostrovski}, {McMahon}, {Connolly}, {Lemon},
  {Auger}, {Banerji}, {Hung}, {Koposov}, {Lidman}, {Reed}, {Allam},
  {Benoit-L{\'e}vy}, {Bertin}, {Brooks}, {Buckley-Geer}, {Carnero Rosell},
  {Carrasco Kind}, {Carretero}, {Cunha}, {da Costa}, {Desai}, {Diehl},
  {Dietrich}, {Evrard}, {Finley}, {Flaugher}, {Fosalba}, {Frieman}, {Gerdes},
  {Goldstein}, {Gruen}, {Gruendl}, {Gutierrez}, {Honscheid}, {James}, {Kuehn},
  {Kuropatkin}, {Lima}, {Lin}, {Maia}, {Marshall}, {Martini}, {Melchior},
  {Miquel}, {Ogand o}, {Plazas Malag{\'o}n}, {Reil}, {Romer}, {Sanchez},
  {Santiago}, {Scarpine}, {Sevilla-Noarbe}, {Soares-Santos}, {Sobreira},
  {Suchyta}, {Tarle}, {Thomas}, {Tucker}, and {Walker}]{Ostrovski17}
Fernanda {Ostrovski}, Richard~G. {McMahon}, Andrew~J. {Connolly}, Cameron~A.
  {Lemon}, Matthew~W. {Auger}, Manda {Banerji}, Johnathan~M. {Hung}, Sergey~E.
  {Koposov}, Christopher~E. {Lidman}, Sophie~L. {Reed}, Sahar {Allam},
  Aur{\'e}lien {Benoit-L{\'e}vy}, Emmanuel {Bertin}, David {Brooks}, Elizabeth
  {Buckley-Geer}, Aurelio {Carnero Rosell}, Matias {Carrasco Kind}, Jorge
  {Carretero}, Carlos~E. {Cunha}, Luiz~N. {da Costa}, Shantanu {Desai},
  H.~Thomas {Diehl}, J{\"o}rg~P. {Dietrich}, August~E. {Evrard}, David~A.
  {Finley}, Brenna {Flaugher}, Pablo {Fosalba}, Josh {Frieman}, David~W.
  {Gerdes}, Daniel~A. {Goldstein}, Daniel {Gruen}, Robert~A. {Gruendl}, Gaston
  {Gutierrez}, Klaus {Honscheid}, David~J. {James}, Kyler {Kuehn}, Nikolay
  {Kuropatkin}, Marcos {Lima}, Huan {Lin}, Marcio A.~G. {Maia}, Jennifer~L.
  {Marshall}, Paul {Martini}, Peter {Melchior}, Ramon {Miquel}, Ricardo {Ogand
  o}, Andr{\'e}s {Plazas Malag{\'o}n}, Kevin {Reil}, Kathy {Romer}, Eusebio
  {Sanchez}, Basilio {Santiago}, Vic {Scarpine}, Ignacio {Sevilla-Noarbe},
  Marcelle {Soares-Santos}, Flavia {Sobreira}, Eric {Suchyta}, Gregory {Tarle},
  Daniel {Thomas}, Douglas~L. {Tucker}, and Alistair~R. {Walker}.
\newblock {VDES J2325-5229 a z = 2.7 gravitationally lensed quasar discovered
  using morphology-independent supervised machine learning}.
\newblock \emph{MNRAS}, 465:\penalty0 4325--4334, Mar 2017.
\newblock \doi{10.1093/mnras/stw2958}.

\bibitem[Owens et~al.(1996)Owens, Griffiths, and
  Ratnatunga]{owensUsingObliqueDecision1996}
E.~A. Owens, R.~E. Griffiths, and K.~U. Ratnatunga.
\newblock Using oblique decision trees for the morphological classification of
  galaxies.
\newblock \emph{MNRAS}, 281\penalty0 (1):\penalty0 153--157, July 1996.
\newblock ISSN 0035-8711.
\newblock \doi{10.1093/mnras/281.1.153}.

\bibitem[{Pang} et~al.(2018){Pang}, {Goseva-Popstojanova}, {Devine}, and
  {McLaughlin}]{Pang18}
Di~{Pang}, Katerina {Goseva-Popstojanova}, Thomas {Devine}, and Maura
  {McLaughlin}.
\newblock {A novel single-pulse search approach to detection of dispersed radio
  pulses using clustering and supervised machine learning}.
\newblock \emph{MNRAS}, 480:\penalty0 3302--3323, Nov 2018.
\newblock \doi{10.1093/mnras/sty1992}.

\bibitem[{Papageorgiou} et~al.(2018){Papageorgiou}, {Catelan}, {Christopoulou},
  {Drake}, and {Djorgovski}]{Papageorgiou18}
Athanasios {Papageorgiou}, M{\'a}rcio {Catelan}, Panagiota-Eleftheria
  {Christopoulou}, Andrew~J. {Drake}, and S.~G. {Djorgovski}.
\newblock {An Updated Catalog of 4680 Northern Eclipsing Binaries with
  Algol-type Light-curve Morphology in the Catalina Sky Surveys}.
\newblock \emph{ApJSS}, 238:\penalty0 4, Sep 2018.
\newblock \doi{10.3847/1538-4365/aad8a9}.

\bibitem[{Pashchenko} et~al.(2018){Pashchenko}, {Sokolovsky}, and
  {Gavras}]{Paschenko18}
Ilya~N. {Pashchenko}, Kirill~V. {Sokolovsky}, and Panagiotis {Gavras}.
\newblock {Machine learning search for variable stars}.
\newblock \emph{MNRAS}, 475:\penalty0 2326--2343, April 2018.
\newblock \doi{10.1093/mnras/stx3222}.

\bibitem[{Pe{\~n}a} et~al.(2018){Pe{\~n}a}, {Fuentes}, {F{\"o}rster},
  {Maureira}, {San Mart{\'\i}n}, {Litt{\'\i}n}, {Huijse}, {Cabrera-Vives},
  {Est{\'e}vez}, {Galbany}, {Gonz{\'a}lez-Gait{\'a}n}, {Mart{\'\i}nez}, {de
  Jaeger}, and {Hamuy}]{Pena18}
J.~{Pe{\~n}a}, C.~{Fuentes}, F.~{F{\"o}rster}, J.~C. {Maureira}, J.~{San
  Mart{\'\i}n}, J.~{Litt{\'\i}n}, P.~{Huijse}, G.~{Cabrera-Vives}, P.~A.
  {Est{\'e}vez}, L.~{Galbany}, S.~{Gonz{\'a}lez-Gait{\'a}n},
  J.~{Mart{\'\i}nez}, Th. {de Jaeger}, and M.~{Hamuy}.
\newblock {Asteroids in the High Cadence Transient Survey}.
\newblock \emph{AJ}, 155:\penalty0 135, Mar 2018.
\newblock \doi{10.3847/1538-3881/aaaaed}.

\bibitem[{Pearson} et~al.(2018){Pearson}, {Palafox}, and {Griffith}]{Pearson18}
Kyle~A. {Pearson}, Leon {Palafox}, and Caitlin~A. {Griffith}.
\newblock {Searching for exoplanets using artificial intelligence}.
\newblock \emph{MNRAS}, 474:\penalty0 478--491, February 2018.
\newblock \doi{10.1093/mnras/stx2761}.

\bibitem[{Peng} et~al.(2018){Peng}, {English}, {Silva}, {Davis}, and
  {Hayes}]{Peng18}
Tianrui~Rae {Peng}, John~Edward {English}, Pedro {Silva}, Darren~R. {Davis},
  and Wayne~B. {Hayes}.
\newblock {SpArcFiRe: morphological selection effects due to reduced visibility
  of tightly winding arms in distant spiral galaxies}.
\newblock \emph{MNRAS}, 479:\penalty0 5532--5543, Oct 2018.
\newblock \doi{10.1093/mnras/sty546}.

\bibitem[Perreault~Levasseur et~al.(2017)Perreault~Levasseur, Hezaveh, and
  Wechsler]{perreaultlevasseurUncertaintiesParametersEstimated2017}
Laurence Perreault~Levasseur, Yashar~D. Hezaveh, and Risa~H. Wechsler.
\newblock Uncertainties in {{Parameters Estimated}} with {{Neural Networks}}:
  {{Application}} to {{Strong Gravitational Lensing}}.
\newblock \emph{ApJ}, 850\penalty0 (1):\penalty0 L7, November 2017.
\newblock \doi{10.3847/2041-8213/aa9704}.

\bibitem[{Pesnell} et~al.(2012){Pesnell}, {Thompson}, and
  {Chamberlin}]{Pesnell12}
W.~Dean {Pesnell}, B.~J. {Thompson}, and P.~C. {Chamberlin}.
\newblock {The Solar Dynamics Observatory (SDO)}.
\newblock \emph{Sol Phys}, 275:\penalty0 3--15, Jan 2012.
\newblock \doi{10.1007/s11207-011-9841-3}.

\bibitem[Petrillo et~al.(2017)Petrillo, Tortora, Chatterjee, Vernardos,
  Koopmans, Verdoes~Kleijn, Napolitano, Covone, Schneider, Grado, and
  McFarland]{petrilloFindingStrongGravitational2017}
C.~E. Petrillo, C.~Tortora, S.~Chatterjee, G.~Vernardos, L.~V.~E. Koopmans,
  G.~Verdoes~Kleijn, N.~R. Napolitano, G.~Covone, P.~Schneider, A.~Grado, and
  J.~McFarland.
\newblock Finding strong gravitational lenses in the {{Kilo Degree Survey}}
  with {{Convolutional Neural Networks}}.
\newblock \emph{MNRAS}, 472\penalty0 (1):\penalty0 1129--1150, November 2017.
\newblock ISSN 0035-8711.
\newblock \doi{10.1093/mnras/stx2052}.

\bibitem[Petrillo et~al.(2019)Petrillo, Tortora, Vernardos, Koopmans, Kleijn,
  Bilicki, Napolitano, Chatterjee, Covone, Dvornik, Erben, Getman, Giblin,
  Heymans, {de Jong}, Kuijken, Schneider, Shan, Spiniello, and
  Wright]{petrilloLinKSDiscoveringGalaxyscale2019}
C.~E. Petrillo, C.~Tortora, G.~Vernardos, L.~V.~E. Koopmans, G.~Verdoes Kleijn,
  M.~Bilicki, N.~R. Napolitano, S.~Chatterjee, G.~Covone, A.~Dvornik, T.~Erben,
  F.~Getman, B.~Giblin, C.~Heymans, J.~T.~A. {de Jong}, K.~Kuijken,
  P.~Schneider, H.~Shan, C.~Spiniello, and A.~H. Wright.
\newblock {{LinKS}}: {{Discovering}} galaxy-scale strong lenses in the
  {{Kilo}}-{{Degree Survey}} using {{Convolutional Neural Networks}}.
\newblock \emph{MNRAS}, 484:\penalty0 3879, January 2019.

\bibitem[Polsterer et~al.(2014)Polsterer, Gieseke, Igel, and
  Goto]{polstererImprovingPerformancePhotometric2014}
K.~L. Polsterer, F.~Gieseke, C~Igel, and T.~Goto.
\newblock Improving the {{Performance}} of {{Photometric Regression Models}}
  via {{Massive Parallel Feature Selection}}.
\newblock \emph{Astronomical Society of the Pacific Conference Series},
  485:\penalty0 425, 2014.

\bibitem[{Pourrahmani} et~al.(2018){Pourrahmani}, {Nayyeri}, and
  {Cooray}]{Pourrahmani18}
Milad {Pourrahmani}, Hooshang {Nayyeri}, and Asantha {Cooray}.
\newblock {LensFlow: A Convolutional Neural Network in Search of Strong
  Gravitational Lenses}.
\newblock \emph{ApJ}, 856:\penalty0 68, Mar 2018.
\newblock \doi{10.3847/1538-4357/aaae6a}.

\bibitem[{Powell} et~al.(2017){Powell}, {Torres-Forn{\'e}}, {Lynch},
  {Trifir{\`o}}, {Cuoco}, {Cavagli{\`a}}, {Heng}, and {Font}]{Powell17}
Jade {Powell}, Alejandro {Torres-Forn{\'e}}, Ryan {Lynch}, Daniele
  {Trifir{\`o}}, Elena {Cuoco}, Marco {Cavagli{\`a}}, Ik~Siong {Heng}, and
  Jos{\'e}~A. {Font}.
\newblock {Classification methods for noise transients in advanced
  gravitational-wave detectors II: performance tests on Advanced LIGO data}.
\newblock \emph{Classical and Quantum Gravity}, 34:\penalty0 034002, Feb 2017.
\newblock \doi{10.1088/1361-6382/34/3/034002}.

\bibitem[Qahwaji and Colak(2007)]{qahwajiAutomaticShortTermSolar2007}
R.~Qahwaji and T.~Colak.
\newblock Automatic {{Short}}-{{Term Solar Flare Prediction Using Machine
  Learning}} and {{Sunspot Associations}}.
\newblock \emph{Sol Phys}, 241\penalty0 (1):\penalty0 195--211, March 2007.
\newblock ISSN 1573-093X.
\newblock \doi{10.1007/s11207-006-0272-5}.
\newblock URL \url{https://doi.org/10.1007/s11207-006-0272-5}.

\bibitem[Quinlan(1986)]{Quinlan1986}
J.~R. Quinlan.
\newblock Induction of decision trees.
\newblock \emph{Mach Learn}, 1\penalty0 (1):\penalty0 81--106, March 1986.
\newblock ISSN 1573-0565.
\newblock \doi{10.1007/BF00116251}.

\bibitem[{Rafieferantsoa} et~al.(2018){Rafieferantsoa}, {Andrianomena}, and
  {Dav{\'e}}]{Rafieferantsoa18}
Mika {Rafieferantsoa}, Sambatra {Andrianomena}, and Romeel {Dav{\'e}}.
\newblock {Predicting the neutral hydrogen content of galaxies from optical
  data using machine learning}.
\newblock \emph{MNRAS}, 479:\penalty0 4509--4525, Oct 2018.
\newblock \doi{10.1093/mnras/sty1777}.

\bibitem[{Reiman} and {G{\"o}hre}(2019)]{Reiman19}
David~M. {Reiman} and Brett~E. {G{\"o}hre}.
\newblock {Deblending galaxy superpositions with branched generative
  adversarial networks}.
\newblock \emph{MNRAS}, 485:\penalty0 2617--2627, Feb 2019.
\newblock \doi{10.1093/mnras/stz575}.

\bibitem[{Reis} et~al.(2018){Reis}, {Poznanski}, {Baron}, {Zasowski}, and
  {Shahaf}]{Reis18}
Itamar {Reis}, Dovi {Poznanski}, Dalya {Baron}, Gail {Zasowski}, and Sahar
  {Shahaf}.
\newblock {Detecting outliers and learning complex structures with large
  spectroscopic surveys - a case study with APOGEE stars}.
\newblock \emph{MNRAS}, 476:\penalty0 2117--2136, May 2018.
\newblock \doi{10.1093/mnras/sty348}.

\bibitem[Reis et~al.(2018)Reis, Poznanski, Baron, Zasowski, and
  Shahaf]{reisDetectingOutliersLearning2018a}
Itamar Reis, Dovi Poznanski, Dalya Baron, Gail Zasowski, and Sahar Shahaf.
\newblock Detecting outliers and learning complex structures with large
  spectroscopic surveys \textendash{} a case study with {{APOGEE}} stars.
\newblock \emph{MNRAS}, 476\penalty0 (2):\penalty0 2117--2136, May 2018.
\newblock ISSN 0035-8711.
\newblock \doi{10.1093/mnras/sty348}.

\bibitem[{Reis} et~al.(2019){Reis}, {Baron}, and {Shahaf}]{Reis19}
Itamar {Reis}, Dalya {Baron}, and Sahar {Shahaf}.
\newblock {Probabilistic Random Forest: A Machine Learning Algorithm for Noisy
  Data Sets}.
\newblock \emph{AJ}, 157:\penalty0 16, Jan 2019.
\newblock \doi{10.3847/1538-3881/aaf101}.

\bibitem[{Rodr{\'\i}guez} et~al.(2018){Rodr{\'\i}guez}, {Kacprzak}, {Lucchi},
  {Amara}, {Sgier}, {Fluri}, {Hofmann}, and {R{\'e}fr{\'e}gier}]{Rodriguez18}
Andres~C. {Rodr{\'\i}guez}, Tomasz {Kacprzak}, Aurelien {Lucchi}, Adam {Amara},
  Rapha{\"e}l {Sgier}, Janis {Fluri}, Thomas {Hofmann}, and Alexandre
  {R{\'e}fr{\'e}gier}.
\newblock {Fast cosmic web simulations with generative adversarial networks}.
\newblock \emph{Computational Astrophysics and Cosmology}, 5:\penalty0 4, Nov
  2018.
\newblock \doi{10.1186/s40668-018-0026-4}.

\bibitem[Rosenblatt(1957)]{Rosenblatt1957}
Frank Rosenblatt.
\newblock The {{Perceptron}}-a perceiving and recognizing automaton.
\newblock \emph{Cornell Aeronautical Lab}, 1957.

\bibitem[Rosenthal(1988)]{rosenthalApplyingArtificialIntelligence1988}
D.~A. Rosenthal.
\newblock Applying artificial intelligence to astronomical databases - a
  surveyof applicable technology.
\newblock \emph{European Southern Observatory Conference and Workshop
  Proceedings}, 28:\penalty0 245, 1988.

\bibitem[{Roy} et~al.(2018){Roy}, {Napolitano}, {La Barbera}, {Tortora},
  {Getman}, {Radovich}, {Capaccioli}, {Brescia}, {Cavuoti}, {Longo}, {Raj},
  {Puddu}, {Covone}, {Amaro}, {Vellucci}, {Grado}, {Kuijken}, {Verdoes Kleijn},
  and {Valentijn}]{Roy18}
N.~{Roy}, N.~R. {Napolitano}, F.~{La Barbera}, C.~{Tortora}, F.~{Getman},
  M.~{Radovich}, M.~{Capaccioli}, M.~{Brescia}, S.~{Cavuoti}, G.~{Longo}, M.~A.
  {Raj}, E.~{Puddu}, G.~{Covone}, V.~{Amaro}, C.~{Vellucci}, A.~{Grado},
  K.~{Kuijken}, G.~{Verdoes Kleijn}, and E.~{Valentijn}.
\newblock {Evolution of galaxy size-stellar mass relation from the Kilo-Degree
  Survey}.
\newblock \emph{MNRAS}, 480:\penalty0 1057--1080, Oct 2018.
\newblock \doi{10.1093/mnras/sty1917}.

\bibitem[{Ruiz} et~al.(2018){Ruiz}, {Corral}, {Mountrichas}, and
  {Georgantopoulos}]{Ruiz18}
A.~{Ruiz}, A.~{Corral}, G.~{Mountrichas}, and I.~{Georgantopoulos}.
\newblock {XMMPZCAT: A catalogue of photometric redshifts for X-ray sources}.
\newblock \emph{A\&A}, 618:\penalty0 A52, Oct 2018.
\newblock \doi{10.1051/0004-6361/201833117}.

\bibitem[{Sabin} et~al.(2013){Sabin}, {Parker}, {Contreras}, {Olgu{\'\i}n},
  {Frew}, {Stupar}, {V{\'a}zquez}, {Wright}, {Corradi}, and {Morris}]{Sabin13}
L.~{Sabin}, Q.~A. {Parker}, M.~E. {Contreras}, L.~{Olgu{\'\i}n}, D.~J. {Frew},
  M.~{Stupar}, R.~{V{\'a}zquez}, N.~J. {Wright}, R.~L.~M. {Corradi}, and
  R.~A.~H. {Morris}.
\newblock {New Galactic supernova remnants discovered with IPHAS}.
\newblock \emph{MNRAS}, 431\penalty0 (1):\penalty0 279--291, May 2013.
\newblock \doi{10.1093/mnras/stt160}.

\bibitem[{Sadeh} et~al.(2016){Sadeh}, {Abdalla}, and {Lahav}]{Sadeh16}
I.~{Sadeh}, F.~B. {Abdalla}, and O.~{Lahav}.
\newblock {ANNz2: Photometric Redshift and Probability Distribution Function
  Estimation using Machine Learning}.
\newblock \emph{PASP}, 128\penalty0 (968):\penalty0 104502, Oct 2016.
\newblock \doi{10.1088/1538-3873/128/968/104502}.

\bibitem[{Saha} et~al.(2018){Saha}, {Basak}, {Safonova}, {Bora}, {Agrawal},
  {Sarkar}, and {Murthy}]{Saha18}
S.~{Saha}, S.~{Basak}, M.~{Safonova}, K.~{Bora}, S.~{Agrawal}, P.~{Sarkar}, and
  J.~{Murthy}.
\newblock {Theoretical validation of potential habitability via analytical and
  boosted tree methods: An optimistic study on recently discovered exoplanets}.
\newblock \emph{A\&C}, 23:\penalty0 141--150, Apr 2018.
\newblock \doi{10.1016/j.ascom.2018.03.003}.

\bibitem[{Salvato} et~al.(2018){Salvato}, {Ilbert}, and {Hoyle}]{Salvato19}
Mara {Salvato}, Olivier {Ilbert}, and Ben {Hoyle}.
\newblock {The many flavours of photometric redshifts}.
\newblock \emph{Nature Astronomy}, 3:\penalty0 212--222, Jun 2018.
\newblock \doi{10.1038/s41550-018-0478-0}.

\bibitem[{Sarro} et~al.(2018){Sarro}, {Ordieres-Mer{\'e}}, {Bello-Garc{\'\i}a},
  {Gonz{\'a}lez-Marcos}, and {Solano}]{Sarro18}
L.~M. {Sarro}, J.~{Ordieres-Mer{\'e}}, A.~{Bello-Garc{\'\i}a},
  A.~{Gonz{\'a}lez-Marcos}, and E.~{Solano}.
\newblock {Estimates of the atmospheric parameters of M-type stars: a
  machine-learning perspective}.
\newblock \emph{MNRAS}, 476:\penalty0 1120--1139, May 2018.
\newblock \doi{10.1093/mnras/sty165}.

\bibitem[{Schindler} et~al.(2017){Schindler}, {Fan}, {McGreer}, {Yang}, {Wu},
  {Jiang}, and {Green}]{Schindler17}
Jan-Torge {Schindler}, Xiaohui {Fan}, Ian~D. {McGreer}, Qian {Yang}, Jin {Wu},
  Linhua {Jiang}, and Richard {Green}.
\newblock {The Extremely Luminous Quasar Survey in the SDSS Footprint. I.
  Infrared-based Candidate Selection}.
\newblock \emph{ApJ}, 851:\penalty0 13, Dec 2017.
\newblock \doi{10.3847/1538-4357/aa9929}.

\bibitem[{Sedaghat} and {Mahabal}(2018)]{Sedaghat18}
Nima {Sedaghat} and Ashish {Mahabal}.
\newblock {Effective image differencing with convolutional neural networks for
  real-time transient hunting}.
\newblock \emph{MNRAS}, 476:\penalty0 5365--5376, June 2018.
\newblock \doi{10.1093/mnras/sty613}.

\bibitem[Selvaraju et~al.(2017)Selvaraju, Cogswell, Das, Vedantam, Parikh, and
  Batra]{selvarajuGradCAMVisualExplanations2017}
Ramprasaath~R. Selvaraju, Michael Cogswell, Abhishek Das, Ramakrishna Vedantam,
  Devi Parikh, and Dhruv Batra.
\newblock Grad-{{CAM}}: {{Visual Explanations From Deep Networks}} via
  {{Gradient}}-{{Based Localization}}.
\newblock In \emph{Proceedings of the {{IEEE International Conference}} on
  {{Computer Vision}}}, volume 2017-October, 2017.

\bibitem[{Sevilla-Noarbe} and
  {Etayo-Sotos}(2015)]{sevilla-noarbeEffectTrainingCharacteristics2015}
I.~{Sevilla-Noarbe} and P.~{Etayo-Sotos}.
\newblock Effect of training characteristics on object classification: {{An}}
  application using {{Boosted Decision Trees}}.
\newblock \emph{A\&C}, 11:\penalty0 64--72, June 2015.
\newblock ISSN 2213-1337.
\newblock \doi{10.1016/j.ascom.2015.03.010}.

\bibitem[{Shallue} and {Vanderburg}(2018)]{Shallue18}
Christopher~J. {Shallue} and Andrew {Vanderburg}.
\newblock {Identifying Exoplanets with Deep Learning: A Five-planet Resonant
  Chain around Kepler-80 and an Eighth Planet around Kepler-90}.
\newblock \emph{AJ}, 155:\penalty0 94, Feb 2018.
\newblock \doi{10.3847/1538-3881/aa9e09}.

\bibitem[Singh et~al.(1998)Singh, Gulati, and
  Gupta]{singhStellarSpectralClassification1998}
Harinder~P. Singh, Ravi~K. Gulati, and Ranjan Gupta.
\newblock Stellar spectral classification using principal component analysis
  and artificial neural networks.
\newblock \emph{MNRAS}, 295\penalty0 (2):\penalty0 312--318, April 1998.
\newblock ISSN 0035-8711.
\newblock \doi{10.1046/j.1365-8711.1998.01255.x}.

\bibitem[{Smirnov} and {Markov}(2017)]{Smirnov17}
Evgeny~A. {Smirnov} and Alexey~B. {Markov}.
\newblock {Identification of asteroids trapped inside three-body mean motion
  resonances: a machine-learning approach}.
\newblock \emph{MNRAS}, 469:\penalty0 2024--2031, Aug 2017.
\newblock \doi{10.1093/mnras/stx999}.

\bibitem[{Speagle} and {Eisenstein}(2017)]{Speagle17}
Joshua~S. {Speagle} and Daniel~J. {Eisenstein}.
\newblock {Deriving photometric redshifts using fuzzy archetypes and
  self-organizing maps - II. Implementation}.
\newblock \emph{MNRAS}, 469:\penalty0 1205--1224, Jul 2017.
\newblock \doi{10.1093/mnras/stx510}.

\bibitem[{Stensbo-Smidt} et~al.(2017){Stensbo-Smidt}, {Gieseke}, {Igel},
  {Zirm}, and {Steenstrup Pedersen}]{Stensbo-Smidt17}
Kristoffer {Stensbo-Smidt}, Fabian {Gieseke}, Christian {Igel}, Andrew {Zirm},
  and Kim {Steenstrup Pedersen}.
\newblock {Sacrificing information for the greater good: how to select
  photometric bands for optimal accuracy}.
\newblock \emph{MNRAS}, 464\penalty0 (3):\penalty0 2577--2596, Jan 2017.
\newblock \doi{10.1093/mnras/stw2476}.

\bibitem[{Storrie-Lombardi} et~al.(1992){Storrie-Lombardi}, Lahav, Sodr{\'e},
  and
  {Storrie-Lombardi}]{storrie-lombardiMorphologicalClassificationGalaxies1992}
M.~C. {Storrie-Lombardi}, O.~Lahav, L.~Sodr{\'e}, and L.~J. {Storrie-Lombardi}.
\newblock Morphological {{Classification}} of galaxies by {{Artificial Neural
  Networks}}.
\newblock \emph{MNRAS}, 259\penalty0 (1):\penalty0 8P--12P, November 1992.
\newblock ISSN 0035-8711.
\newblock \doi{10.1093/mnras/259.1.8P}.

\bibitem[{Stoughton} and {et al.}(2002)]{Stoughton02}
Chris {Stoughton} and {et al.}
\newblock {Sloan Digital Sky Survey: Early Data Release}.
\newblock \emph{AJ}, 123:\penalty0 485--548, Jan 2002.
\newblock \doi{10.1086/324741}.

\bibitem[{S{\"u}veges} et~al.(2017){S{\"u}veges}, {Barblan},
  {Lecoeur-Ta{\"\i}bi}, {Pr{\v{s}}a}, {Holl}, {Eyer}, {Kochoska}, {Mowlavi},
  and {Rimoldini}]{Suveges17}
M.~{S{\"u}veges}, F.~{Barblan}, I.~{Lecoeur-Ta{\"\i}bi}, A.~{Pr{\v{s}}a},
  B.~{Holl}, L.~{Eyer}, A.~{Kochoska}, N.~{Mowlavi}, and L.~{Rimoldini}.
\newblock {Gaia eclipsing binary and multiple systems. Supervised
  classification and self-organizing maps}.
\newblock \emph{A\&A}, 603:\penalty0 A117, Jul 2017.
\newblock \doi{10.1051/0004-6361/201629710}.

\bibitem[{Szalay} and {Gray}(2001)]{Szalay2001}
Alexander {Szalay} and Jim {Gray}.
\newblock {The World-Wide Telescope}.
\newblock \emph{Science}, 293:\penalty0 2037--2040, Sep 2001.
\newblock \doi{10.1126/science.293.5537.2037}.

\bibitem[Szalay and Gray(2006)]{Szalay-Gray-2006}
Alexander Szalay and Jim Gray.
\newblock 2020 computing: Science in an exponential world.
\newblock \emph{Nature}, 440\penalty0 (7083):\penalty0 413--414, 03 2006.

\bibitem[{Tachibana} and {Miller}(2018)]{Tachibana18}
Yutaro {Tachibana} and A.~A. {Miller}.
\newblock {A Morphological Classification Model to Identify Unresolved
  PanSTARRS1 Sources: Application in the ZTF Real-time Pipeline}.
\newblock \emph{PASP}, 130:\penalty0 128001, Dec 2018.
\newblock \doi{10.1088/1538-3873/aae3d9}.

\bibitem[{Tan} et~al.(2018){Tan}, {Lyon}, {Stappers}, {Cooper}, {Hessels},
  {Kondratiev}, {Michilli}, and {Sanidas}]{Tan18}
C.~M. {Tan}, R.~J. {Lyon}, B.~W. {Stappers}, S.~{Cooper}, J.~W.~T. {Hessels},
  V.~I. {Kondratiev}, D.~{Michilli}, and S.~{Sanidas}.
\newblock {Ensemble candidate classification for the LOTAAS pulsar survey}.
\newblock \emph{MNRAS}, 474:\penalty0 4571--4583, Mar 2018.
\newblock \doi{10.1093/mnras/stx3047}.

\bibitem[{Timlin} et~al.(2018){Timlin}, {Ross}, {Richards}, {Myers},
  {Pellegrino}, {Bauer}, {Lacy}, {Schneider}, {Wollack}, and
  {Zakamska}]{Timlin18}
John~D. {Timlin}, Nicholas~P. {Ross}, Gordon~T. {Richards}, Adam~D. {Myers},
  Andrew {Pellegrino}, Franz~E. {Bauer}, Mark {Lacy}, Donald~P. {Schneider},
  Edward~J. {Wollack}, and Nadia~L. {Zakamska}.
\newblock {The Clustering of High-redshift (2.9 {\ensuremath{\leq}} z
  {\ensuremath{\leq}} 5.1) Quasars in SDSS Stripe 82}.
\newblock \emph{ApJ}, 859:\penalty0 20, May 2018.
\newblock \doi{10.3847/1538-4357/aab9ac}.

\bibitem[{Ucci} et~al.(2017){Ucci}, {Ferrara}, {Gallerani}, and
  {Pallottini}]{Ucci17}
G.~{Ucci}, A.~{Ferrara}, S.~{Gallerani}, and A.~{Pallottini}.
\newblock {Inferring physical properties of galaxies from their emission-line
  spectra}.
\newblock \emph{MNRAS}, 465:\penalty0 1144--1156, Feb 2017.
\newblock \doi{10.1093/mnras/stw2836}.

\bibitem[{Ucci} et~al.(2018){Ucci}, {Ferrara}, {Pallottini}, and
  {Gallerani}]{Ucci18}
G.~{Ucci}, A.~{Ferrara}, A.~{Pallottini}, and S.~{Gallerani}.
\newblock {GAME: GAlaxy Machine learning for Emission lines}.
\newblock \emph{MNRAS}, 477:\penalty0 1484--1494, Jun 2018.
\newblock \doi{10.1093/mnras/sty804}.

\bibitem[Utama and Piekarewicz(2017)]{utamaRefiningMassFormulas2017}
R.~Utama and J.~Piekarewicz.
\newblock Refining mass formulas for astrophysical applications: {{A Bayesian}}
  neural network approach.
\newblock \emph{Phys. Rev. C}, 96\penalty0 (4):\penalty0 044308, October 2017.
\newblock \doi{10.1103/PhysRevC.96.044308}.

\bibitem[{Vafaei Sadr} et~al.(2018){Vafaei Sadr}, {Farhang}, {Movahed},
  {Bassett}, and {Kunz}]{Vafaei18}
A.~{Vafaei Sadr}, M.~{Farhang}, S.~M.~S. {Movahed}, B.~{Bassett}, and
  M.~{Kunz}.
\newblock {Cosmic string detection with tree-based machine learning}.
\newblock \emph{MNRAS}, 478:\penalty0 1132--1140, Jul 2018.
\newblock \doi{10.1093/mnras/sty1055}.

\bibitem[{Vafaei Sadr} et~al.(2019){Vafaei Sadr}, {Vos}, {Bassett}, {Hosenie},
  {Oozeer}, and {Lochner}]{Vafaei19}
A.~{Vafaei Sadr}, Etienne~E. {Vos}, Bruce~A. {Bassett}, Zafiirah {Hosenie},
  N.~{Oozeer}, and Michelle {Lochner}.
\newblock {DEEPSOURCE: point source detection using deep learning}.
\newblock \emph{MNRAS}, 484:\penalty0 2793--2806, Apr 2019.
\newblock \doi{10.1093/mnras/stz131}.

\bibitem[{Valenzuela} and {Pichara}(2018)]{Valenzuela18}
Lucas {Valenzuela} and Karim {Pichara}.
\newblock {Unsupervised classification of variable stars}.
\newblock \emph{MNRAS}, 474:\penalty0 3259--3272, Mar 2018.
\newblock \doi{10.1093/mnras/stx2913}.

\bibitem[van~der Maaten and Hinton(2008)]{maatenVisualizingDataUsing2008}
Laurens van~der Maaten and Geoffrey Hinton.
\newblock Visualizing data using t-{{SNE}}.
\newblock \emph{Journal of machine learning research}, 9\penalty0
  (Nov):\penalty0 2579--2605, 2008.

\bibitem[{van Haarlem} and {et al.}(2013)]{VanHaarlem13}
M.~P. {van Haarlem} and {et al.}
\newblock {LOFAR: The LOw-Frequency ARray}.
\newblock \emph{A\&A}, 556:\penalty0 A2, Aug 2013.
\newblock \doi{10.1051/0004-6361/201220873}.

\bibitem[{van Roestel} et~al.(2018){van Roestel}, {Kupfer}, {Ruiz-Carmona},
  {Groot}, {Prince}, {Burdge}, {Laher}, {Shupe}, and {Bellm}]{vanRoestel18}
J.~{van Roestel}, T.~{Kupfer}, R.~{Ruiz-Carmona}, P.~J. {Groot}, T.~A.
  {Prince}, K.~{Burdge}, R.~{Laher}, D.~L. {Shupe}, and E.~{Bellm}.
\newblock {Discovery of 36 eclipsing EL CVn binaries found by the Palomar
  Transient Factory}.
\newblock \emph{MNRAS}, 475:\penalty0 2560--2590, Apr 2018.
\newblock \doi{10.1093/mnras/stx3291}.

\bibitem[Vanzella et~al.(2004)Vanzella, Cristiani, Fontana, Nonino, Arnouts,
  Giallongo, Grazian, Fasano, Popesso, Saracco, and
  Zaggia]{vanzellaPhotometricRedshiftsMultilayer2004}
E.~Vanzella, S.~Cristiani, A.~Fontana, M.~Nonino, S.~Arnouts, E.~Giallongo,
  A.~Grazian, G.~Fasano, P.~Popesso, P.~Saracco, and S.~Zaggia.
\newblock Photometric redshifts with the {{Multilayer Perceptron Neural
  Network}}: {{Application}} to the {{HDF}}-{{S}} and {{SDSS}}.
\newblock \emph{A\&A}, 423\penalty0 (2):\penalty0 761--776, August 2004.
\newblock ISSN 0004-6361, 1432-0746.
\newblock \doi{10.1051/0004-6361:20040176}.

\bibitem[{Vavilova} et~al.(2018){Vavilova}, {Elyiv}, and
  {Vasylenko}]{Vavilova18}
I.~B. {Vavilova}, A.~A. {Elyiv}, and M.~Yu. {Vasylenko}.
\newblock {Behind the Zone of Avoidance of the Milky Way: what can we Restore
  by Direct and Indirect Methods?}
\newblock \emph{Russian Radio Physics and Radio Astronomy}, 23:\penalty0
  244--257, Dec 2018.
\newblock \doi{10.15407/rpra23.04.244}.

\bibitem[{Vida} and {Roettenbacher}(2018)]{Vida18}
Kriszti{\'a}n {Vida} and Rachael~M. {Roettenbacher}.
\newblock {Finding flares in Kepler data using machine-learning tools}.
\newblock \emph{A\&A}, 616:\penalty0 A163, Sep 2018.
\newblock \doi{10.1051/0004-6361/201833194}.

\bibitem[{Viironen} et~al.(2009){Viironen}, {Mampaso}, {Corradi},
  {Rodr{\'\i}guez}, {Greimel}, {Sabin}, {Sale}, {Unruh}, {Delgado-Inglada},
  {Drew}, {Giammanco}, {Groot}, {Parker}, {Sokoloski}, and
  {Zijlstra}]{Viironen09}
K.~{Viironen}, A.~{Mampaso}, R.~L.~M. {Corradi}, M.~{Rodr{\'\i}guez},
  R.~{Greimel}, L.~{Sabin}, S.~E. {Sale}, Y.~{Unruh}, G.~{Delgado-Inglada},
  J.~E. {Drew}, C.~{Giammanco}, P.~{Groot}, Q.~A. {Parker}, J.~{Sokoloski}, and
  A.~{Zijlstra}.
\newblock {New young planetary nebulae in IPHAS}.
\newblock \emph{A\&A}, 502\penalty0 (1):\penalty0 113--129, Jul 2009.
\newblock \doi{10.1051/0004-6361/200811575}.

\bibitem[Vincent et~al.(2008)Vincent, Larochelle, Bengio, and
  Manzagol]{Vincent08}
Pascal Vincent, Hugo Larochelle, Yoshua Bengio, and Pierre-Antoine Manzagol.
\newblock Extracting and composing robust features with denoising autoencoders.
\newblock In \emph{Proceedings of the 25th International Conference on Machine
  Learning}, ICML '08, pages 1096--1103, New York, NY, USA, 2008. ACM.
\newblock ISBN 978-1-60558-205-4.
\newblock \doi{10.1145/1390156.1390294}.

\bibitem[{Vink} et~al.(2008){Vink}, {Drew}, {Steeghs}, {Wright}, {Martin},
  {G{\"a}nsicke}, {Greimel}, and {Drake}]{Vink08}
Jorick~S. {Vink}, Janet~E. {Drew}, Danny {Steeghs}, Nick~J. {Wright},
  Eduardo~L. {Martin}, Boris~T. {G{\"a}nsicke}, Robert {Greimel}, and Jeremy
  {Drake}.
\newblock {IPHAS discoveries of young stars towards Cyg OB2 and its southern
  periphery}.
\newblock \emph{MNRAS}, 387\penalty0 (1):\penalty0 308--318, Jun 2008.
\newblock \doi{10.1111/j.1365-2966.2008.13220.x}.

\bibitem[Wadadekar(2004)]{wadadekarEstimatingPhotometricRedshifts2004}
Yogesh Wadadekar.
\newblock Estimating {{Photometric Redshifts Using Support Vector Machines}}.
\newblock \emph{PASP}, 117\penalty0 (827):\penalty0 79, December 2004.
\newblock ISSN 1538-3873.
\newblock \doi{10.1086/427710}.

\bibitem[{Walsh} et~al.(2007){Walsh}, {Jerjen}, and {Wiallman}]{Walsh07}
S.~M. {Walsh}, H.~{Jerjen}, and B.~{Wiallman}.
\newblock {A Pair of Bo{\"o}tes: A New Milky Way Satellite}.
\newblock \emph{ApJL}, 662\penalty0 (2):\penalty0 L83--L86, Jun 2007.
\newblock \doi{10.1086/519684}.

\bibitem[{Wan} et~al.(2018){Wan}, {Kafle}, {Lewis}, {Mackey}, {Sharma}, and
  {Ibata}]{Wan18}
Zhen {Wan}, Prajwal~R. {Kafle}, Geraint~F. {Lewis}, Dougal {Mackey}, Sanjib
  {Sharma}, and Rodrigo~A. {Ibata}.
\newblock {Galactic cartography with SkyMapper - I. Population substructure and
  the stellar number density of the inner halo}.
\newblock \emph{MNRAS}, 480:\penalty0 1218--1228, Oct 2018.
\newblock \doi{10.1093/mnras/sty1880}.

\bibitem[{Wang} et~al.(2017){Wang}, {Guo}, and {Luo}]{Wang17}
Ke~{Wang}, Ping {Guo}, and A.~Li {Luo}.
\newblock {A new automated spectral feature extraction method and its
  application in spectral classification and defective spectra recovery}.
\newblock \emph{MNRAS}, 465:\penalty0 4311--4324, Mar 2017.
\newblock \doi{10.1093/mnras/stw2894}.

\bibitem[{Way} et~al.(2012){Way}, {Scargle}, {Ali}, and {Srivastava}]{Way2012}
Michael~J. {Way}, Jeffrey~D. {Scargle}, Kamal~M. {Ali}, and Ashok~N.
  {Srivastava}.
\newblock \emph{{Advances in Machine Learning and Data Mining for Astronomy}}.
\newblock 2012.

\bibitem[Weir et~al.(1995)Weir, Fayyad, Djorgovski, and
  Roden]{weirSKICATSystemProcessing1995}
Nicholas Weir, Usama~M. Fayyad, S.~G. Djorgovski, and Joseph Roden.
\newblock The {{SKICAT System}} for {{Processing}} and {{Analyzing Digital
  Imaging Sky Surveys}}.
\newblock \emph{PASP}, 107:\penalty0 1243, December 1995.
\newblock \doi{10.1086/133683}.
\newblock URL
  \url{https://ui.adsabs.harvard.edu/abs/1995PASP..107.1243W/abstract}.

\bibitem[Whitmore(1984)]{whitmoreObjectiveClassificationSystem1984}
B.~C. Whitmore.
\newblock An objective classification system for spiral galaxies. {{I}}.
  {{The}} two dominant dimensions.
\newblock \emph{ApJ}, 278:\penalty0 61, March 1984.
\newblock ISSN 0004-637X.
\newblock \doi{10.1086/161768}.

\bibitem[{Wils} et~al.(2010){Wils}, {G{\"a}nsicke}, {Drake}, and
  {Southworth}]{Wils10}
Patrick {Wils}, Boris~T. {G{\"a}nsicke}, Andrew~J. {Drake}, and John
  {Southworth}.
\newblock {Data mining for dwarf novae in SDSS, GALEX and astrometric
  catalogues}.
\newblock \emph{MNRAS}, 402\penalty0 (1):\penalty0 436--446, Feb 2010.
\newblock \doi{10.1111/j.1365-2966.2009.15894.x}.

\bibitem[{Wo{\'z}niak} et~al.(2004){Wo{\'z}niak}, {Williams}, {Vestrand}, and
  {Gupta}]{p.r.wozniakIdentifyingRedVariables2004}
P.~R. {Wo{\'z}niak}, S.~J. {Williams}, W.~T. {Vestrand}, and V.~{Gupta}.
\newblock {Identifying Red Variables in the Northern Sky Variability Survey}.
\newblock \emph{AJ}, 128\penalty0 (6):\penalty0 2965--2976, Dec 2004.
\newblock \doi{10.1086/425526}.

\bibitem[{Xin} et~al.(2017){Xin}, {Di}, {Wang}, {Wan}, and {Yue}]{Xin17}
Xin {Xin}, Kaichang {Di}, Yexin {Wang}, Wenhui {Wan}, and Zongyu {Yue}.
\newblock {Automated detection of new impact sites on Martian surface from
  HiRISE images}.
\newblock \emph{Advances in Space Research}, 60:\penalty0 1557--1569, Oct 2017.
\newblock \doi{10.1016/j.asr.2017.06.044}.

\bibitem[{Xu} and {Offner}(2017)]{Xu17a}
Duo {Xu} and Stella S.~R. {Offner}.
\newblock {Assessing the Performance of a Machine Learning Algorithm in
  Identifying Bubbles in Dust Emission}.
\newblock \emph{ApJ}, 851:\penalty0 149, Dec 2017.
\newblock \doi{10.3847/1538-4357/aa9a42}.

\bibitem[{Yan} et~al.(2018){Yan}, {Xu}, {Walsh}, {Macquart}, {MacLeod},
  {Zhang}, {Hancock}, {Chen}, and {Tang}]{Yan18}
Qing-Zeng {Yan}, Ye~{Xu}, A.~J. {Walsh}, J.~P. {Macquart}, G.~C. {MacLeod},
  Bo~{Zhang}, P.~J. {Hancock}, Xi~{Chen}, and Zheng-Hong {Tang}.
\newblock {Improved selection criteria for H II regions, based on IRAS
  sources}.
\newblock \emph{MNRAS}, 476:\penalty0 3981--3990, May 2018.
\newblock \doi{10.1093/mnras/sty518}.

\bibitem[Yang and Li(2015)]{yangAutoencoderStellarSpectra2015}
Tan Yang and Xiangru Li.
\newblock An autoencoder of stellar spectra and its application in
  automatically estimating atmospheric parameters.
\newblock \emph{MNRAS}, 452\penalty0 (1):\penalty0 158--168, September 2015.
\newblock ISSN 0035-8711.
\newblock \doi{10.1093/mnras/stv1210}.

\bibitem[{Yang} et~al.(2018){Yang}, {Yang}, {Bai}, {Zhou}, {Feng}, and
  {Liang}]{Yang18}
Yunfei {Yang}, Hongjuan {Yang}, Xianyong {Bai}, Huituan {Zhou}, Song {Feng},
  and Bo~{Liang}.
\newblock {Automatic Detection of Sunspots on Full-disk Solar Images using the
  Simulated Annealing Genetic Method}.
\newblock \emph{PASP}, 130:\penalty0 104503, Oct 2018.
\newblock \doi{10.1088/1538-3873/aadbfa}.

\bibitem[{Yeakel} et~al.(2018){Yeakel}, {Vandegriff}, {Mitchell}, {Hamilton},
  {Jackman}, {Delamere}, and {Roussos}]{Yeakel18}
K.~{Yeakel}, J.~D. {Vandegriff}, D.~G. {Mitchell}, D.~C. {Hamilton}, C.~M.
  {Jackman}, P.~A. {Delamere}, and E.~{Roussos}.
\newblock {Automatic Detection of Magnetospheric Regions around Saturn using
  Cassini Data}.
\newblock In \emph{AGU Fall Meeting Abstracts}, volume 2018, pages P41D--3769,
  Dec 2018.

\bibitem[Yip et~al.(2004)Yip, Connolly, Berk, Ma, Frieman, SubbaRao, Szalay,
  Richards, Hall, Schneider, Hopkins, Trump, and
  Brinkmann]{yipSpectralClassificationQuasars2004}
C.~W. Yip, A.~J. Connolly, D.~E.~Vanden Berk, Z.~Ma, J.~A. Frieman,
  M.~SubbaRao, A.~S. Szalay, G.~T. Richards, P.~B. Hall, D.~P. Schneider, A.~M.
  Hopkins, J.~Trump, and J.~Brinkmann.
\newblock Spectral {{Classification}} of {{Quasars}} in the {{Sloan Digital Sky
  Survey}}: {{Eigenspectra}}, {{Redshift}}, and {{Luminosity Effects}}.
\newblock \emph{AJ}, 128\penalty0 (6):\penalty0 2603, December 2004.
\newblock ISSN 1538-3881.
\newblock \doi{10.1086/425626}.

\bibitem[{Yong} et~al.(2018){Yong}, {King}, {Webster}, {Bate}, {O'Dowd}, and
  {Labrie}]{Yong18}
Suk~Yee {Yong}, Anthea~L. {King}, Rachel~L. {Webster}, Nicholas~F. {Bate},
  Matthew~J. {O'Dowd}, and Kathleen {Labrie}.
\newblock {Using the Properties of Broad Absorption Line Quasars to Illuminate
  Quasar Structure}.
\newblock \emph{MNRAS}, 479:\penalty0 4153--4171, Sep 2018.
\newblock \doi{10.1093/mnras/sty1540}.

\bibitem[{York} and {et al.}(2000)]{York00}
Donald~G. {York} and {et al.}
\newblock {The Sloan Digital Sky Survey: Technical Summary}.
\newblock \emph{AJ}, 120:\penalty0 1579--1587, Sep 2000.
\newblock \doi{10.1086/301513}.

\bibitem[{Zevin} et~al.(2017){Zevin}, {Coughlin}, {Bahaadini}, {Besler},
  {Rohani}, {Allen}, {Cabero}, {Crowston}, {Katsaggelos}, {Larson}, {Lee},
  {Lintott}, {Littenberg}, {Lundgren}, {{\O}sterlund}, {Smith}, {Trouille}, and
  {Kalogera}]{Zevin17}
M.~{Zevin}, S.~{Coughlin}, S.~{Bahaadini}, E.~{Besler}, N.~{Rohani},
  S.~{Allen}, M.~{Cabero}, K.~{Crowston}, A.~K. {Katsaggelos}, S.~L. {Larson},
  T.~K. {Lee}, C.~{Lintott}, T.~B. {Littenberg}, A.~{Lundgren},
  C.~{{\O}sterlund}, J.~R. {Smith}, L.~{Trouille}, and V.~{Kalogera}.
\newblock {Gravity Spy: integrating advanced LIGO detector characterization,
  machine learning, and citizen science}.
\newblock \emph{Classical and Quantum Gravity}, 34:\penalty0 064003, Mar 2017.
\newblock \doi{10.1088/1361-6382/aa5cea}.

\bibitem[{Zhang} et~al.(2018){Zhang}, {Zhang}, and {Zhao}]{Zhang18}
Jingyi {Zhang}, Yanxia {Zhang}, and Yongheng {Zhao}.
\newblock {Imbalanced Learning for RR Lyrae Stars Based on SDSS and GALEX
  Databases}.
\newblock \emph{AJ}, 155:\penalty0 108, Mar 2018.
\newblock \doi{10.3847/1538-3881/aaa5b1}.

\bibitem[{Zhang} and {Zhao}(2015)]{Zhang15}
Yanxia {Zhang} and Yongheng {Zhao}.
\newblock {Astronomy in the Big Data Era}.
\newblock \emph{Data Science Journal}, 14:\penalty0 11, May 2015.
\newblock \doi{10.5334/dsj-2015-011}.

\bibitem[{Zhao} et~al.(2018){Zhao}, {Peng}, {Wang}, {Qiao}, {Guo}, {Xiao}, and
  {Wang}]{Zhao18}
Hao {Zhao}, Wen-Xi {Peng}, Huan-Yu {Wang}, Rui {Qiao}, Dong-Ya {Guo}, Hong
  {Xiao}, and Zhao-Min {Wang}.
\newblock {A machine learning method to separate cosmic ray electrons from
  protons from 10 to 100 GeV using DAMPE data}.
\newblock \emph{Research in Astronomy and Astrophysics}, 18:\penalty0 071, Jun
  2018.
\newblock \doi{10.1088/1674-4527/18/6/71}.

\end{thebibliography}



\end{document}